\documentclass{rescience}

\usepackage{caption}
\usepackage{subcaption}
\usepackage{graphicx}
\usepackage{multirow}
\usepackage[capitalize,noabbrev]{cleveref}

\graphicspath{{figures/}}

\begin{document}

\renewcommand{\arraystretch}{1.2}


\begin{tikzpicture}[remember picture, overlay]
  \node [shape=rectangle, draw=darkred, fill=darkred, yshift=-34mm,
        anchor=north west, minimum width=3.75cm, minimum height=10mm]
        at (current page.north west) {};
  \node [text=white, anchor=center, yshift=-39mm, xshift=1.875cm]
         at (current page.north west) {\small 
         RESCIENCE C};
\end{tikzpicture}

{\let\newpage\relax\maketitle} \maketitle

\marginnote{
  \footnotesize \sffamily
  \textbf{Edited by}\\
  \ifdefempty{\editorNAME}{\textcolor{darkgray}{(Editor)}}
             {\editorNAME\ifdefempty{\editorORCID}{}{$^{\orcid{\editorORCID}}$}}\\
  ~\\
  \ifdefempty{\reviewerINAME}{}
  {
  \textbf{Reviewed by}\\
  \ifdefempty{\reviewerINAME}{\textcolor{darkgray}{}}
             {\reviewerINAME\ifdefempty{\reviewerIORCID}{}{$^{\orcid{\reviewerIORCID}}$}\\}
  \ifdefempty{\reviewerIINAME}{\textcolor{darkgray}{}}
             {\reviewerIINAME\ifdefempty{\reviewerIIORCID}{}{$^{\orcid{\reviewerIIORCID}}$}\\}
  ~\\
  }
  \textbf{Received}\\
  \ifdefempty{\dateRECEIVED}{---}{\dateRECEIVED}\\
  ~\\
  \textbf{Published}\\
  \ifdefempty{\datePUBLISHED}{---}{\datePUBLISHED}\\
  ~\\
  \textbf{DOI}\\
  \ifdefempty{\articleDOI}{---}{\articleDOI}
}

\newcommand{\container}[1]{\def\@container{#1}}
\begin{container}
  \afterpage {
    \begin{statement}
      \scriptsize \sffamily
      \hrule \vskip .5em
      Copyright © \articleYEAR~\authorsABBRV,
      released under a Creative Commons Attribution 4.0 International license.
      
      Correspondence should be addressed to
      \contactNAME~(\href{mailto:\contactEMAIL}{\contactEMAIL})
  
      The authors have declared that no competing interests exist.

      \ifdefempty{\codeURL}{}
      {Code is available at
      \href{\codeURL}{\detokenize\expandafter{\codeURL}}\ifdefempty{\codeDOI}{.}{ -- DOI \doi{\codeDOI}.}\ifdefempty{\codeSWH}{.}{ -- SWH  \href{https://archive.softwareheritage.org/\codeSWH/}{\detokenize\expandafter{\codeSWH}}.}}

      \ifdefempty{\dataURL}{}
      {Data is available at
      \href{\dataURL}{\detokenize\expandafter{\dataURL}}\ifdefempty{\dataDOI}{.}{ -- DOI \doi{\dataDOI}.}}

      \ifdefempty{\reviewURL}{}
     {Open peer review is available at \href{\reviewURL}{\detokenize\expandafter{\reviewURL}}.}
    \end{statement}
  }
\end{container}

\ifdefempty{\articleABSTRACT}{}{
  {\small \sffamily \textbf{Abstract} \articleABSTRACT \par}
}


\section{Introduction}

Reproducibility and replicability are essential for the progress of science.
In 2017, the US Congress directed the National Science Foundation to contract with the National Academies of Sciences, Engineering, and Medicine (NASEM) to assess reproducibility and replicability in scientific and engineering research.
NASEM published a consensus report\supercite{nasem_2019} in May 2019, containing findings and recommendations to improve rigor and transparency in research.
The report also provides clear definitions of ``reproducibility'' and ``replicability'', intended to apply across all fields of science:

\begin{itemize}
  \item[] \textbf{Reproducibility} is obtaining consistent results using the same input data; computational steps, methods, and code; and conditions of analysis.
  \item[] \textbf{Replicability} is obtaining consistent results across studies aimed at answering the same scientific question, each of which had obtained its own data. Two studies may be considered to have replicated if they obtain consistent results given the level of uncertainty inherent in the system under study.
\end{itemize}

The minimum requirement for computational research to be reproducible is to make code and data available to others.
\citet{peng_2011} introduced the concept of a reproducibility spectrum, in which reproducible research is a ``minimum standard for judging scientific claims when full independent replication of a study is not possible'' or not available.
The two extremes on the reproducibility spectrum are ``not reproducible'' (when a published manuscript is the sole deliverable from a study) and ``fully replicated'' (the gold standard for a study).

This paper addresses reproducibility and replicability in computational fluid dynamics, a mature field and one of the oldest branches of computational science. 
At the center of this field are the Navier-Stokes equations, which are notoriously difficult to solve numerically, with computational experiments often taking a long time, even on parallel compute clusters.
Both this difficulty and the history of the field (with early progress done in secret defense laboratories) contribute to rather poor standards of reproducibility.
Research results are regularly communicated via published articles without accompanying software or data.

In the past, we have undertaken a replication of results from our own research group on unsteady fluid dynamics\supercite{krishnan_et_al_2014} and published the outputs and lessons learned from this exercise.\supercite{mesnard_barba_2017}
We aim here to assess the effort needed to replicate the computational results from another research group and set our sights on a computational fluid dynamics study from Li and Dong\supercite{li_dong_2016} that investigated the dynamics of pitching and rolling wings.
While many prior studies have focused on the pitching and/or heaving motion of three-dimensional wings, only few studies have looked at the combined rolling and pitching motion.
As explained by \citet{li_dong_2016}, the pitching-rolling kinematics has the potential to serve as a better canonical model for the hydrodynamics of bio-inspired flapping propulsors.
The authors carried out a parametric study, using their own research code, to quantify the effects of the Reynolds number, Strouhal number, aspect ratio, and rolling/pitching phase difference on the wake topology and the propulsive performance of flapping wings.
\cref{tab:parameters} lists the parameters and values considered in the original study.
The deliverable from this study was the journal publication itself; the computational code, input data, and conditions of analysis used to produce the numerical results were not made available.
Thus, referring back to Peng's reproducibility spectrum, we consider the study to be not reproducible.
Our objective was to replicate the scientific findings claimed in the original study and to do it in a reproducible way.
We have re-implemented the three-dimensional rolling and pitching kinematics in an open-source code shared on GitHub\footnote{PetIBM-rollingpitching: \url{github.com/barbagroup/petibm-rollingpitching}} and prepared extensive reproducibility packages for all results.

\begin{table}
  \centering
  \begin{tabular}{ll}
    \hline\hline
    Parameter & Values \\
    \hline
    Wing aspect ratio $AR$ & $1.27$, $1.91$, $2.55$ \\
    Reynolds number $Re$ & $100$, $200$, $400$ \\
    Strouhal number $St$ & $0.4$, $0.6$, $0.8$, $1.0$, $1.2$ \\
    Rolling amplitude $A_\phi$ & $45^o$ \\
    Pitching amplitude $A_\theta$ & $45^o$ \\
    Phase-difference angle $\psi$ & $60^o$, $70^o$, $80^o$, $90^o$, $100^o$, $110^o$, $120^o$ \\
    \hline\hline
  \end{tabular}
  \caption{Parameter values used in \citet{li_dong_2016} and in the present replication study.}
  \label{tab:parameters}
\end{table}

\section{Numerical methods and problem setup}

The original study modeled the wing kinematics with an elliptical disk that undergoes a rolling motion around the streamwise $x$-axis and a pitching motion around its spanwise axis.
The wing is characterized by the chord length $c$ and the spanwise length $S$.
The aspect ratio of the wing is given by $AR = S^2 / A_\text{plan}$, where $A_\text{plan}$ is the planform area of the plate ($A_\text{plan} = \pi c S / 4$).

The rolling motion is defined by the instantaneous rolling position:

\begin{equation}
  \phi (t) = -A_\phi \cos \left( 2 \pi f t \right)
\end{equation}

where $t$ is the time, $f$ is the flapping frequency, and $A_\phi$ is the rolling amplitude.

The pitching motion along the spanwise axis is governed by the instantaneous pitching position:

\begin{equation}
  \theta (t) = -A_\theta \cos(2 \pi f t + \psi)
\end{equation}

where $A_\theta$ is the pitching amplitude and $\psi$ is the phase-difference angle between the pitching and rolling motions.

For the present replication study, we use the same wing kinematics and numerically solve the three-dimensional Navier-Stokes equations (velocity/pressure formulation) for an incompressible viscous flow.
The Reynolds number is defined as $Re = \frac{U_\infty c}{\nu}$, where $U_\infty$ is the incoming freestream speed and $\nu$ is the kinematic viscosity.
The convective and diffusion terms of the partial differential equations are time-integrated using second-order accurate Adams-Bashforth and Crank-Nicolson methods, respectively.
We enforce a Dirichlet condition (streamwise velocity set to the freestream speed $U_\infty$) on all boundaries, except at the outlet where we use a convective boundary condition (to carry vortical structures outside the computational domain).

Our code base, PetIBM, solves the incompressible Navier-Stokes equations using a projection method, seen as an approximate block-LU decomposition of the fully discretized equations.\supercite{perot_1993}
To compute the flow around a moving object (e.g., a pitching-rolling wing), we use an immersed boundary technique.
The fluid equations are solved over an extended domain that includes the interior of the immersed object.
The boundary of the object is represented by a collection of Lagrangian markers that moves with a prescribed rigid kinematics and on which we enforce a no-slip condition.
The presence of the body in the domain is taken into account by modifying the fluid equations in the vicinity of its boundary.
This approach enables us to solve the equations on a simple fixed structured Cartesian grid.

Different near-boundary treatments lead to different immersed boundary methods.
The original study used a sharp-interface method with a ghost-cell methodology.\supercite{mittal_et_al_2008}
PetIBM employs regularized delta functions to transfer data between the Lagrangian markers and the Eulerian grid points (on which the fluid equations are solved).
Our code base includes several implementations of the immersed-boundary projection method;\supercite{taira_colonius_2007} we use the formulation of \citet{li_et_al_2016} for all computations of the present study.
These methods fall into the category of diffuse-interface methods, as the discrete delta function smears the solution over a few grid cells around the boundary.

Each time step, we successively solve three linear systems for an intermediate velocity field, the Lagrangian forces, and the pressure field.
The system for the velocity is solved using a stabilized bi-conjugate gradient method (from the PETSc library) with a Jacobi preconditioner and a convergence criterion based on the absolute $L_2$-norm of the residual set to $atol = 10^{-6}$.
We solve the system for the Lagrangian forces with a direct solver (SuperLU\_dist library).
The pressure Poisson system is solved with a conjugate-gradient method using a classical algebraic multigrid technique (via the NVIDIA AmgX library); here, too, convergence is reached when the absolute $L_2$-norm of the residual is $10^{-6}$.

To quantify aerodynamic performance of the wing, we report the thrust, lift, and spanwise force coefficients, defined as

\begin{equation}
  C_{T, L, Z} = \frac{\left( T, L, Z \right)}{\frac{1}{2} \rho U_\infty^2 A_\text{plan}},
\end{equation}

where $T$, $L$, and $Z$ are the thrust, lift, and spanwise forces, obtained by integrating the $x$, $y$, and $z$ components of the Lagrangian forces along the surface of the immersed boundary.
The forces are a primary output variable of PetIBM.

Following the original study, we also define the propulsive efficiency of the wing:

\begin{equation}
  \eta = \frac{\overline{T} U_\infty}{\overline{P}},
\end{equation}

where $P$ is the hydrodynamic power, and the overline symbol represents a cycle-averaged quantity.
(In line with the original study, we only consider the positive power to compute the average.)

While \citet{li_dong_2016} do not report how they computed the hydrodynamic power, we feel the need to elaborate on it in the context of a diffuse-interface immersed boundary.
The regularized delta kernel smears the solution near the body.
It means that the solution within the support of the kernel may contain spurious artifacts.
To avoid this problem, the surface pressure is defined as the fluid pressure interpolated at a distance of $3\%$ of the chord length along the wing normal on each side of the flat surface.
In other words, the hydrodynamic power is defined as the integral over an expanded surface of the inner product between the interpolated fluid pressure and the body velocity.

\section{Reproducible computational workflow}

The final product of the original study is a published manuscript in the journal Physics of Fluids.
Although the manuscript is well detailed, the code and input data used to produce the computational results were not made publicly available by the authors.
In that regard, we consider the study to not be reproducible.
Thus, we aim to replicate the scientific findings claimed in the original study with our own research software stack and deliver reproducible results.

PetIBM\supercite{chuang_et_al_2018} is developed in the open under the permissive (non-copyleft) 3-Clause BSD license, version-controlled with Git, and hosted on a public GitHub repository.\footnote{PetIBM: \url{github.com/barbagroup/petibm}}
Each major release of the software is archived on the data repository Zenodo.

Our implementation of the three-dimensional rolling and pitching wing, which relies on PetIBM, is also open source and available on GitHub\footnote{PetIBM-rollingpitching: \url{github.com/barbagroup/petibm-rollingpitching}} under the same license.
The repository contains all input data and processing scripts that were used to produce the computational results reported in the next section.
This allows anyone to inspect the code, to verify the steps that were taken to produce computational results, and to modify and re-use it for other applications.
The repository also includes \textsc{README} files to guide readers that may be interested in re-running the analysis.
Upon submission of the present manuscript, the application repository, as well as the data needed to reproduce the figures, have been archived on Zenodo.

We leveraged our University high-performance-computing (HPC) cluster, called Pegasus, to run all simulations reported here.
(We used computational nodes with Dual 20-Core 3.70GHz Intel Xeon Gold 6148 processors and NVIDIA V100 GPU devices.)
To reduce the burden of building PetIBM and its applications on the cluster, we used the container technology from Docker\supercite{merkel_2014} and Singularity.\supercite{kurtzer_et_al_2017}
Containers allow us to capture the conditions of analysis in a formatted image that can be shared with others.
We have already used Docker containers in the past to create a reproducible workflow for scientific applications on the public cloud provider Microsoft Azure.\supercite{mesnard_barba_2020}
Here, we aim to adopt a similar workflow on our local HPC cluster.
Early in this replication study, we hit a snag: Docker is not available to users on Pegasus.
Indeed, Docker is not available at most HPC centers for security reasons.
Submitting container-based jobs with Docker implies running a Docker daemon (a background process) that requires root privileges that users do not and should not have on shared production clusters.
Thus, we decided to leverage the Singularity container technology to conduct the replication study on Pegasus.
Singularity is more recent than Docker, was designed from the ground up to prevent escalation of user privileges, and is compatible with Docker images.

Our reproducible workflow starts with creating a Docker image that installs PetIBM and its applications, as well as all their dependencies.
We then push the image to a public registry on DockerHub.\footnote{DockerHub registry: \url{hub.docker.com/repository/docker/mesnardo/petibm-rollingpitching}}
Anyone interested in using the application code can now pull the image from the registry and spin up a Docker container to get a faithfully reproduced computational environment.
Next, we use the cloud service Singularity Hub to build a Singularity image\footnote{Singularity Hub registry: \url{singularity-hub.org/collections/2855}} out of the Docker image.
We finally pull the Singularity image on the cluster where we run container-based jobs.

\section{Results}

Following the original study, we set out to replicate the investigation of the wake topology and aerodynamic performance of low-aspect-ratio wings undergoing a pitching-rolling motion.
We conducted the same parametric study to quantify the effect of the Reynolds number, Strouhal number, wing's aspect ratio, and phase-difference angle (between the rolling and pitching motions).
These parameters govern the near-body and wake topology of the flow, and thus impact the performance of the modeled wing.
We initiated this replication exercise with the hope to confirm the general trends of the original results, while using a different immersed-boundary solver and modeling the wing with a flat plate (instead of disk with non-zero thickness).
In this section, we first present a grid-independence study, which allows us to freeze some simulation parameters (such as the grid size or the time-step size) for the remainder of the parametric study.
As in the original investigation, we report the results of the baseline case (for a circular plate) and assess the effect of the flow parameters.
We then present the results from the baseline case (for a circular wing) and look at the effect of the viscosity, geometry, and kinematics on the wake topology and aerodynamic performances of the wing.

\subsection{Grid-independence study}

In the original study, the authors reported the results of a grid-independence study to justify the spatial and temporal grid resolutions used for the parametric study.
They compared force coefficients, profiles of the velocity components, profiles of the fluctuating kinetic energy, and distances between vortical structures in the near wake, obtained with different grid resolutions.
Here, we also report the results of our grid-independence study before moving on to the results of the parametric study.

We use the same domain size as in the original study: $30c \times 25c \times 25c$ (where $c$ is the chord length of the wing).
The root of the wing (around which the plate undergoes the rolling/pitching motion) is located at the center of the computational domain.
We keep the spatial grid uniform (with highest resolution) in the sub-area of the domain that covers the motion of the wing.
Outside this area, we also add an extra uniform layer with grid-spacing size $\Delta x = 0.05c$, in the sub-domain $\left[ -2c, 6c \right] \times \left[ -3c, 3c \right] \times \left[ -1c, 2c \right]$, which covers the near-wake region.
(We opted for a smooth transition between the two uniform regions, in which the grid-cell widths are stretched with a constant ratio of $1.1$ in all directions, except in the streamwise direction behind the wing where we used a ratio of $1.03$.)
Finally, the grid-cell width is stretched to the external boundaries with a constant ratio of $1.2$.
To the readers interested in further inspecting the geometric characteristics of the grids used in the present study: we used Python scripts to codify the grid parameters and saved them into PetIBM-readable \texttt{yaml} files (available on the GitHub repository).

In the present study, we model the wing with a flat elliptical surface, discretized with Lagrangian markers uniformly distributed on its surface (with a similar resolution as the grid-spacing size of the background Eulerian grid).

As in the original study, we consider the case of a circular wing ($AR = 1.27$) with Reynolds number $Re = 200$, Strouhal number $St = 0.6$, and phase-difference angle $\psi = 90^o$, to assess independence in the numerical results.
We investigated the effect of the grid-spacing size, the time-step size, and the convergence criterion of the iterative solvers, on the numerical solution.

\begin{table}[!h]
  \centering
  \begin{tabular}{lcccc}
    \hline\hline
    Grid & $(\Delta x / c)_\text{min}$ & $n_x \times n_y \times n_z$ & \# grid cells ($\times 10^6$) & \# body markers \\
    \hline
    Coarse & $0.03$ & $226 \times 185 \times 122$ & $5.1$ & $1005$ \\
    Nominal & $0.01$ & $331 \times 302 \times 211$ & $21.1$ & $9042$ \\
    Fine & $0.005$ & $467 \times 465 \times 333$ & $72.3$ & $36158$ \\
    \hline\hline
  \end{tabular}
  \caption{Characteristics of the computational grids used for in the grid-independence study. Body markers were uniformly distributed on the surface of the wing using a similar resolution as the background fluid grid.}
  \label{tab:independence_grid_charateristics}
\end{table}

\begin{figure}[!h]
  \centering
  \includegraphics[width=\textwidth]{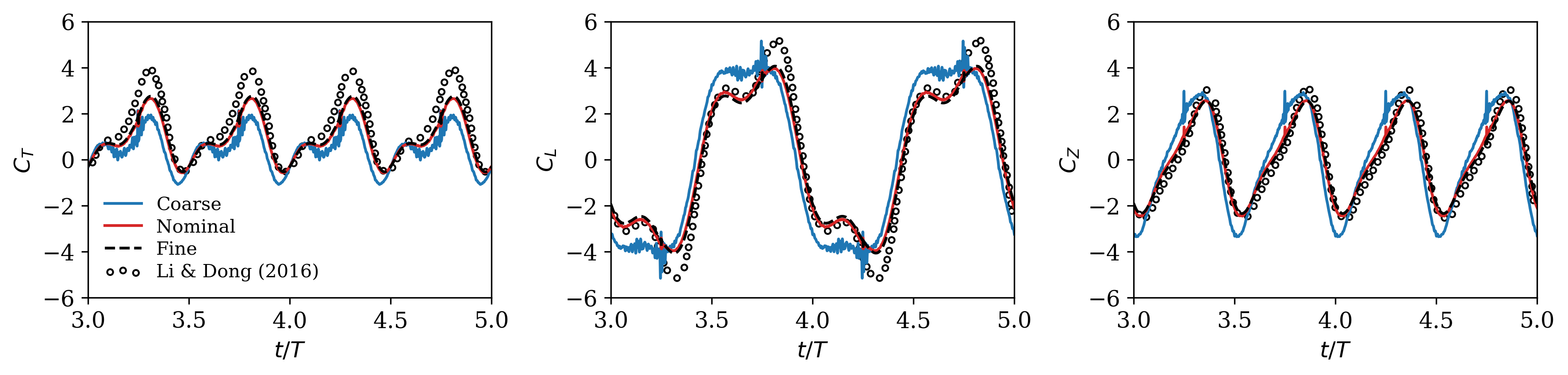}
  \caption{History of the thrust ($C_T$), lift ($C_L$), and spanwise ($C_Z$) coefficients over two flapping cycles of a circular plate ($AR = 1.27$) at Reynolds number $Re = 200$ and Strouhal number $St = 0.6$. We compare the solution obtained with PetIBM on three spatial grids: coarse ($\Delta x = 0.03c$), nominal ($\Delta x = 0.01c$), and fine ($\Delta x = 0.005c$). We also report the force coefficients published in \citet{li_dong_2016}.}
  \label{fig:independence_force_coefficients_dx}
\end{figure}

To assess the effect of the grid-spacing size $\Delta x$ in the vicinity of the wing on the solution, we computed five flapping cycles on three grids: coarse ($\Delta x = 0.03c$), nominal ($\Delta x = 0.01c$) and fine ($\Delta x = 0.005c$).
\cref{tab:independence_grid_charateristics} reports characteristics of the spatial grids used for the independence study.
Note that the ``nominal'' grid in the original study contains $\sim 6.2 \times 10^6$ cells with smallest grid spacing $\Delta x = 0.03c$ near the wing.
This is significantly coarser than our nominal grid, which contains $\sim 21.1 \times 10^6$ cells with $\Delta x = 0.01c$ in the vicinity of the wing.
As seen in \cref{fig:independence_force_coefficients_dx}, the computed aerodynamic forces on our coarse mesh (similar resolution as their nominal grid) contains spurious artifacts.
Spurious force oscillations exist in almost all immersed-boundary methods, although their magnitudes depend on the approach used.
We reduced the noise by increasing the grid resolution near the wing.

\begin{table}[!h]
  \centering
  \begin{tabular}{lcccc}
    \hline\hline
    Case & $\overline{C_T}$ & $\left( C_L \right)_\text{r.m.s.}$ & $\left( C_Z \right)_\text{r.m.s.}$ & $\eta$ \\
    \hline
    Li \& Dong (2016) & $1.462$ & $3.417$ & $1.912$ & $0.1927$ \\
    Coarse grid ($\Delta x = 0.03c$) & $0.442$ & $3.323$ & $2.097$ & $0.1002$ \\
    Nominal grid ($\Delta x = 0.01c$) & $0.914$ & $2.852$ & $1.650$ & $0.1635$ \\
    Fine grid ($\Delta x = 0.005c$) & $0.986$ & $2.817$ & $1.600$ & $0.1727$ \\
    Disk ($3\%$ thickness) & $0.762$ & $3.013$ & $1.885$ & $-$ \\
    $1000$ steps/cycle & $0.919$ & $2.853$ & $1.643$ & $0.1636$ \\
    Tighter solvers ($atol = 10^{-9}$) & $0.914$ & $2.852$ & $1.650$ & $0.1635$ \\
    \hline\hline
  \end{tabular}
  \caption{Results of the grid-independence study for a flapping wing with $AR = 1.27$, $St = 0.6$, $Re = 200$, and $\psi = 90^o$. We also report the results from \citet{li_dong_2016} (data digitized from figures of the original publication).}
  \label{tab:independence_results}
\end{table}

\cref{fig:independence_force_coefficients_dx} shows the history of the thrust, lift, and spanwise coefficients over two flapping cycles ($2000$ time steps per cycle), obtained on the coarse, nominal, and fine spatial grids.
When using a similar grid resolution as in the original study ($\Delta x = 0.03c$), the instantaneous force components are noisy and we do not clearly distinguish two peaks in the lift force each half cycle.
(Authors of the original study reported two peaks in both the thrust and lift forces, every half cycle.)
Spurious noise can be reduced by refining the spatial grid.
The history of the forces is visually similar for the nominal and fine grids, with the presence of two peaks in the thrust and lift.
\cref{tab:independence_results} reports hydrodynamic quantities and propulsive efficiency obtained from the simulations computed during the independence study.
Refining the computational grid leads to a relative difference of $7.9\%$ in the mean thrust coefficient, $1.2\%$ in the r.m.s. value for the lift coefficient, and $3\%$ in the r.m.s. value of spanwise coefficients.
Compared to \citet{li_dong_2016}, we obtain lower magnitude in the peaks of the force coefficients.
Note that we used the nominal spatial grid ($\Delta x = 0.01c$) for the replication of the parametric study.

\begin{figure}[!h]
  \centering
  \includegraphics[width=\textwidth]{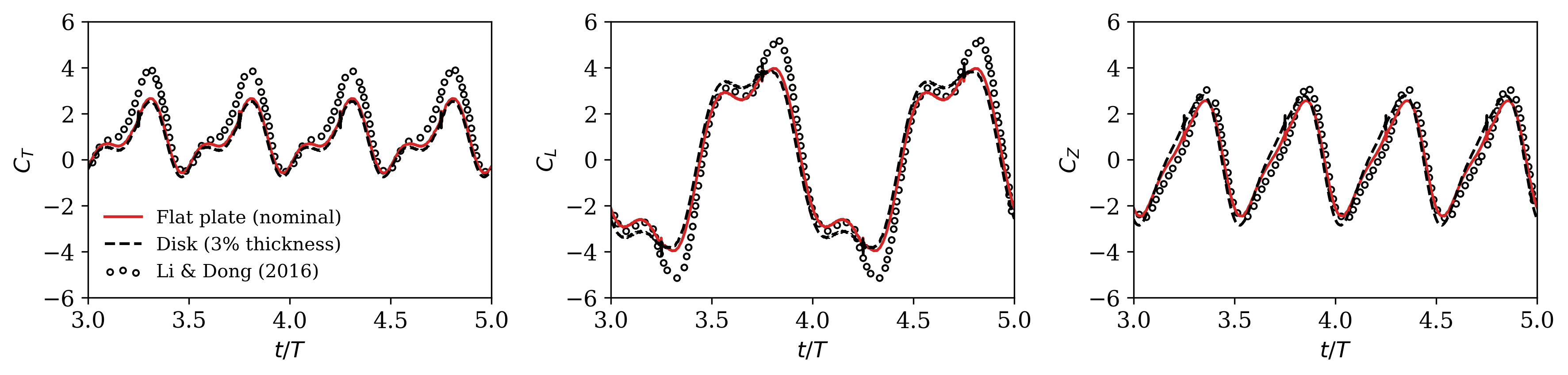}
  \caption{History of the thrust ($C_T$), lift ($C_L$), and spanwise ($C_Z$) coefficients over two flapping cycles of a circular plate ($AR = 1.27$) at Reynolds number $Re = 200$ and Strouhal number $St = 0.6$. We compare solutions on the nominal grid where the wing is modeled as a flat plate and as a disk with a thickness of $3\%$ of the chord length (as done in the original study\supercite{li_dong_2016}).}
  \label{fig:independence_force_coefficients_disk}
\end{figure}

The original study modeled the wing with an elliptical disk of thickness equal to $3\%$ of the chord length.
Since the immersed boundary method used for the present replication allows us to model thin volumes with flat surfaces,
we compared the aerodynamic forces obtained on a disk to those on a flat plate.
The history of the thrust coefficient (\cref{fig:independence_force_coefficients_disk}) is similar whether we use a disk or a flat plate.
We note slightly larger magnitude in the first peak of the lift and in the peaks of the spanwise force when using a disk.
Overall, the disk-based simulation does not really improve the solution and we decided to model the wing with a flat plate for the parametric study.
Also, fewer Lagrangian markers are needed to discretized a flat-plate wing, making simulations computationally cheaper.
(Runtime for the flat-wing simulation was about $8.4$ hours to compute 5 flapping cycles on 2 nodes of our Pegasus cluster; with a disk, the runtime was approximately $10.8$ hours.)

\begin{figure}[!h]
  \centering
  \includegraphics[width=\textwidth]{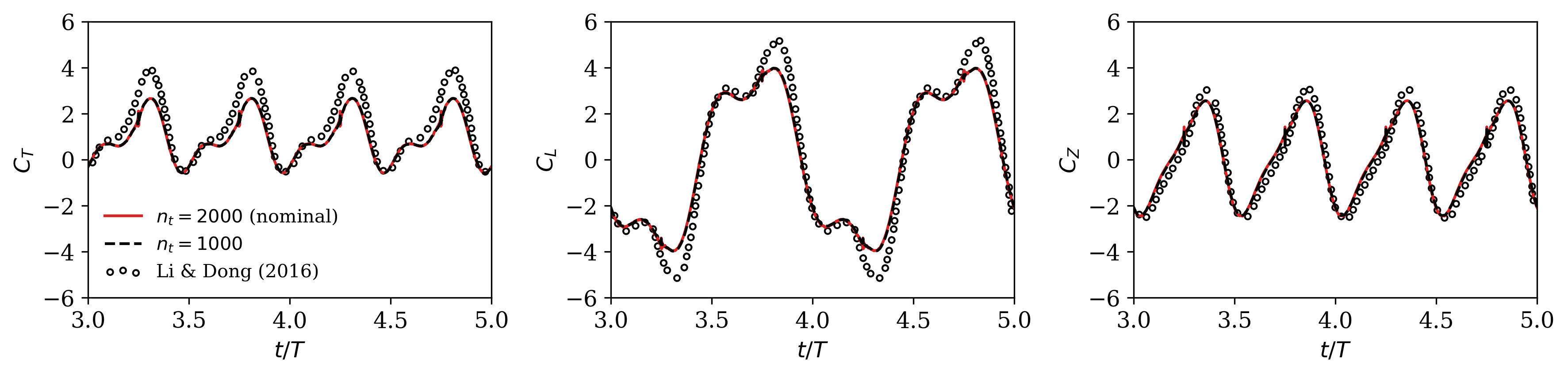}
  \caption{History of the thrust ($C_T$), lift ($C_L$), and spanwise ($C_Z$) coefficients over two flapping cycles of a circular plate ($AR = 1.27$) at Reynolds number $Re = 200$ and Strouhal number $St = 0.6$. We compare the instantaneous force coefficients obtained with $n_t = 2000$ and $n_t = 1000$ time steps per flapping cycle obtained on the nominal grid. Markers show digitized data from Fig.~9 of \citet{li_dong_2016}.}
  \label{fig:independence_force_coefficients_dt}
\end{figure}

Although the authors of the original study reported results to assess independence in the results when refining the temporal grid, they did not mention how many time steps per flapping cycle were computed.
(We know that force statistics and profiles of the velocity components did not significantly change when the time-step size was halved.)
\cref{fig:independence_force_coefficients_dt} shows the history of the force coefficients obtained on two temporal grids ($2000$ and $1000$ time steps per flapping cycle), using the nominal spatial grid.
The force signals are visually identical when doubling the time-step size, and the maximum relative difference in the aerodynamic quantities is less than $1\%$.
For all simulations of the parametric study, we computed $2000$ time steps per flapping cycle.

Finally, we also checked that aerodynamic statistics do not change significantly when setting a tighter convergence criterion ($atol$) for the iterative solvers (based on the absolute size of the residual in the $L_2$-norm).
Mean thrust coefficient, root-mean-square values of the lift and spanwise coefficients, and propulsive efficiency remain identical when reducing the convergence criterion of the iterative solvers by 3 orders of magnitude (reported in \cref{tab:independence_results}).

\begin{figure}[!h]
  \centering
  \begin{subfigure}[c]{0.48\textwidth}
    \centering
    \includegraphics[width=\linewidth]{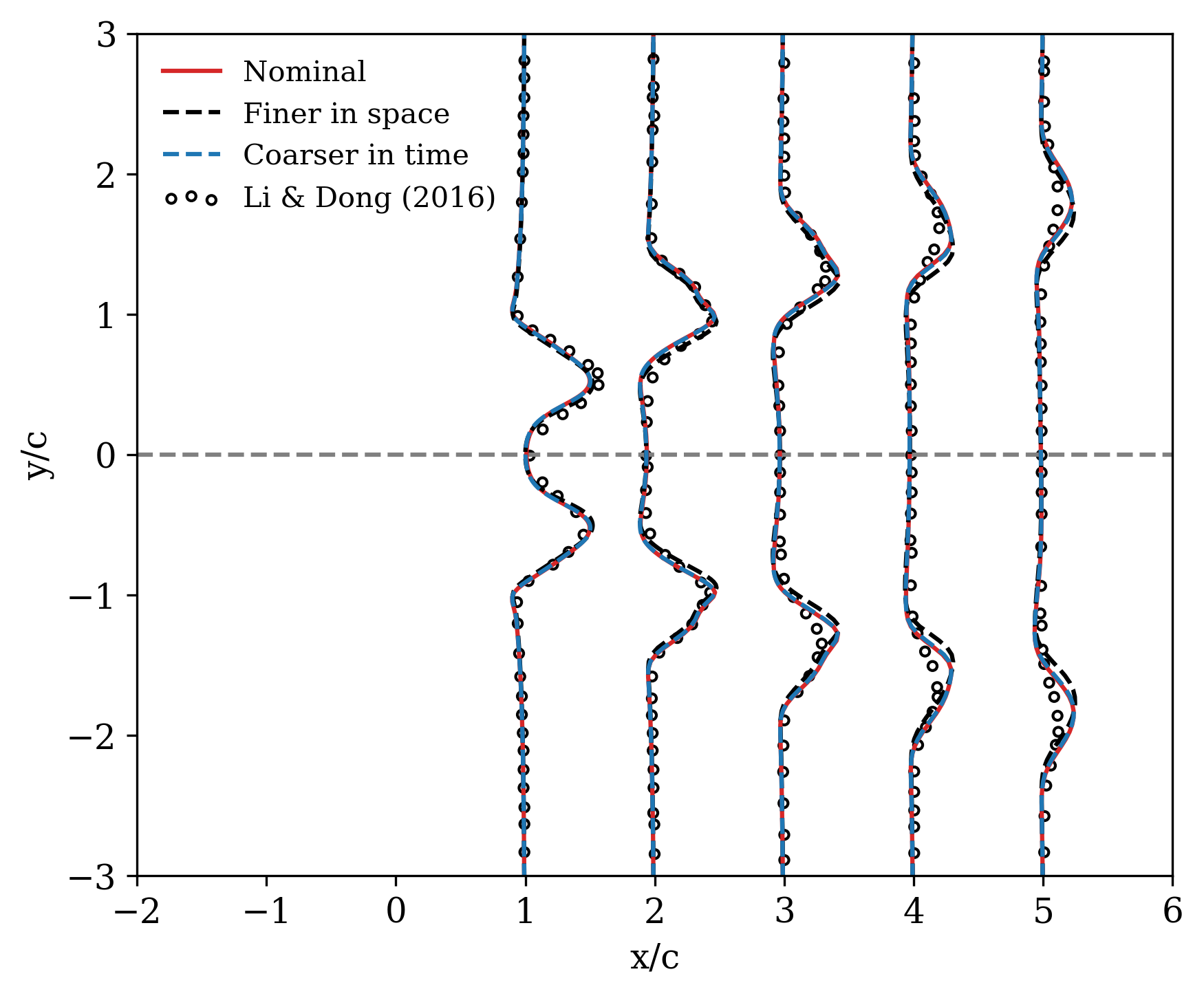}
    \caption{}
    \label{fig:independence_profiles:streamwise}
  \end{subfigure}
  \begin{subfigure}[c]{0.48\textwidth}
    \centering
    \includegraphics[width=\linewidth]{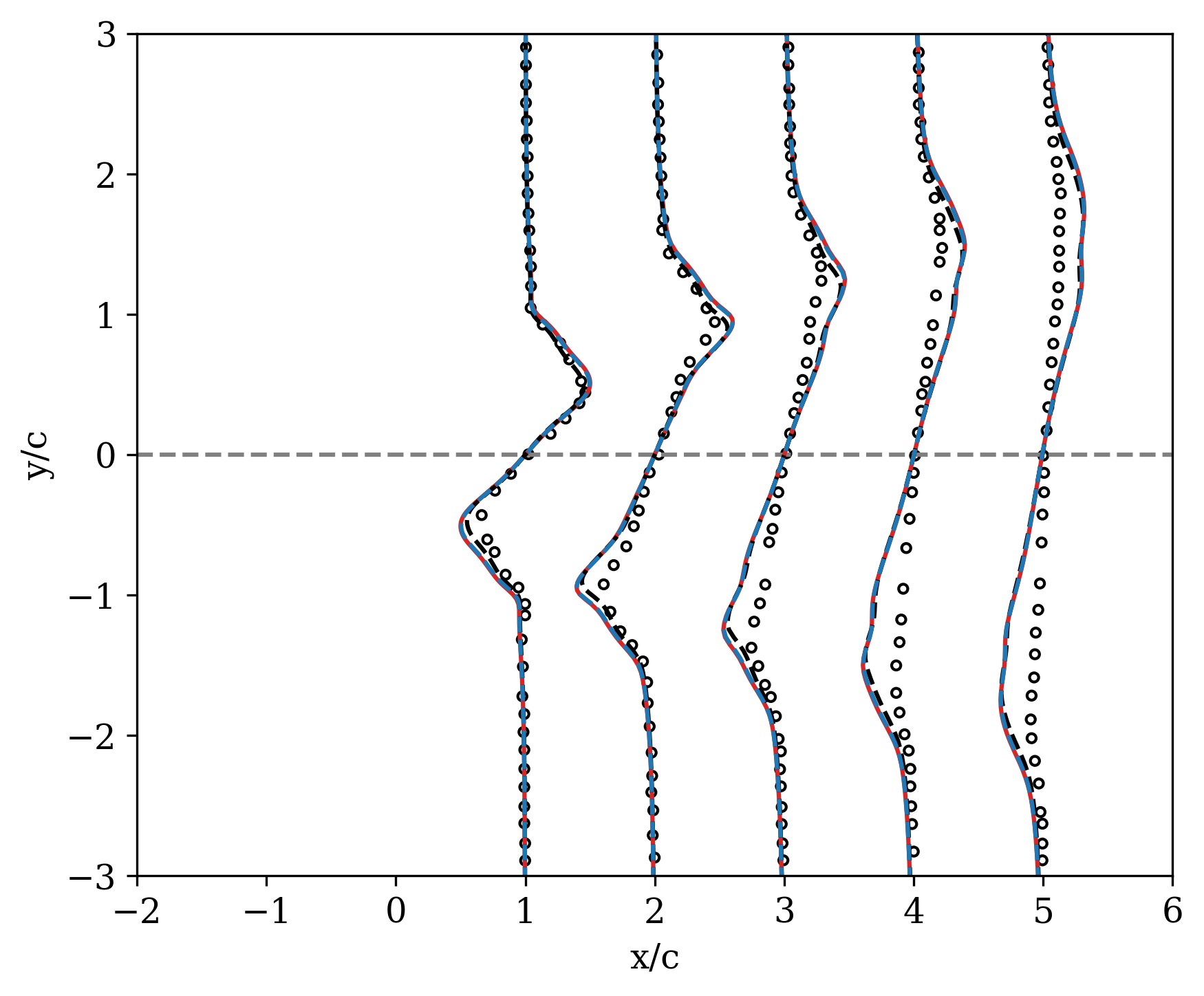}
    \caption{}
    \label{fig:independence_profiles:transverse}
  \end{subfigure}
  \begin{subfigure}[c]{0.48\textwidth}
    \centering
    \includegraphics[width=\linewidth]{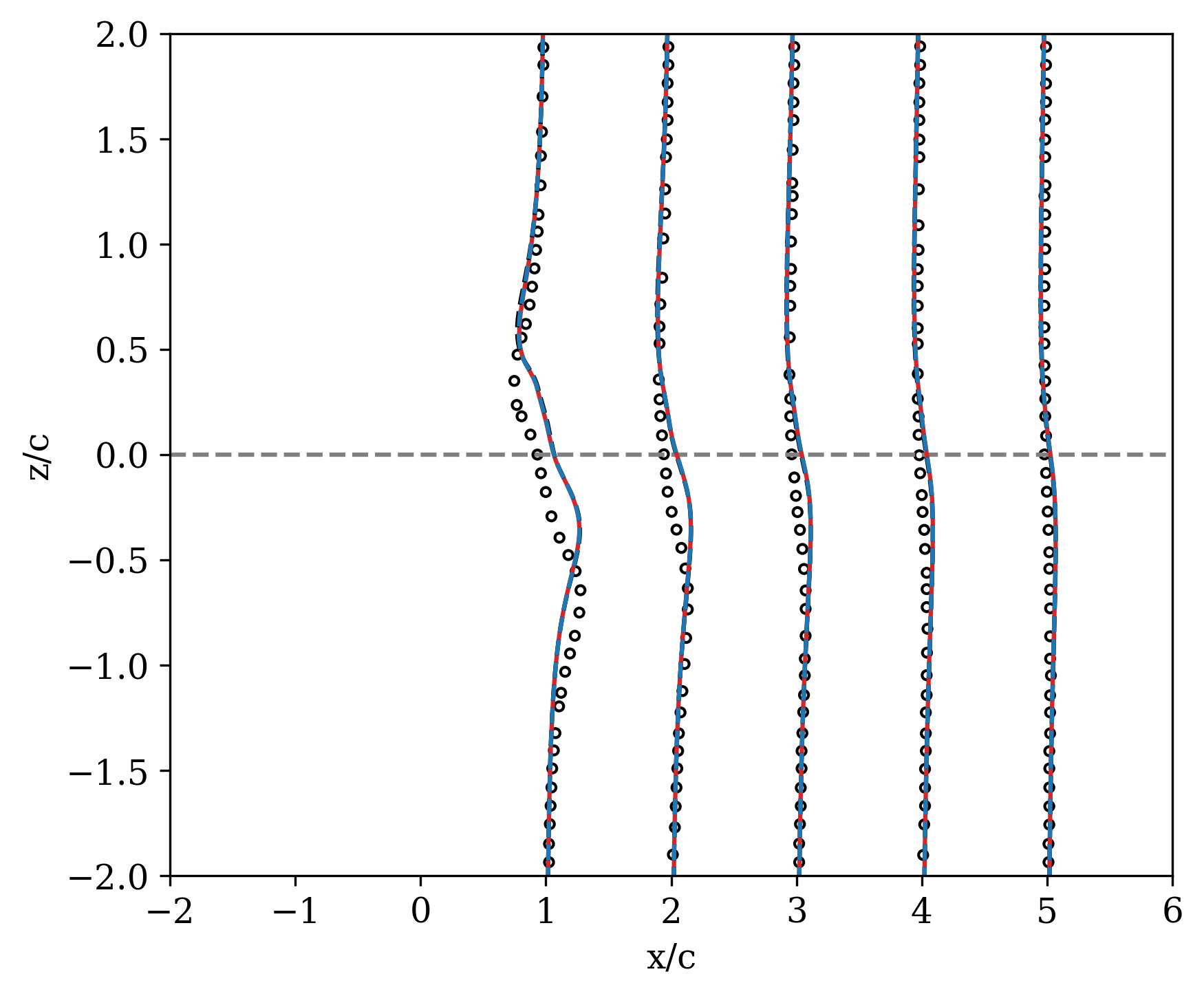}
    \caption{}
    \label{fig:independence_profiles:spanwise}
  \end{subfigure}
  \begin{subfigure}[c]{0.48\textwidth}
    \centering
    \includegraphics[width=\linewidth]{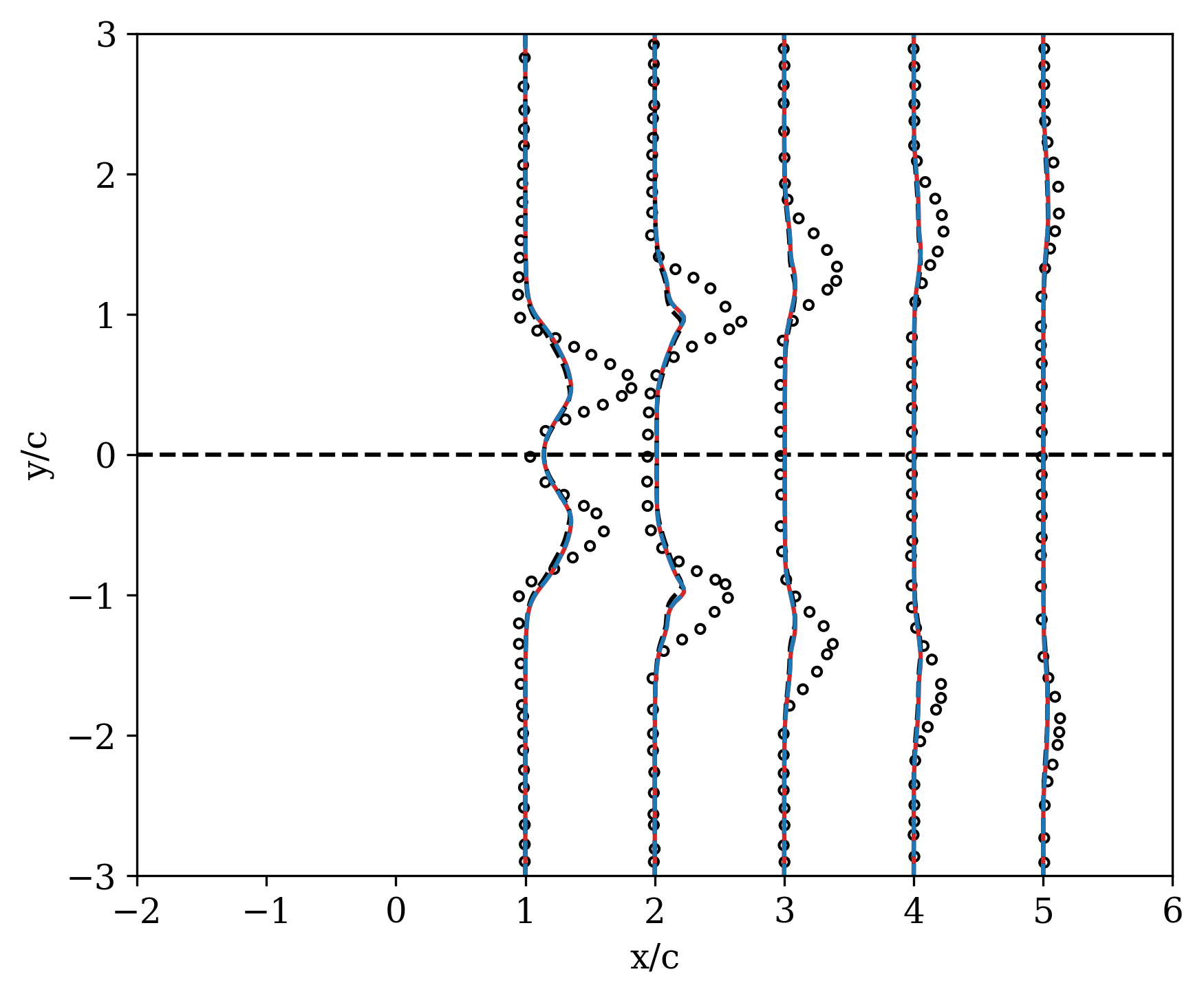}
    \caption{}
    \label{fig:independence_profiles:kinetic}
  \end{subfigure}
  \caption{Comparison of the velocity profiles and profiles of the fluctuation of the kinetic energy in the wake of a flapping circular plate ($AR = 1.27$) at Reynolds number $Re = 200$ and Strouhal number $St = 0.6$. We report profiles at locations $x / c = 1$, $2$, $3$, $4$, and $5$. We compare the profiles obtained on the nominal grid, on the finer grid in space, and on the coarser grid in time. (a) Streamwise velocity ($\overline{u_1} - U_\infty$) profiles in the $x/y$ plane at $z = S / 2$; (b) transverse velocity ($\overline{u_2}$) profiles in the $x/y$ plane at $z = S / 2$; (c) spanwise velocity ($\overline{u_3}$) in the $x/z$ plane at $y = 0$; (d) fluctuation of the kinetic energy $( \overline{{u_1'}^2} + \overline{{u_2'}^2} + \overline{{u_3'}^2} ) / 2$ in the $x/y$ plane at $z = S / 2$. We also report digitized data from Fig.~4 of \citet{li_dong_2016}.}
  \label{fig:independence_profiles}
\end{figure}

We also assess the effect of the grid-spacing and time-step sizes on the flow dynamics by comparing the profiles of the mean velocity components and of the fluctuating kinetic energy at several locations in the wake of the wing.
\cref{fig:independence_profiles:streamwise,fig:independence_profiles:transverse} show the mean streamwise ($\overline{u_1} - u_\infty$) and transverse ($\overline{u_2}$) velocity profiles in the $x/y$ plane at mid-span ($z = S / 2$).
\cref{fig:independence_profiles:spanwise} shows the mean spanwise ($\overline{u_3}$) velocity profiles at the center ($y = 0$) of the $x/z$ plane.
Profiles of the fluctuating kinetic energy ($( \overline{{u_1'}^2} + \overline{{u_2'}^2} + \overline{{u_3'}^2} ) / 2$) at mid-span are reported in \cref{fig:independence_profiles:kinetic}.
We also report digitized data from \citet{li_dong_2016} for comparison.
Although there exist some differences with the results of the original study, we note that the profiles of the present study are visually similar when refining the spatial grid and when using fewer time steps per flapping cycle.

\begin{figure}[!h]
  \centering
  \begin{subfigure}[c]{0.35\textwidth}
    \centering
    \includegraphics[width=\linewidth]{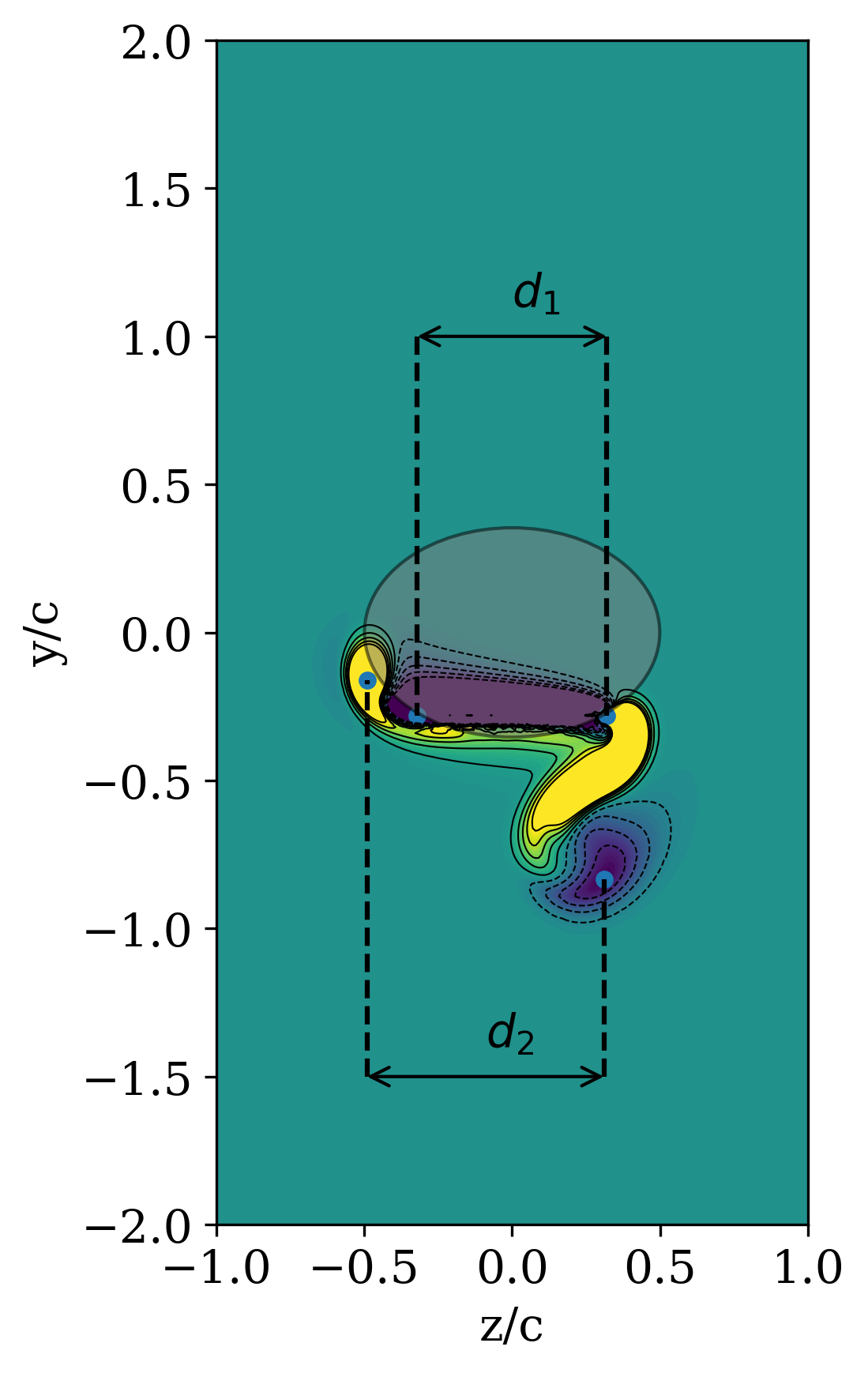}
    \caption{}
  \end{subfigure}
  \hspace{1em}
  \begin{subfigure}[c]{0.35\textwidth}
    \centering
    \includegraphics[width=\linewidth]{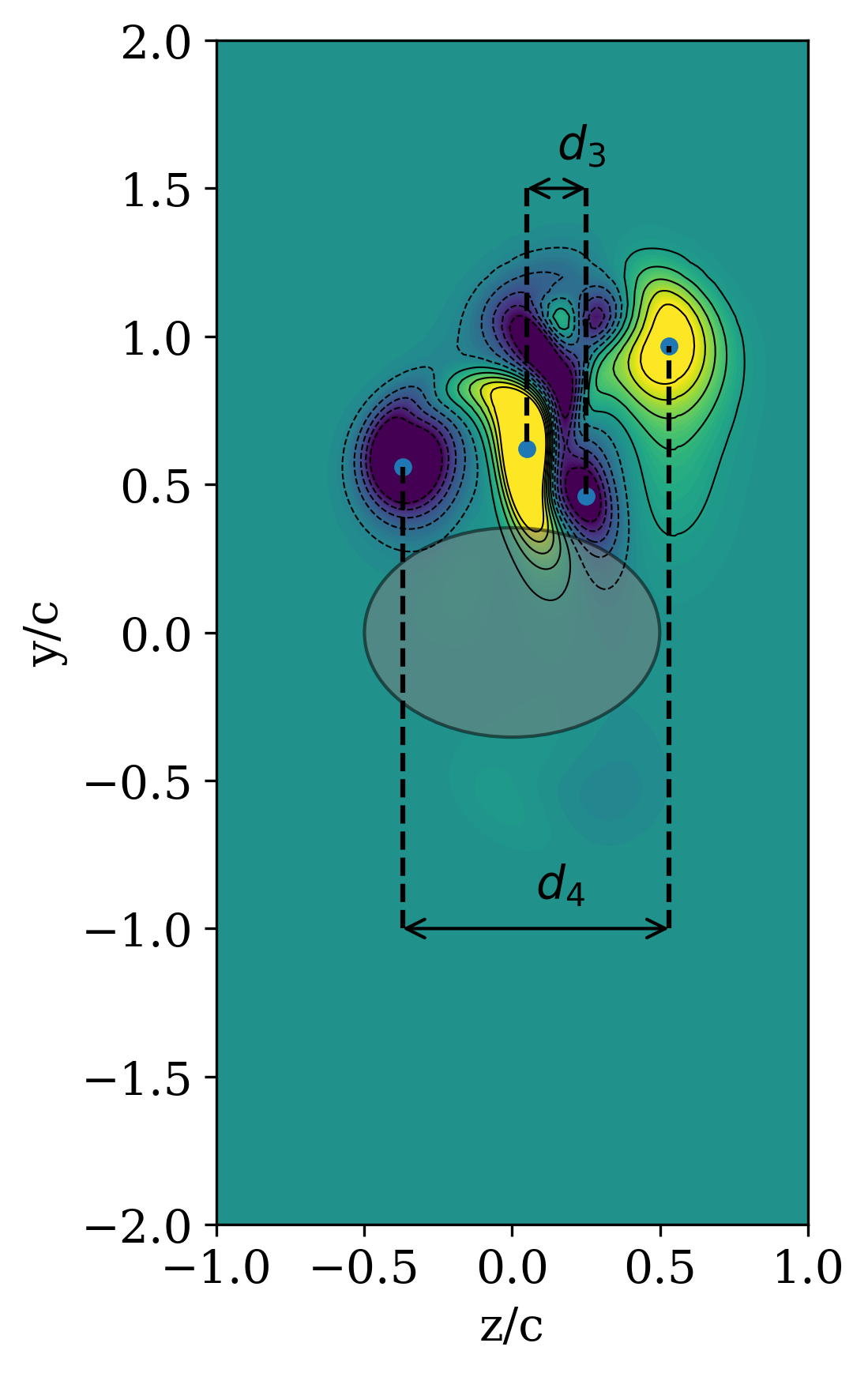}
    \caption{}
  \end{subfigure}
  \caption{Two-dimensional contours of the streamwise vorticity ($-5 \leq w_x \leq 5$) in the $y/z$ plane at $t/T = 4.25$ in the near wake at $x / c = 0.3$ (a) and in the far wake at $x / c = 1.3$ (b). We compare the distances between vortical structures obtained on different grids to assess independence of the numerical solution. (See Fig.~5 of \citet{li_dong_2016} for comparison.)}
  \label{fig:independence_wx_distances}
\end{figure}

\begin{table}[!h]
  \centering
  \begin{tabular}{lcccc}
    \hline\hline
    Case & $d_1$ & $d_2$ & $d_3$ & $d_4$ \\
    \hline
    Nominal & $0.810$ & $0.630$ & $0.200$ & $0.900$ \\
    Finer in space & $0.780$ & $0.575$ & $0.215$ & $0.900$ \\
    Coarser in time & $0.810$ & $0.600$ & $0.190$ & $0.900$ \\
    \hline\hline
  \end{tabular}
  \caption{Comparison of the distances between vortex structures at $t / T = 4.25$ for a flapping wing with $AR = 1.27$, $St = 0.6$, $Re = 200$, and $\psi = 90^o$.}
  \label{tab:independence_wx_distances}
\end{table}

Following the work of \citet{li_dong_2016}, we also tried to assess the effects of different grids on the instantaneous vortical structures.
As in the original study, we quantified the size of the vortex loops at time $t/T = 4.25$ in the near ($x/c = 0.3$) and far ($x/c = 1.3$) wake.
\cref{fig:independence_wx_distances} shows slices of the streamwise vorticity field in the near and far wake, where we also report distances between between each pair of vortices.
(Center of each vortex is chosen as the point with the maximum absolute value of the streamwise vorticity.)
\cref{tab:independence_wx_distances} reports the distances between vortex centers.
Compared to nominal grid, the maximum relative difference occurs for the distance $d_2$ (near wake) on the finer spatial grid and is about $9\%$.
The relatively high difference could be caused by the diffuse-interface immersed boundary method used in the present replication study.
We employ regularized delta kernels, to transfer information between the Eulerian grid and the Lagrangian surface mesh, which have a compact support that is proportional to the Eulerian grid-spacing size $\Delta x$.
These kernels smear the solution near the immersed boundary.
In other words, the solution is not physical over a thin shell around the wing.
As we refine the spatial grid (and the Lagrangian mesh), the distance over which the solution is smeared is reduced.
Nevertheless, results of the parametric study show that we were able to replicate the main scientific findings of the original study using our nominal grid (with resolution $\Delta x = 0.01c$ near the immersed boundary), computing $2000$ time steps per flapping cycles of a wing (modeled as a flat plate).

\subsection{Baseline case}

As in the original study, we start by looking at the wake topology and aerodynamic forces produced by a rolling and pitching circular plate ($AR = 1.27$) at Reynolds number $Re = 200$, with Strouhal number $St = 0.6$ and phase-difference angle $\psi = 90^o$.

\begin{figure}[H]
  \centering
  \begin{subfigure}[b]{0.3\textwidth}
    \centering
    \includegraphics[width=\linewidth]{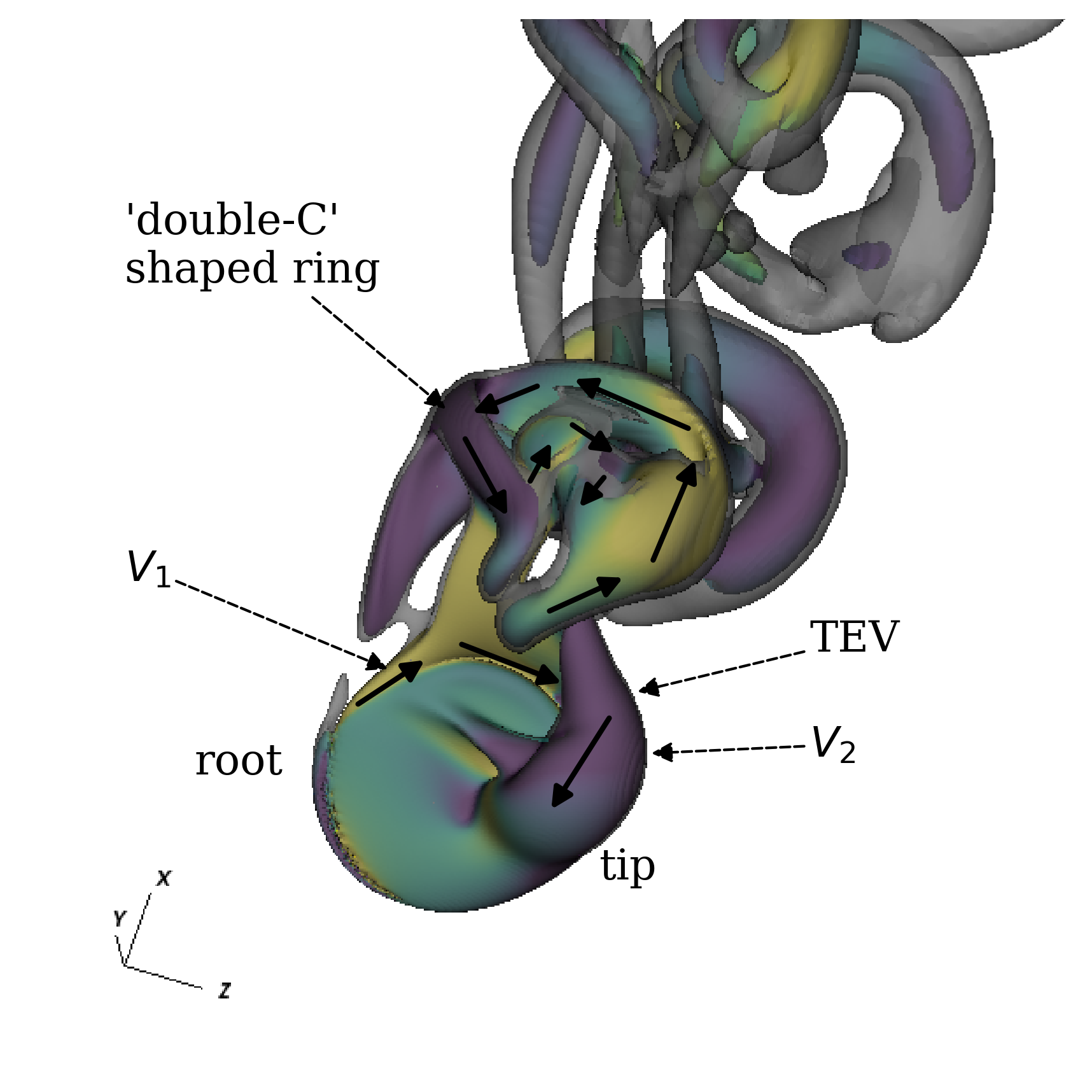}
    \caption{$t / T = 3.875$}
    \label{fig:baseline_qcrit_perspective:0007750}
  \end{subfigure}
  \hspace{0.5em}
  \begin{subfigure}[b]{0.3\textwidth}
    \centering
    \includegraphics[width=\linewidth]{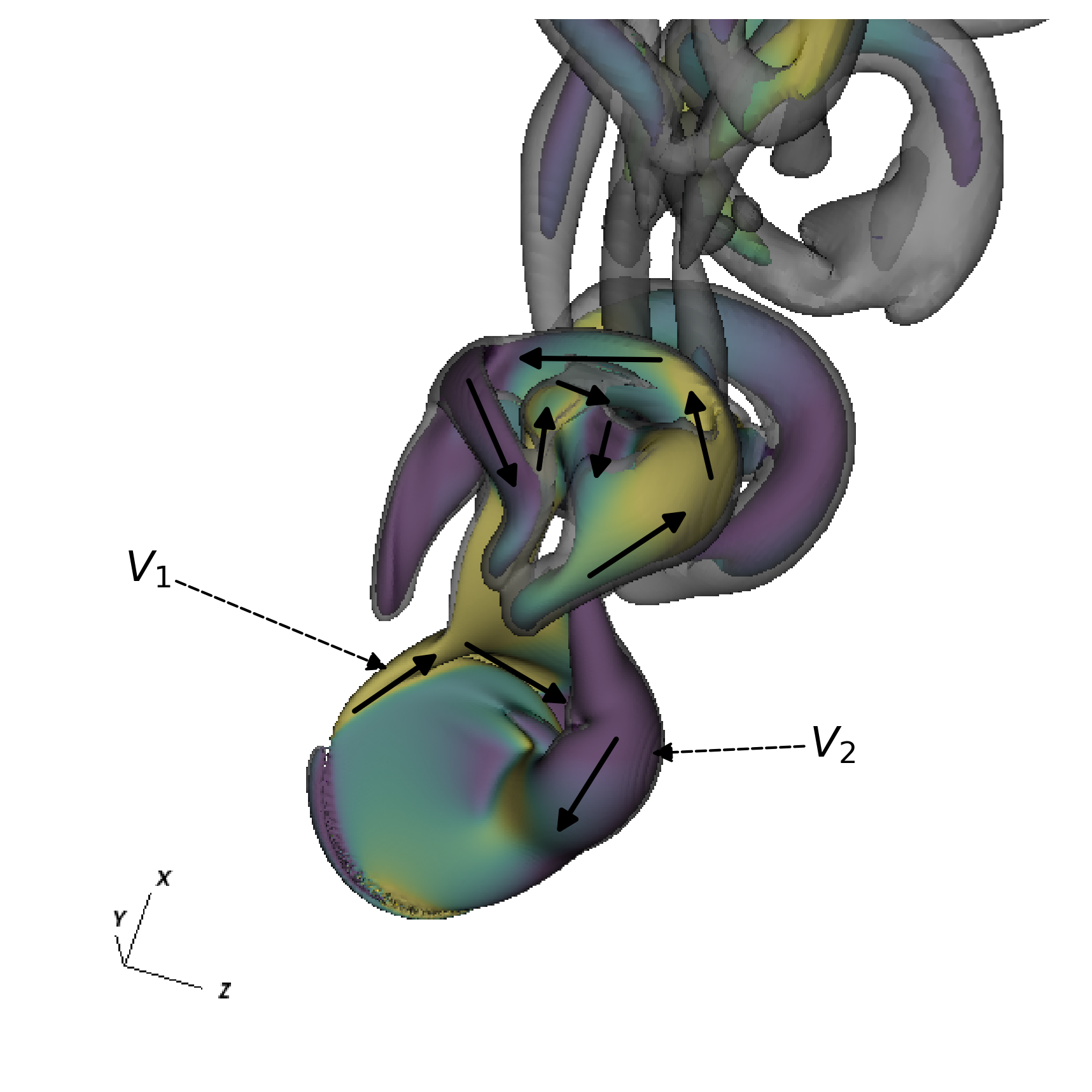}
    \caption{$t / T = 3.9375$}
    \label{fig:baseline_qcrit_perspective:0007875}
  \end{subfigure}
  \hspace{0.5em}
  \begin{subfigure}[b]{0.3\textwidth}
    \centering
    \includegraphics[width=\linewidth]{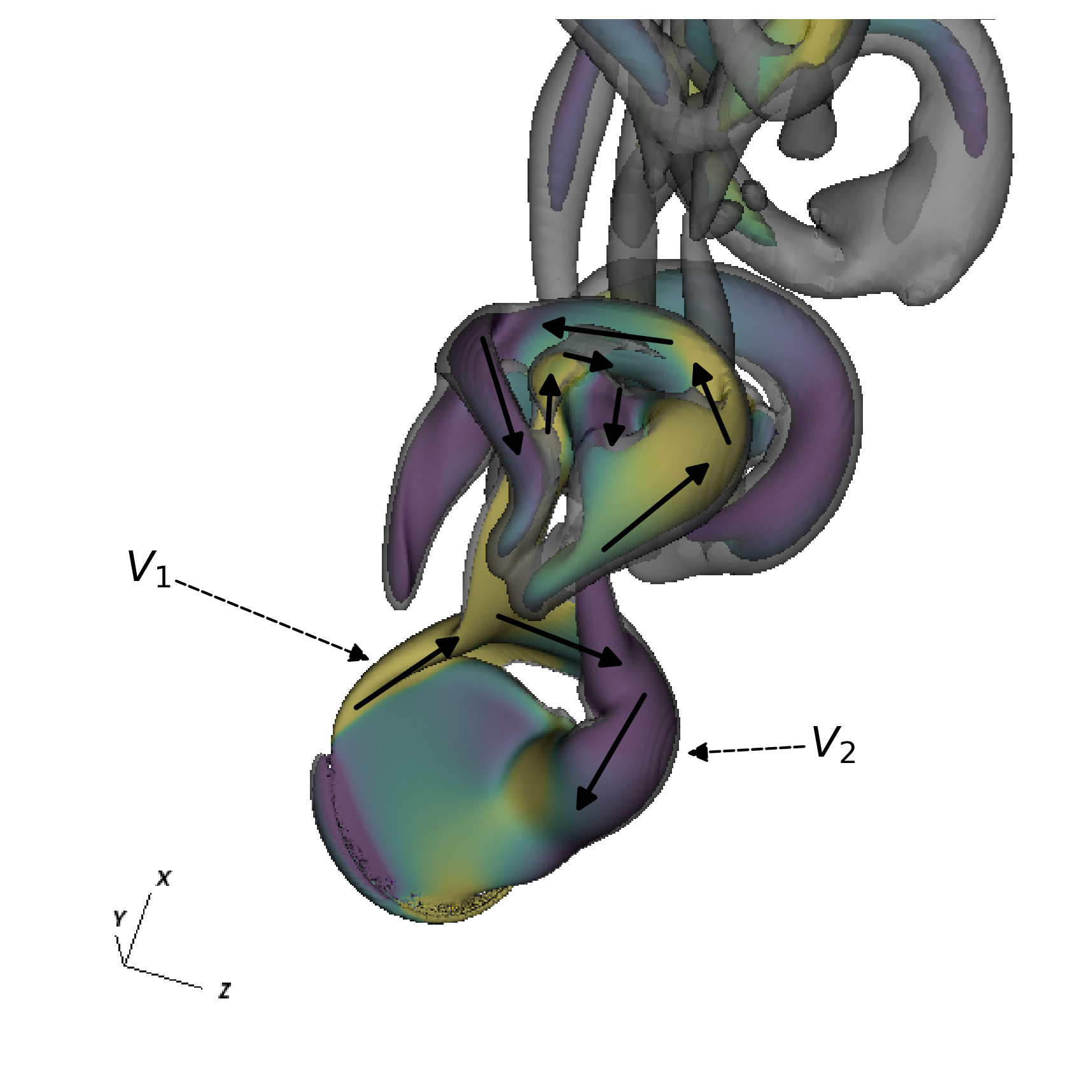}
    \caption{$t / T = 4.0$}
    \label{fig:baseline_qcrit_perspective:0008000}
  \end{subfigure}
  \vspace{0.5em}
  \begin{subfigure}[b]{0.3\textwidth}
    \centering
    \includegraphics[width=\linewidth]{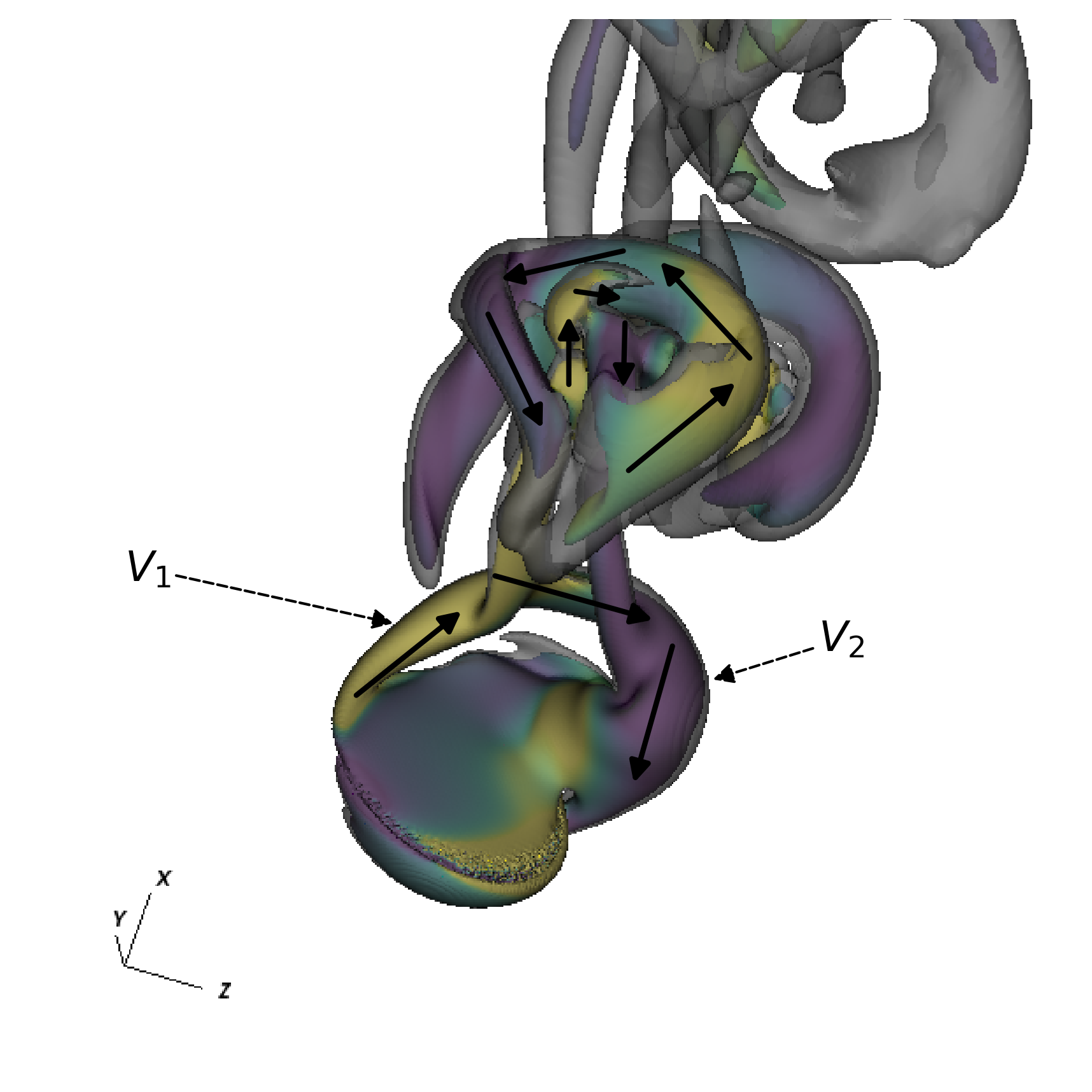}
    \caption{$t / T = 4.125$}
    \label{fig:baseline_qcrit_perspective:0008250}
  \end{subfigure}
  \hspace{0.5em}
  \begin{subfigure}[b]{0.3\textwidth}
    \centering
    \includegraphics[width=\linewidth]{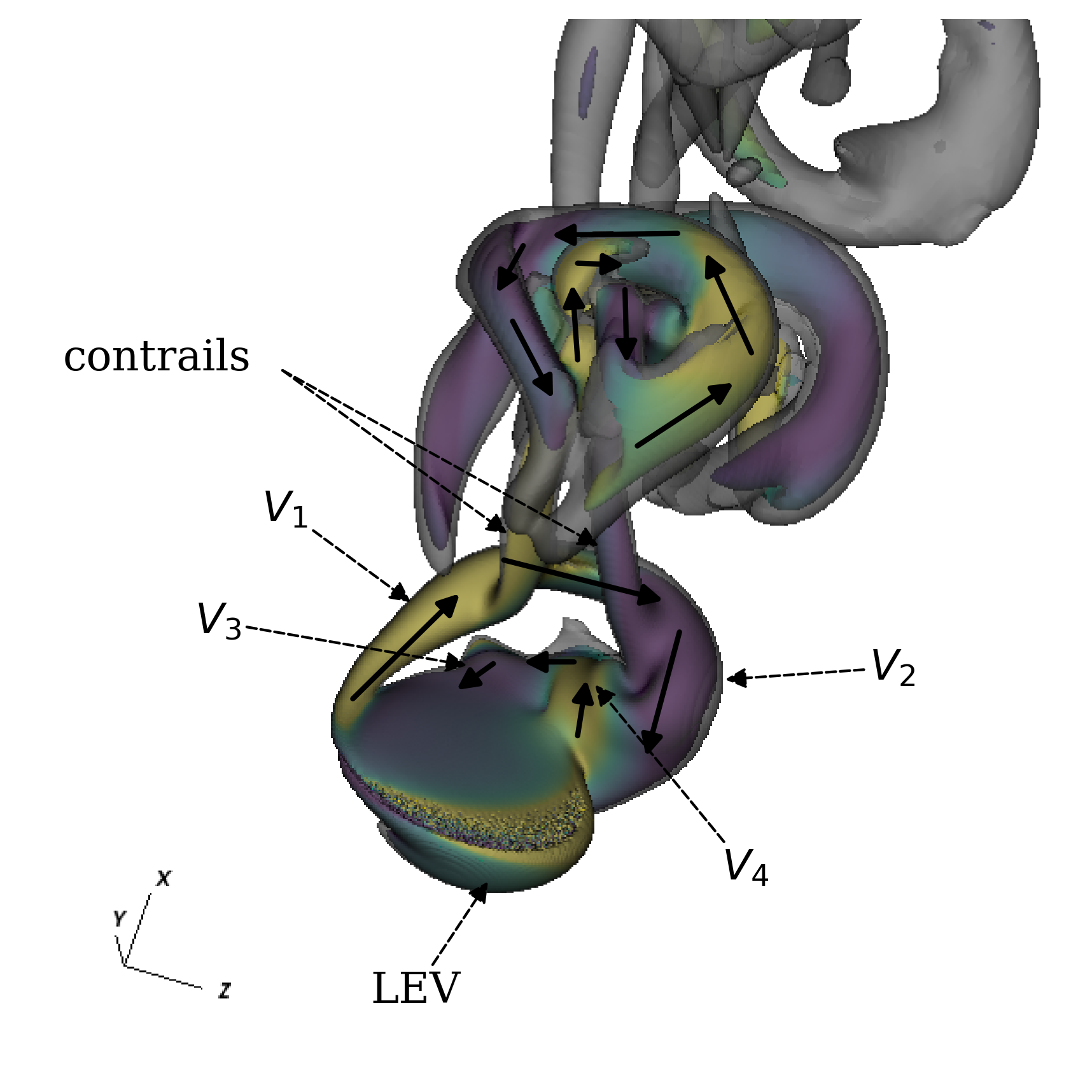}
    \caption{$t / T = 4.1875$}
    \label{fig:baseline_qcrit_perspective:0008375}
  \end{subfigure}
  \hspace{0.5em}
  \begin{subfigure}[b]{0.3\textwidth}
    \centering
    \includegraphics[width=\linewidth]{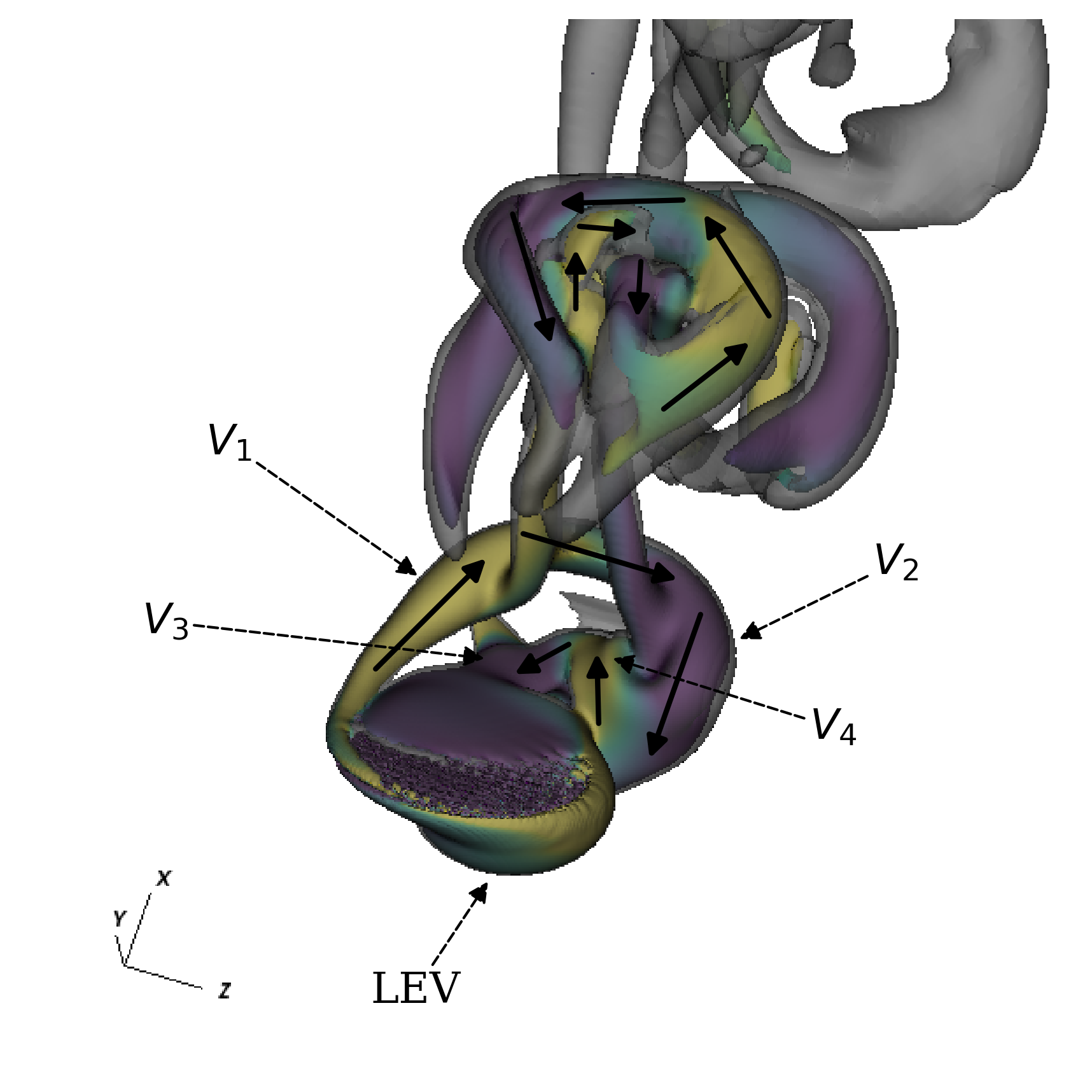}
    \caption{$t / T = 4.25$}
    \label{fig:baseline_qcrit_perspective:0008500}
  \end{subfigure}
  \vspace{0.5em}
  \begin{subfigure}[b]{0.3\textwidth}
    \centering
    \includegraphics[width=\linewidth]{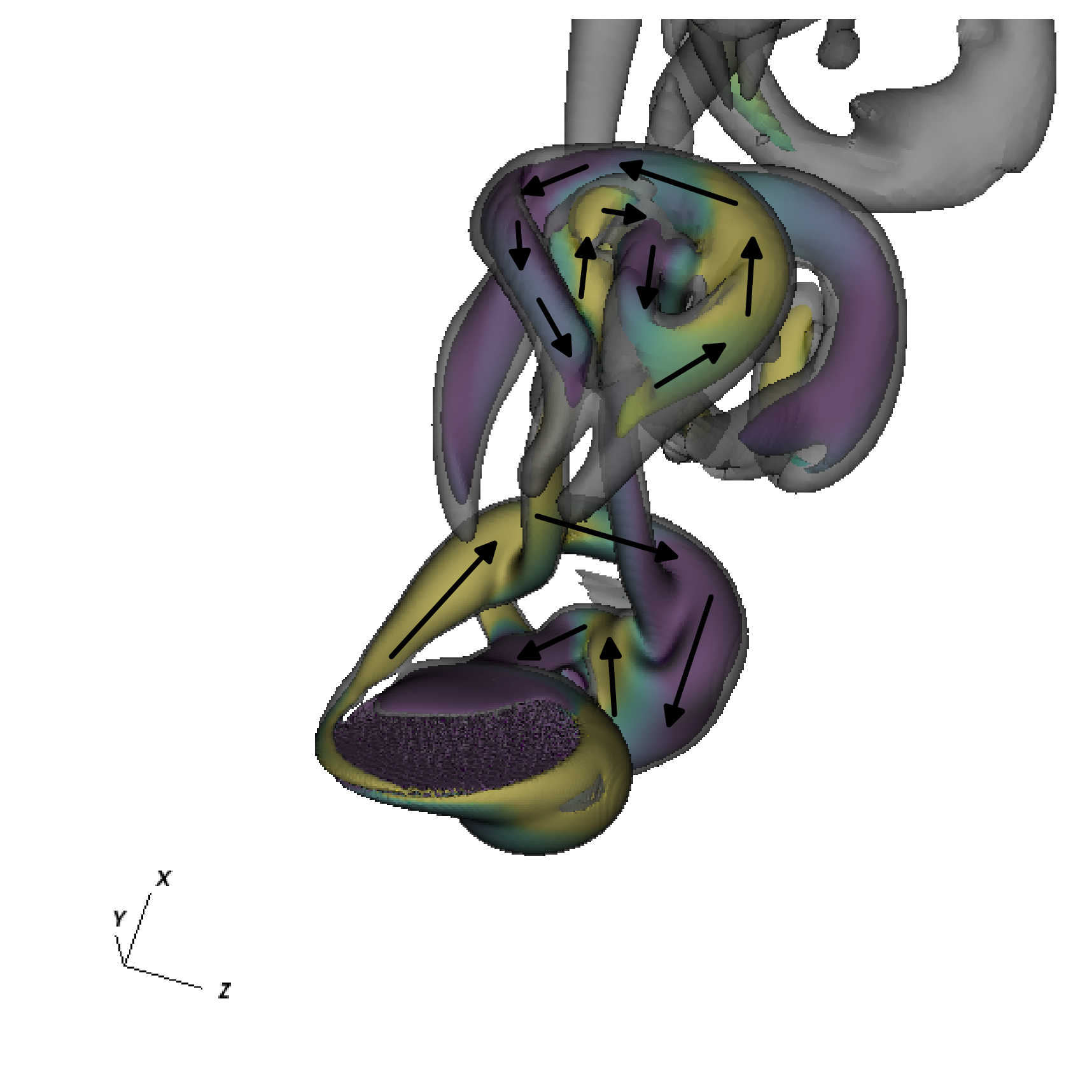}
    \caption{$t / T = 4.3125$}
    \label{fig:baseline_qcrit_perspective:0008625}
  \end{subfigure}
  \hspace{0.5em}
  \begin{subfigure}[b]{0.3\textwidth}
    \centering
    \includegraphics[width=\linewidth]{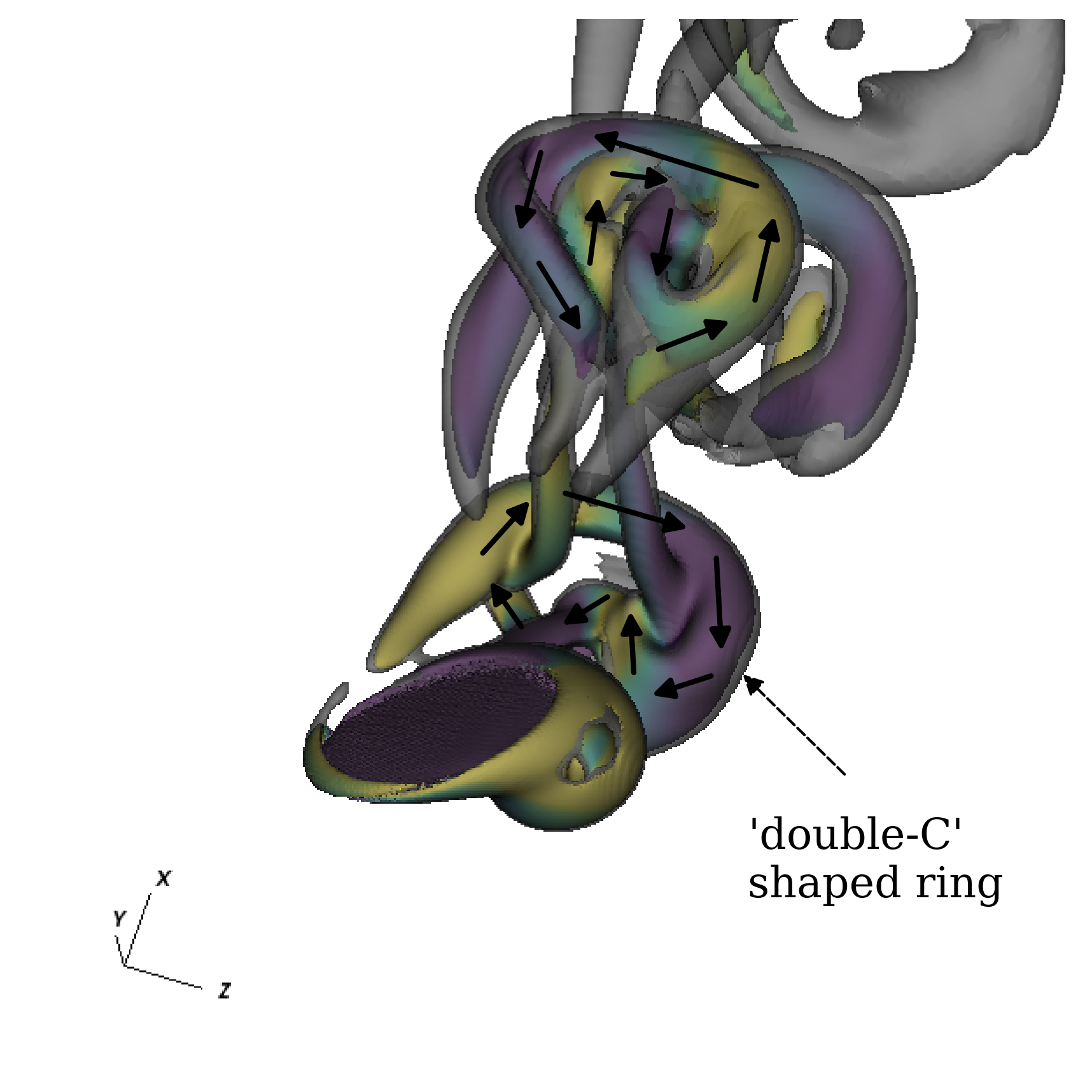}
    \caption{$t / T = 4.375$}
    \label{fig:baseline_qcrit_perspective:0008750}
  \end{subfigure}
  \hspace{0.5em}
  \begin{subfigure}[b]{0.3\textwidth}
    \centering
    \includegraphics[width=\linewidth]{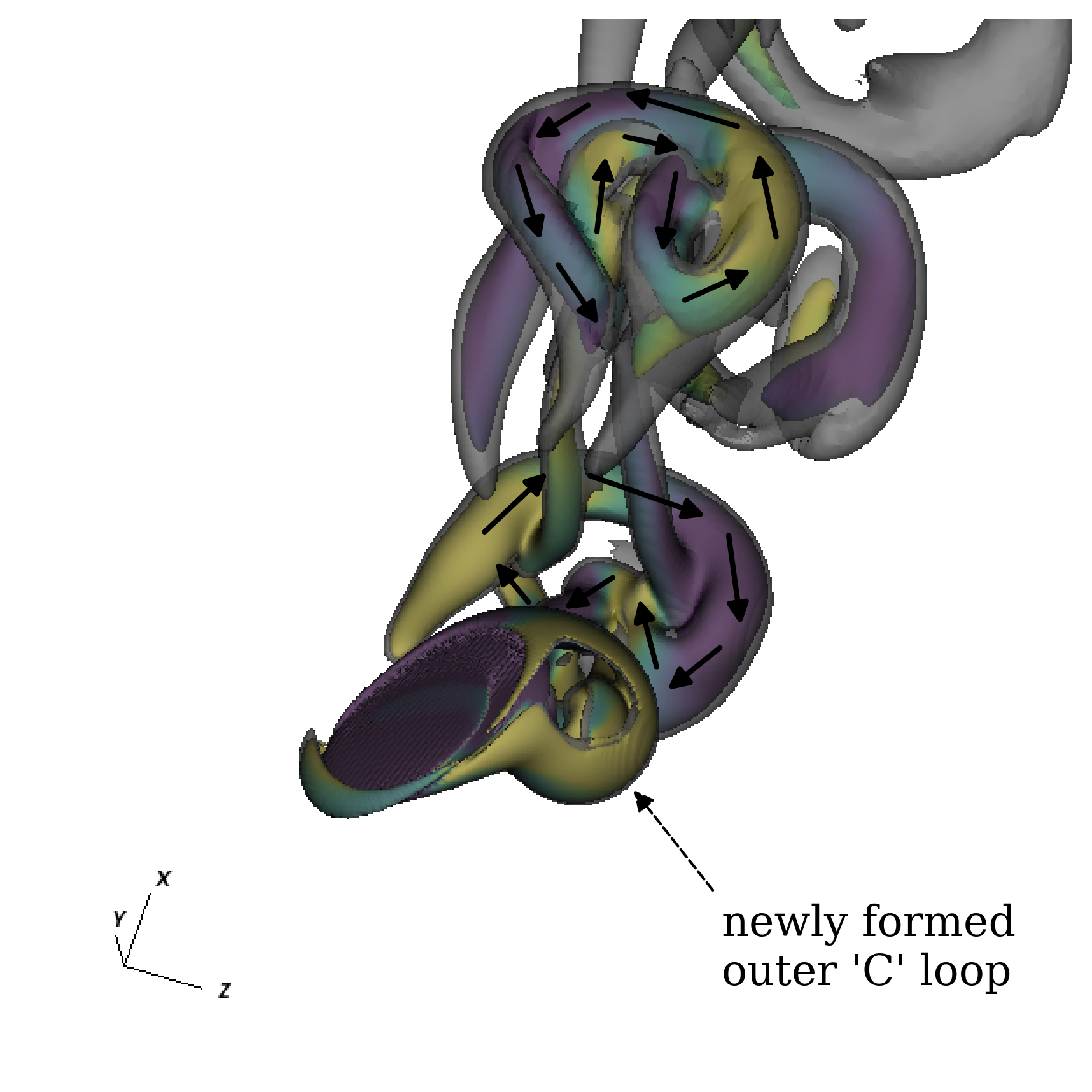}
    \caption{$t / T = 4.4375$}
    \label{fig:baseline_qcrit_perspective:0008875}
  \end{subfigure}
  \caption{Wake topology for a flapping circular plate ($AR = 1.27$) with $St = 0.6$, $Re = 200$, and $\psi = 90^o$ at $t / T = 3.875$, $3.9375$, $4.0$, $4.125$, $4.1875$, $4.25$, $4.3125$, $4.375$, $4.4375$. We show the contours of the $Q$-criterion for $Q = 1$ (gray) and $Q = 6$ (colored by the streamwise vorticity, $-5 \leq w_x \leq 5$). (See Fig.~6 of \citet{li_dong_2016} for comparison.)}
  \label{fig:baseline_qcrit_perspective}
\end{figure}

\begin{figure}[H]
  \centering
  \begin{minipage}{0.55\linewidth}
    \begin{subfigure}[t]{\linewidth}
      \includegraphics[width=\textwidth]{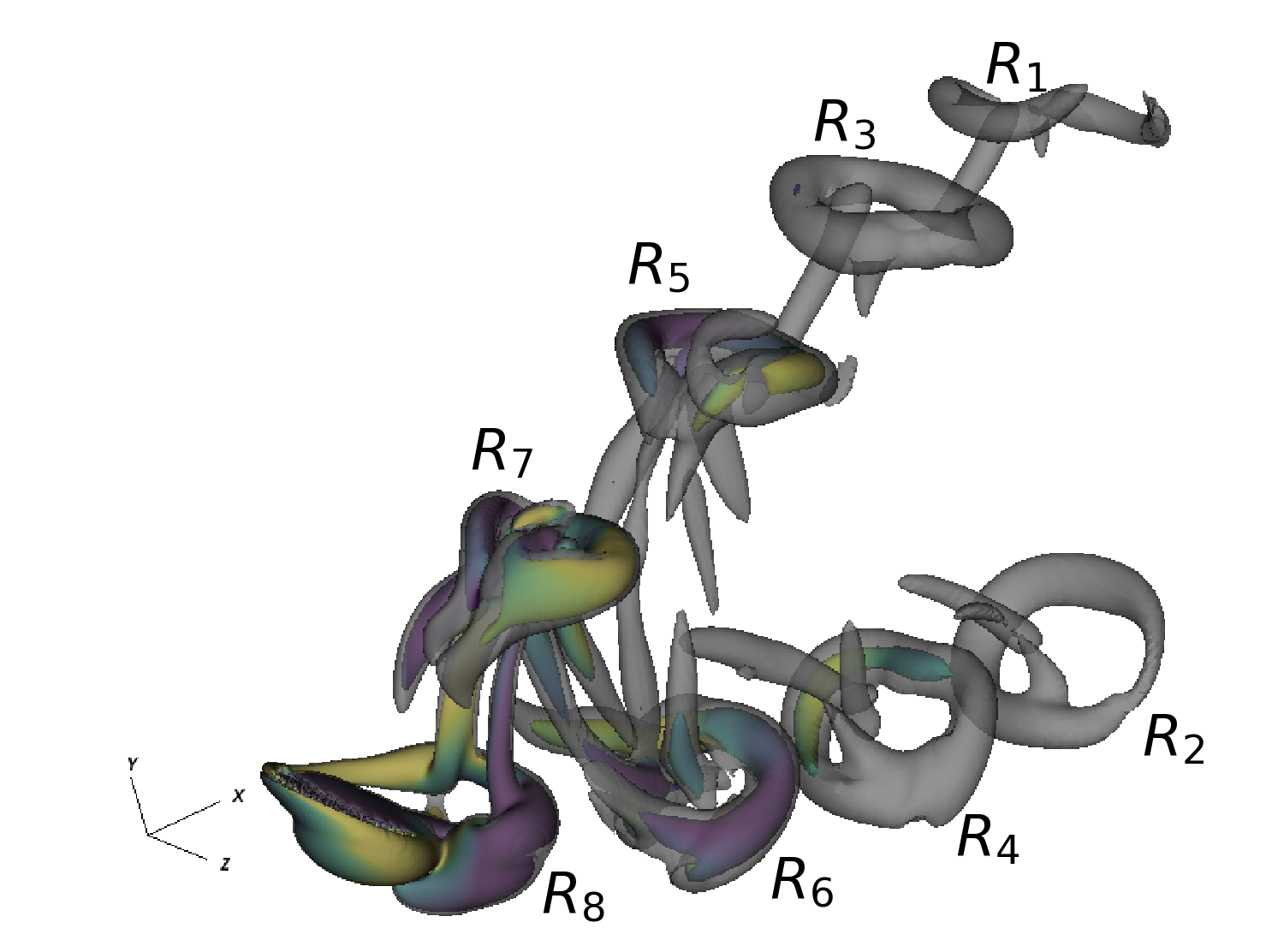}
      \caption{}
      \label{fig:baseline_wake_topology:perspective}
    \end{subfigure}
  \end{minipage}
  \begin{minipage}{0.35\linewidth}
    \begin{subfigure}[t]{\linewidth}
      \includegraphics[width=\textwidth]{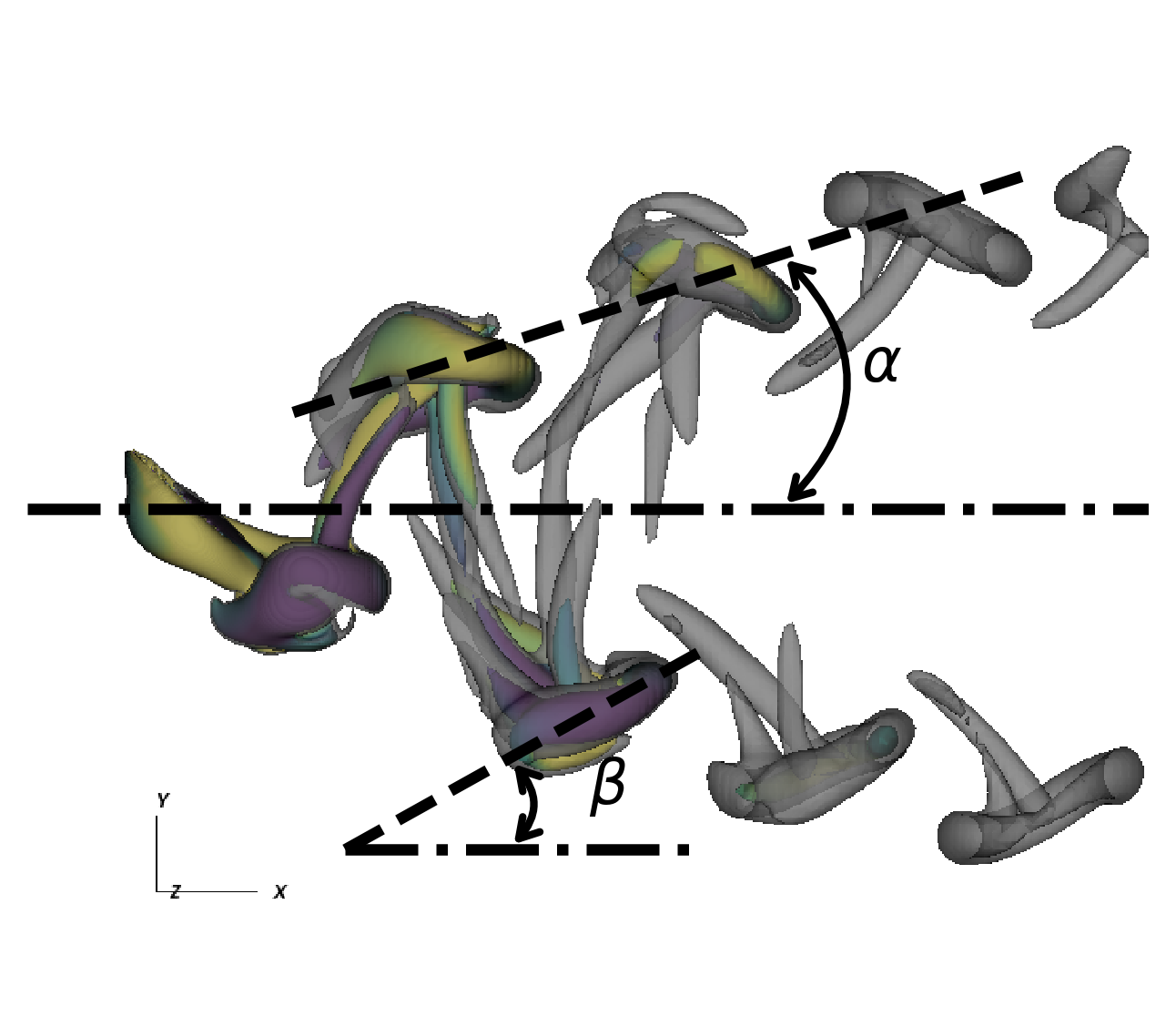}
      \caption{}
      \label{fig:baseline_wake_topology:lateral}
    \end{subfigure}
    \vspace{1cm}
    \begin{subfigure}[b]{\linewidth}
      \includegraphics[width=\textwidth]{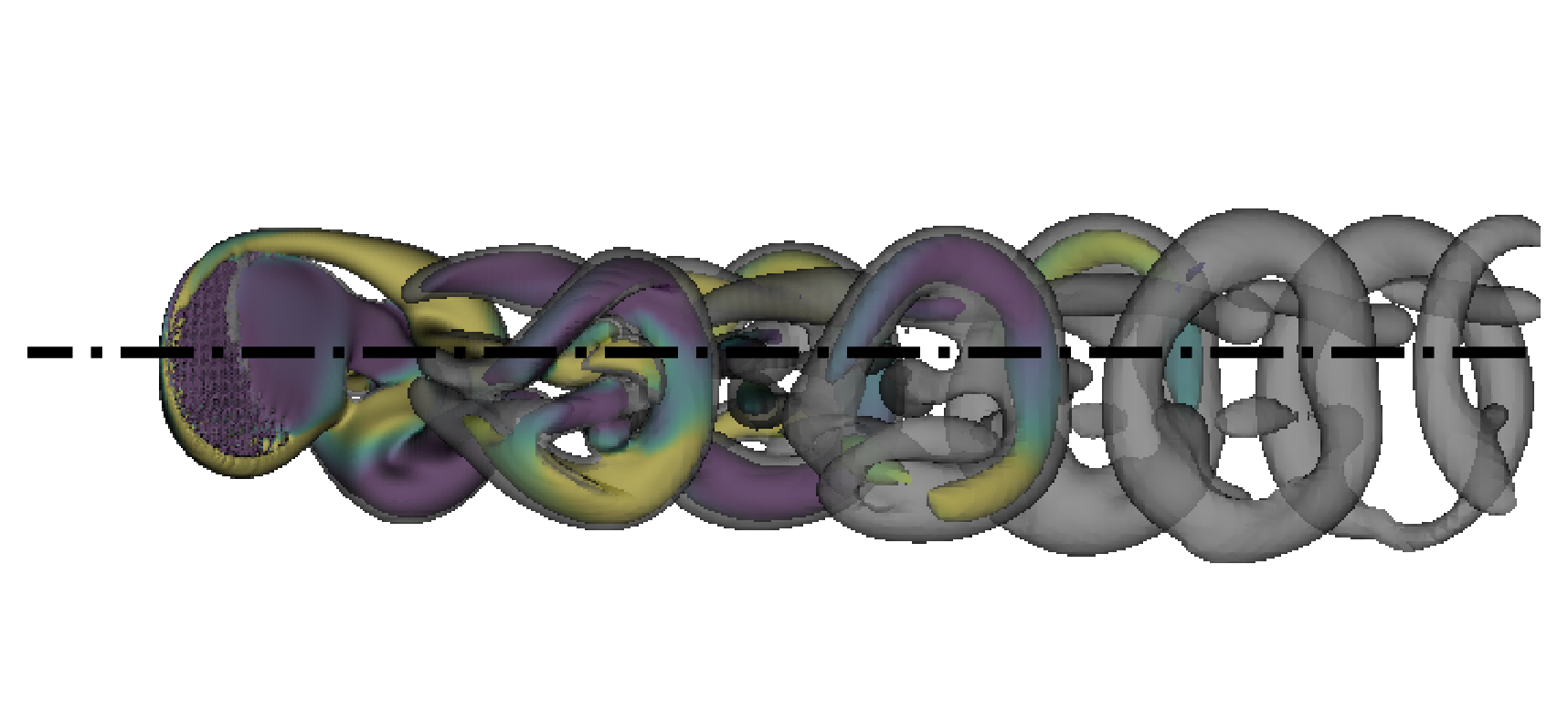}
      \caption{}
      \label{fig:baseline_wake_topology:top}
    \end{subfigure}
  \end{minipage}
  \caption{Wake topology of a pitching-rolling circular plate after four flapping cycles. We show flow structures using the $Q$-criterion for $Q = 1$ (gray) and $Q = 6$ (colored by the streamwise vorticity, $-5 \leq w_x \leq 5$). (a) Perspective view, (b) side view, and (c) top view. (See Fig.~7 of \citet{li_dong_2016} for comparison.)}
  \label{fig:baseline_wake_topology}
\end{figure}

We were able to observe the same flow features as the ones reported in the original study for the baseline case.
\cref{fig:baseline_qcrit_perspective} shows the vortex shedding dynamics in the vicinity of the wing at nine different time values.
We visualize vortices using the $Q$-criterion at $Q = 1$ (in gray) and $Q = 6$ (colored by the streamwise vorticity).
As the wing rolls downward (\crefrange{fig:baseline_qcrit_perspective:0007750}{fig:baseline_qcrit_perspective:0008000}), we observe the formation of a ``C''-shaped vortex loop between the root vortex ($V_1$), the tip vortex ($V_2$), and the trailing-edge vortex (TEV).
The rolling motion induces a strength asymmetry between $V_1$ and $V_2$.
As the plate starts rolling upward while pitching up (\crefrange{fig:baseline_qcrit_perspective:0008250}{fig:baseline_qcrit_perspective:0008500}), we note the formation of new vortices from the trailing edge ($V_3$) and from the leading-edge vortex ($V_4$).
These vortices interact with each other to form an additional ``C''-shaped vortex loop in opposite direction.
As the two vortex loops propagates downstream, they form a ``double-C''-shaped vortex structure, which was reported in the original study.
Each flapping cycle produces a pair of ``double-C''-shaped vortex structures (with opposite direction), leading to a bifurcated wake pattern (\cref{fig:baseline_wake_topology:perspective,fig:baseline_wake_topology:lateral}).
The ``double-C''-shaped vortex structures evolve into single-loop vortex when convected downstream.
As noted in the original study, vortex rings shed in the wake gradually increase in size as they move further downstream, with a slight deflection in the spanwise $z$-direction towards the tip of the wing (\cref{fig:baseline_wake_topology:top}).
Also observable on \cref{fig:baseline_wake_topology:lateral} are the set of ``contrails'' connecting adjacent rings.

\begin{figure}[H]
  \centering
  \begin{subfigure}[t]{0.24\textwidth}
    \centering
    \includegraphics[width=\linewidth]{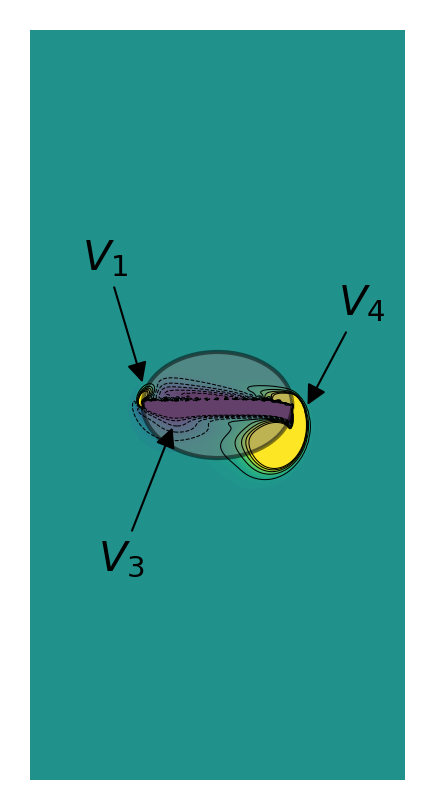}
    \caption{}
    \label{fig:baseline_wx_slices:a}
  \end{subfigure}
  \begin{subfigure}[t]{0.24\textwidth}
    \centering
    \includegraphics[width=\linewidth]{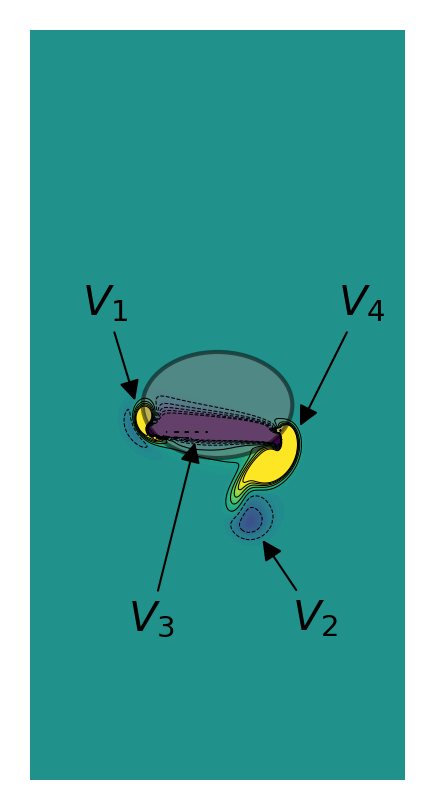}
    \caption{}
    \label{fig:baseline_wx_slices:b}
  \end{subfigure}
  \begin{subfigure}[t]{0.24\textwidth}
    \centering
    \includegraphics[width=\linewidth]{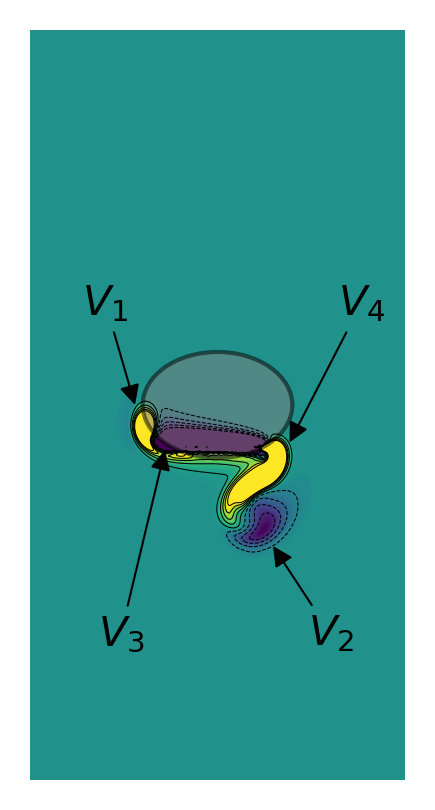}
    \caption{}
    \label{fig:baseline_wx_slices:c}
  \end{subfigure}
  \begin{subfigure}[t]{0.24\textwidth}
    \centering
    \includegraphics[width=\linewidth]{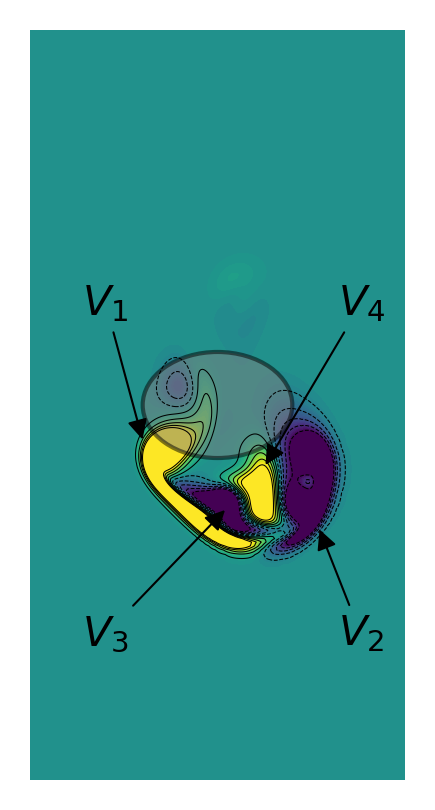}
    \caption{}
    \label{fig:baseline_wx_slices:d}
  \end{subfigure}
  \vspace{0.5cm}
  \begin{subfigure}[t]{0.24\textwidth}
    \centering
    \includegraphics[width=\linewidth]{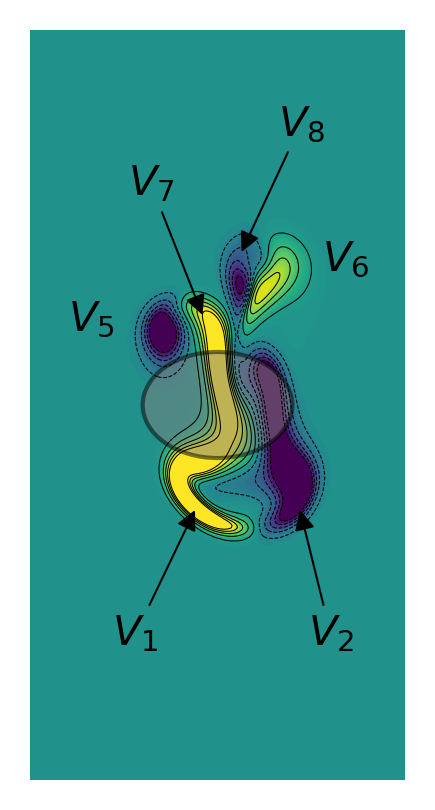}
    \caption{}
    \label{fig:baseline_wx_slices:e}
  \end{subfigure}
  \begin{subfigure}[t]{0.24\textwidth}
    \centering
    \includegraphics[width=\linewidth]{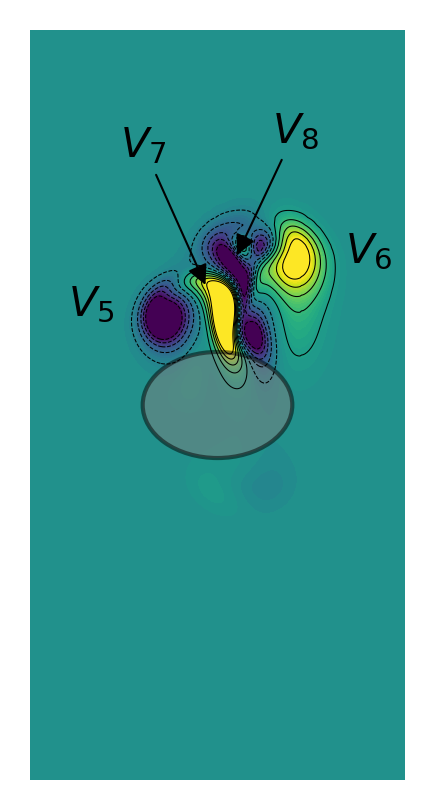}
    \caption{}
    \label{fig:baseline_wx_slices:f}
  \end{subfigure}
  \begin{subfigure}[t]{0.24\textwidth}
    \centering
    \includegraphics[width=\linewidth]{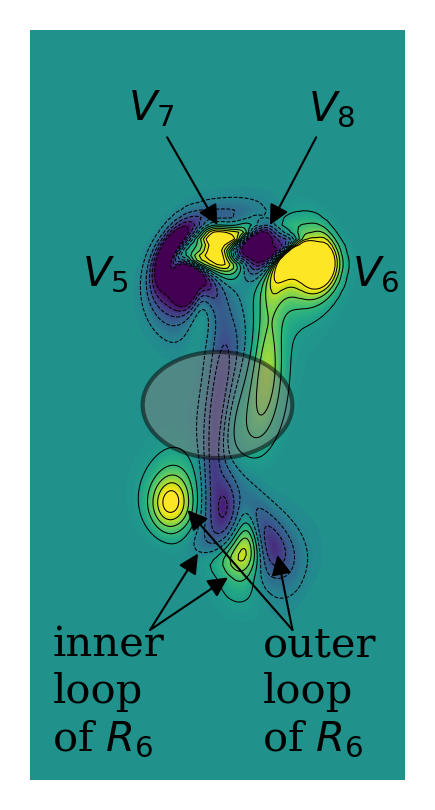}
    \caption{}
    \label{fig:baseline_wx_slices:g}
  \end{subfigure}
  \begin{subfigure}[t]{0.24\textwidth}
    \centering
    \includegraphics[width=\linewidth]{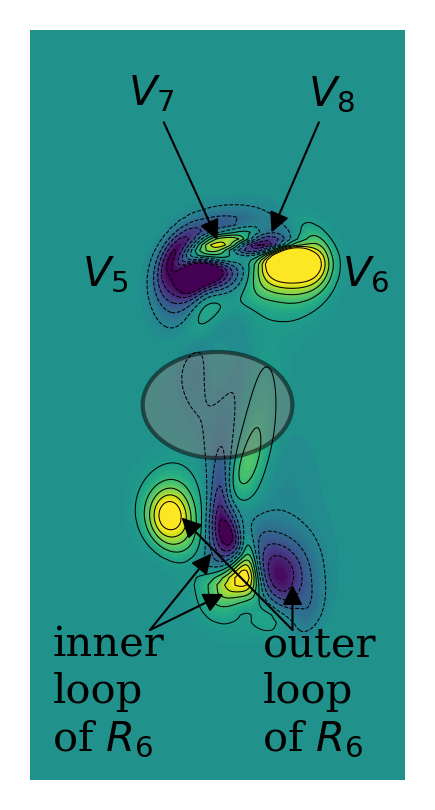}
    \caption{}
    \label{fig:baseline_wx_slices:h}
  \end{subfigure}
  \vspace{0.5cm}
  \begin{subfigure}[t]{0.24\textwidth}
    \centering
    \includegraphics[width=\linewidth]{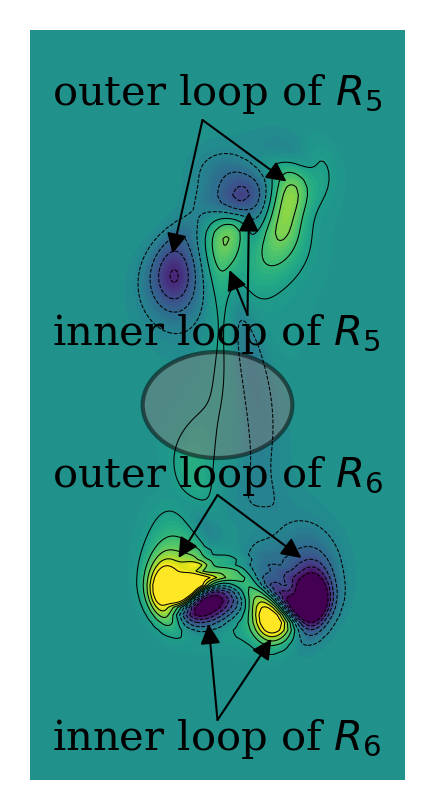}
    \caption{}
    \label{fig:baseline_wx_slices:i}
  \end{subfigure}
  \begin{subfigure}[t]{0.24\textwidth}
    \centering
    \includegraphics[width=\linewidth]{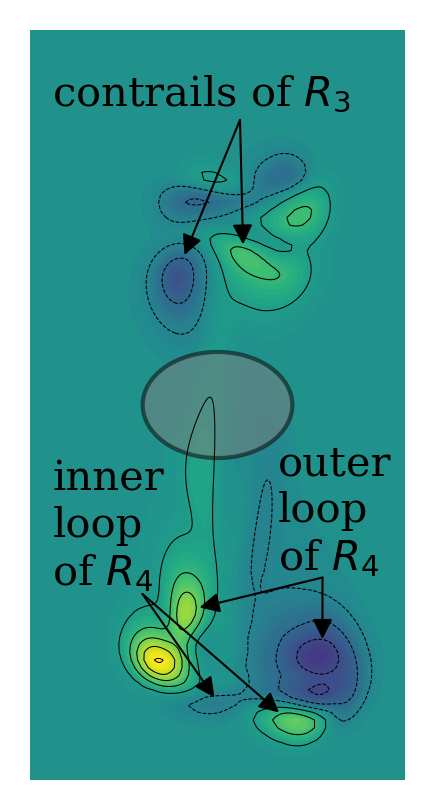}
    \caption{}
    \label{fig:baseline_wx_slices:j}
  \end{subfigure}
  \begin{subfigure}[t]{0.24\textwidth}
    \centering
    \includegraphics[width=\linewidth]{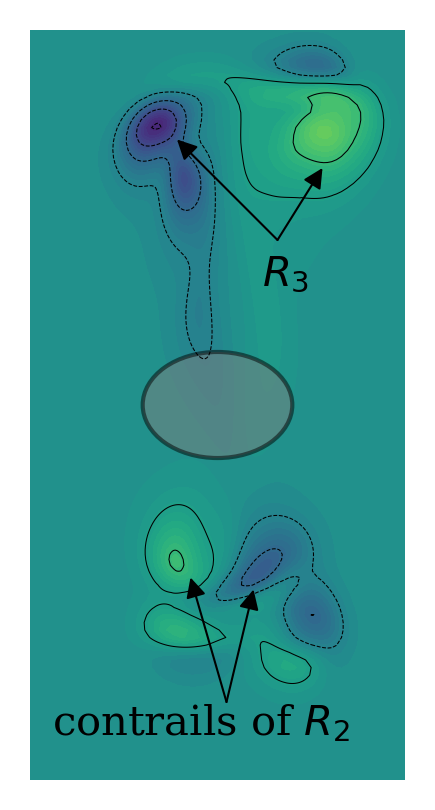}
    \caption{}
    \label{fig:baseline_wx_slices:k}
  \end{subfigure}
  \begin{subfigure}[t]{0.24\textwidth}
    \centering
    \includegraphics[width=\linewidth]{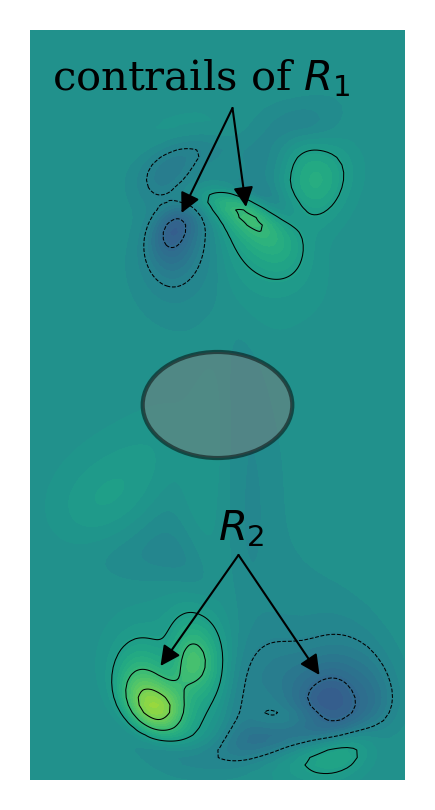}
    \caption{}
    \label{fig:baseline_wx_slices:l}
  \end{subfigure}
  \caption{Two-dimensional slices of the streamwise vorticity ($-5 \leq w_x \leq 5$) at $t / T = 4.25$ in the $y/z$ plane at different $x$ locations in the wake of a flapping circular plate ($AR = 1.27$) with $St = 0.6$, $Re = 200$, and $\psi = 90^o$. (a-l)Wake locations: $x / c = 0$, $0.2$, $0.3$, $0.75$, $1.1$, $1.3$, $1.85$, $2.0$, $2.7$, $3.8$, $4.5$, and $5.25$. (See Fig.~8 of \citet{li_dong_2016} for comparison.)}
  \label{fig:baseline_wx_slices}
\end{figure}

\cref{fig:baseline_wx_slices} shows two-dimensional slices of the streamwise vorticity component, in the $y/z$ plane at various locations along the $x$-axis, at the middle of the upstroke ($t/T = 4.25$)

\begin{figure}[!h]
  \centering
  \includegraphics[width=\linewidth]{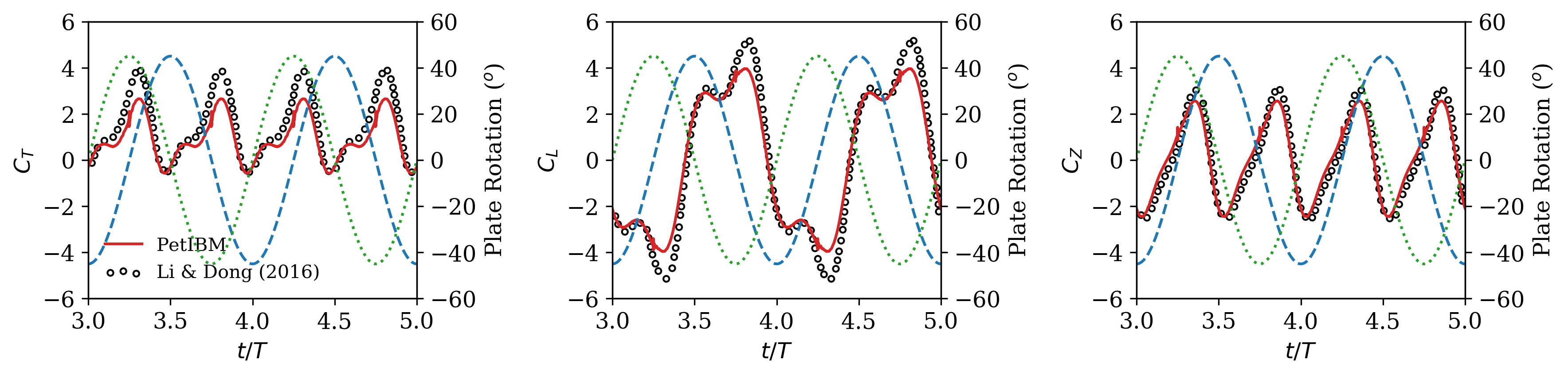}
  \caption{History of the thrust ($C_T$), lift ($C_L$), and spanwise ($C_Z$) coefficients during the fourth and fifth flapping cycles of a circular plate ($AR = 1.27$) with $St = 0.6$, $Re = 200$, and $\psi = 90^o$. We also report the digitized coefficients from Fig.~9 of \cite{li_dong_2016} and the instantaneous values of the rolling and pitching angles (dashed blue line and dotted green line, respectively).}
  \label{fig:baseline_force_coefficients}
\end{figure}

\begin{table}[!h]
  \centering
  \begin{tabular}{lcccccc}
    \hline\hline
    Case & $\left| C_T \right|_\text{max}$ & $\overline{C_T}$ & $\left| C_L \right|_\text{max}$ & $\overline{C_L}$ & $\left| C_Z \right|_\text{max}$ & $\overline{C_Z}$ \\
    \hline
    Present & $2.65$ & $0.91$ & $3.96$ & $0.0$ & $2.56$ & $0.1$ \\
    \citet{li_dong_2016} & $3.98$ & $1.46$ & $5.35$ & $0.0$ & $3.19$ & $0.1$ \\
    \hline\hline
  \end{tabular}
  \caption{Force coefficient statistics for the baseline case. We also report the values from \citet{li_dong_2016}.}
  \label{tab:baseline_force_coefficients}
\end{table}

\cref{fig:baseline_force_coefficients} shows the history of the thrust, lift, and spanwise coefficients over two flapping cycles.
\cref{tab:baseline_force_coefficients} reports statistics about the force coefficients, with comparison to the values reported in the original study.
Although peak values for all force components and mean thrust coefficient are smaller than the values reported in \citet{li_dong_2016}, we observe similar features in the force signals.
Mean lift and spanwise forces are approximately zero.
Peak magnitudes in the lift and spanwise forces are in similar range to the peak thrust.
Each half cycle, we observe two peaks in the thrust and lift forces, when the plate is near the center of its trajectory, with a smaller value for the first peak.
There is also a small production of drag when the plate starts to reverse its rolling direction (e.g., at $t/T = 3.5$ when the wing starts rolling downwards).

\subsection{Effect of the Strouhal number}

Next, we look at the effect of the Strouhal number on the wake topology and aerodynamic performance.
We computed the three-dimensional flow around a circular flat plate ($AR = 1.27$) at Reynolds number $200$ with a fixed phase-difference angle $\psi = 90^o$, while varying the Strouhal number $St$.

\begin{figure}[!h]
  \centering
  \begin{subfigure}[]{0.45\textwidth}
    \centering
    \includegraphics[width=\linewidth]{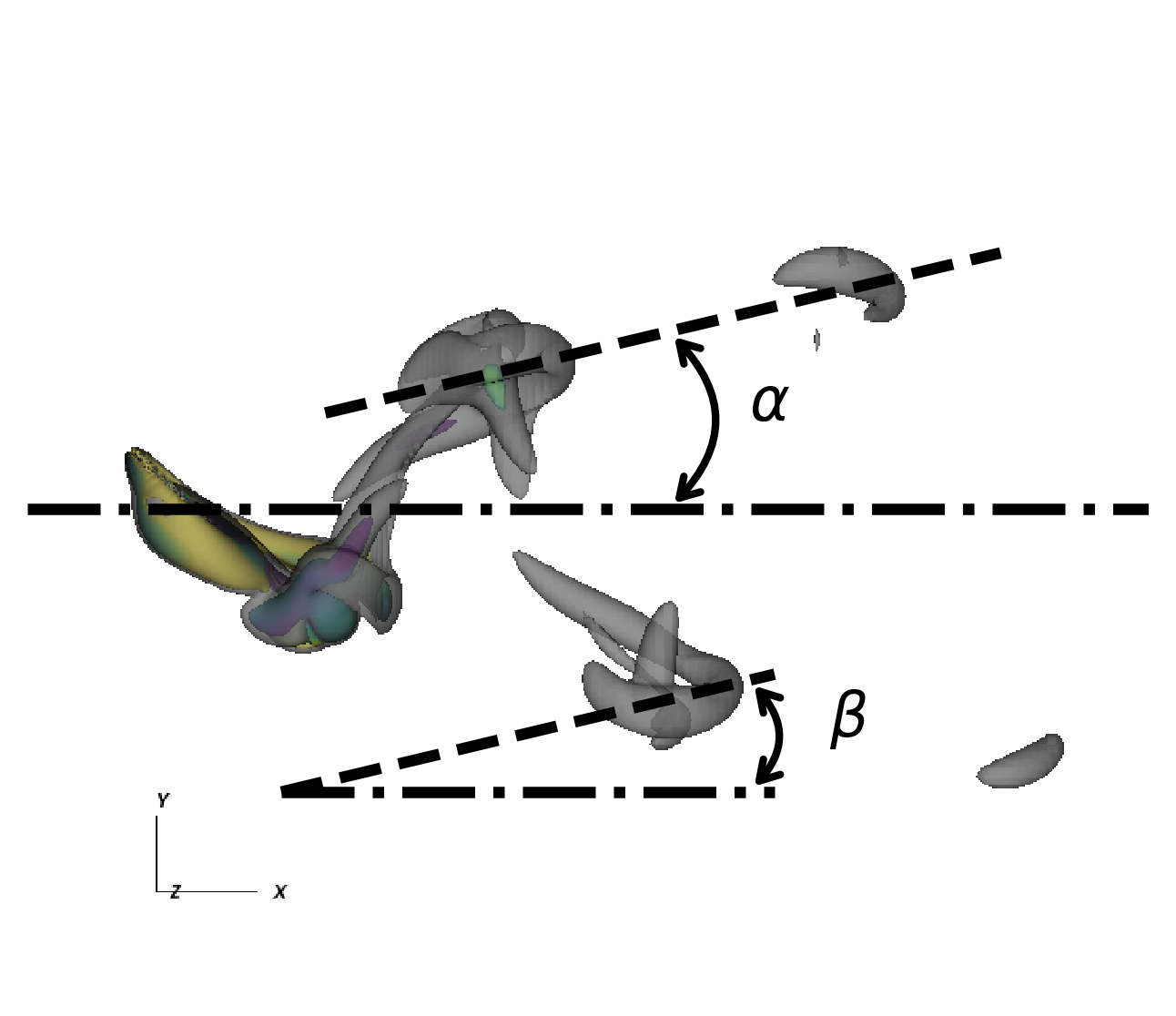}
    \caption{}
    \label{fig:strouhal_wake_topology:0.4_lateral}
  \end{subfigure}
  \hfill
  \begin{subfigure}[]{0.45\textwidth}
    \centering
    \includegraphics[width=\linewidth]{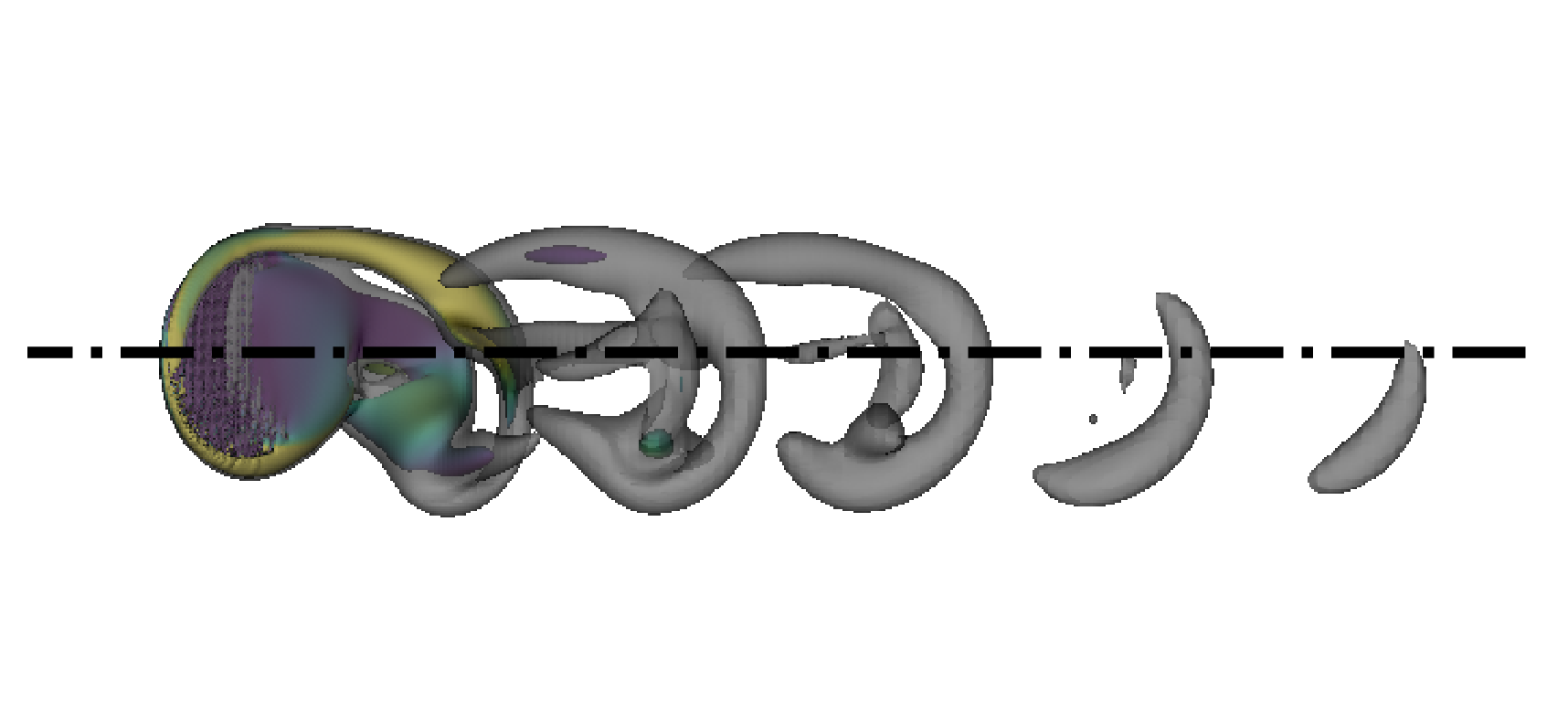}
    \caption{}
    \label{fig:strouhal_wake_topology:0.4_top}
  \end{subfigure}
  \vspace{1cm}
  \begin{subfigure}[]{0.45\textwidth}
    \centering
    \includegraphics[width=\linewidth]{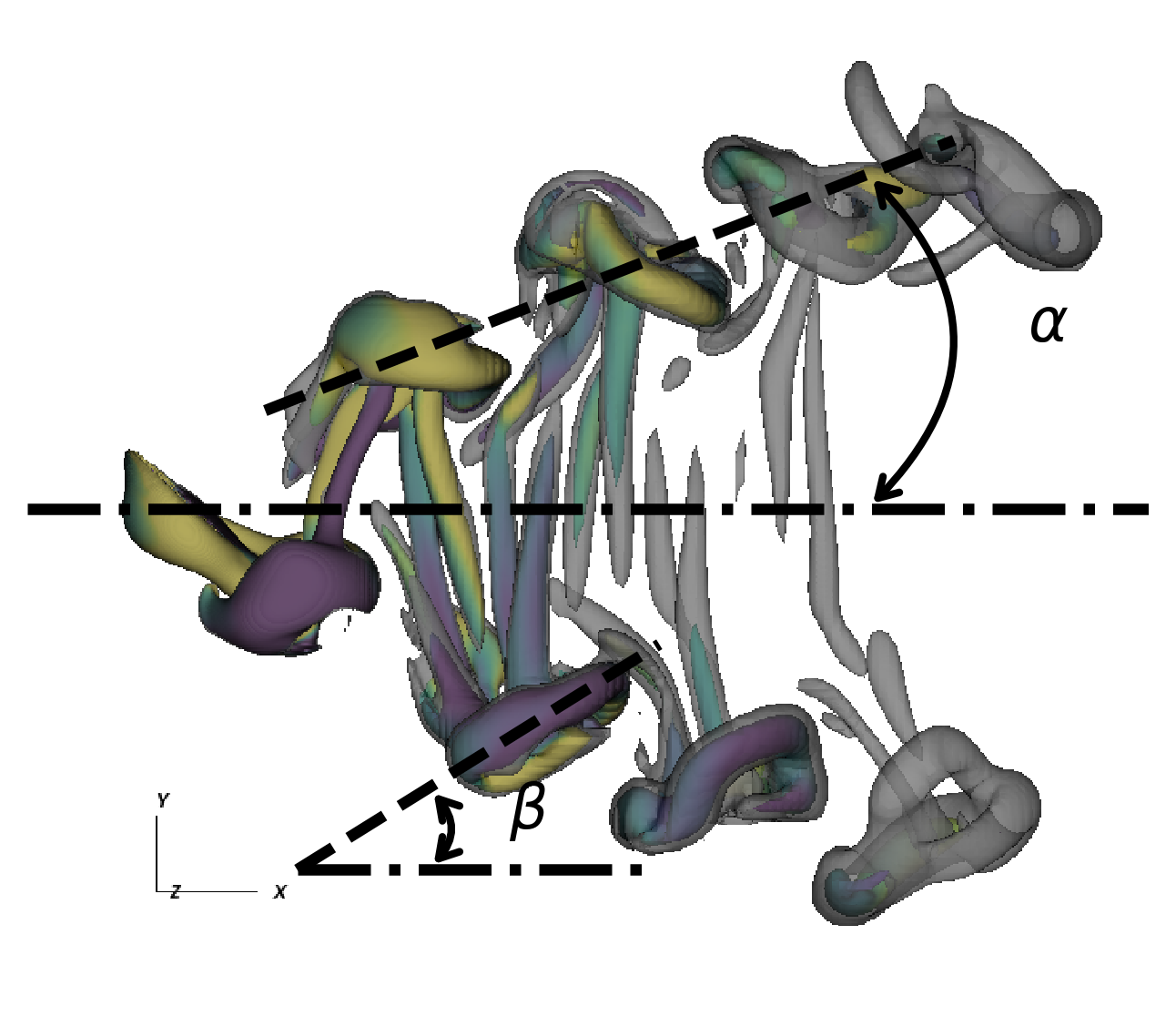}
    \caption{}
    \label{fig:strouhal_wake_topology:0.8_lateral}
  \end{subfigure}
  \hfill
  \begin{subfigure}[]{0.45\textwidth}
    \centering
    \includegraphics[width=\linewidth]{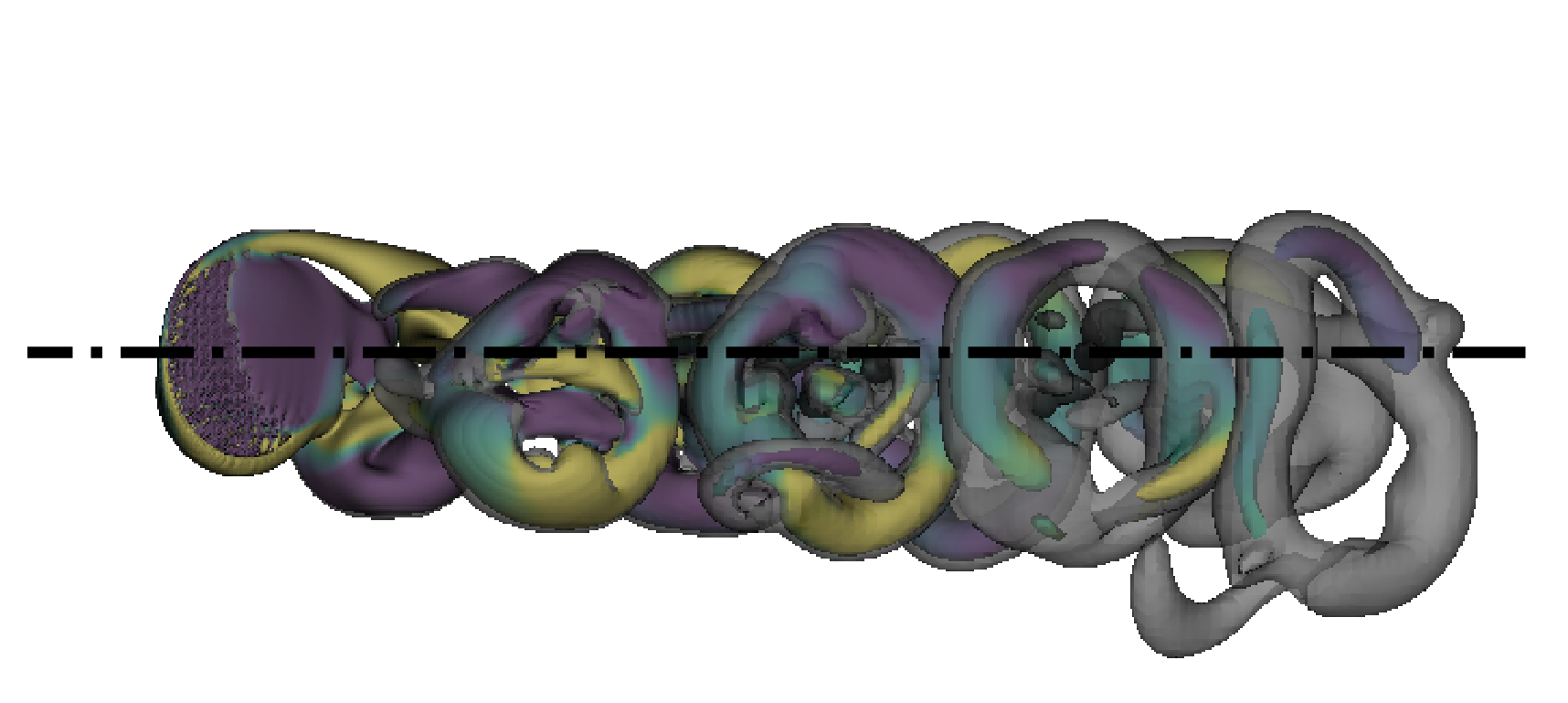}
    \caption{}
    \label{fig:strouhal_wake_topology:0.8_top}
  \end{subfigure}
  \caption{Wake topology captured at $t / T = 4.25$ (when the plate is at middle point during the upstroke) for a circular plate ($AR = 1.27$) at Reynolds number $200$ and Strouhal numbers $St = 0.4$ (top) and $St = 0.8$ (bottom). Wake structures are represented with the Q-criterion for $Q = 1$ (gray) and $Q = 6$ (colored with the streamwise vorticity, $-5 \leq w_x \leq 5$). (a,c) lateral view of the wake; (b,d) top view. (See Fig.~10 \citet{li_dong_2016} for comparison.)}
  \label{fig:strouhal_wake_topology}
\end{figure}

\begin{table}[!h]
  \centering
  \begin{tabular}{ccccc}
    \hline\hline
    \multirow{2}{*}{$St$} &
      \multicolumn{2}{c}{Present} &
      \multicolumn{2}{c}{\citet{li_dong_2016}} \\
    & $\alpha$ ($^o$) & $\beta$ ($^o$) & $\alpha$ ($^o$) & $\beta$ ($^o$) \\
    \hline
    $0.4$ & $13$ & $13$ & $16$ & $20$ \\
    $0.6$ & $18$ & $29$ & $21$ & $24$ \\
    $0.8$ & $21$ & $32$ & $27$ & $35$ \\
    $1.0$ & $26$ & $26$ & $30$ & $40$ \\
    $1.2$ & $24$ & $25$ & $31$ & $28$ \\
    \hline\hline
  \end{tabular}
  \caption{Effect of the Strouhal number on the wake oblique angle ($\alpha$) and vortex ring orientation angle ($\beta$) at $Re = 200$ with $\psi = 90^o$. For comparison, we also report values published in \citet{li_dong_2016}.}
  \label{tab:strouhal_angles}
\end{table}

\cref{fig:strouhal_wake_topology} shows lateral and top views of the shedding vortex pattern, at $t/T = 4.25$, obtained at Strouhal numbers $St = 0.4$ and $0.8$.
(\cref{fig:baseline_wake_topology} shows the wake topology for $St = 0.6$ at the same time instant.)
As reported in the original study, we note a decrease in the vorticity strength for the lower Strouhal number ($St = 0.4$), and a rapid evolution of the ``double-C''-shaped vortex structures into single vortex rings.
At higher Strouhal number ($St = 0.8$), we observe more interaction between adjacent vortex rings.
\citet{li_dong_2016} reported values of the oblique angle ($\alpha$), defined as the angle between the horizontal $x$-axis and the line passing through the first two shed vortex rings adjacent to the trailing-edge of the plate.
They also evaluated the inclination angle ($\beta$) of a near vortex ring with respect to the wake centerline.
\cref{tab:strouhal_angles} reports the oblique and inclination angles obtained with our simulations.
Our angles are different from the values reported in the original study.
The authors observed a monotonic increase in the oblique angle with respect to the Strouhal number and a peak in the inclination angle at Strouhal number $St = 1.0$ (followed by a sudden decrease).
Here, we observe an increase in the oblique angle but until $St = 1.0$, and the inclination angle peaks at $St = 0.8$.
Matching the observations reported in the original study, we also note the wake deflection along the mid-span axis (\cref{fig:strouhal_wake_topology:0.4_top,fig:strouhal_wake_topology:0.8_top,fig:baseline_wake_topology:top}).
The main difference is that the authors reported that for the higher Strouhal number case, the wake starts deflecting towards the tip of the wing and then gradually deflects back in the far wake.
We do not observe the back deflection in our simulation at Strouhal number $St = 0.8$ (\cref{fig:strouhal_wake_topology:0.8_top}), probably because we only computed five flapping cycles, while it looks like the simulation reported in the original study was computed for a longer time (additional vortex rings are present in the wake).

\begin{figure}[!h]
  \centering
  \includegraphics[width=0.5\textwidth]{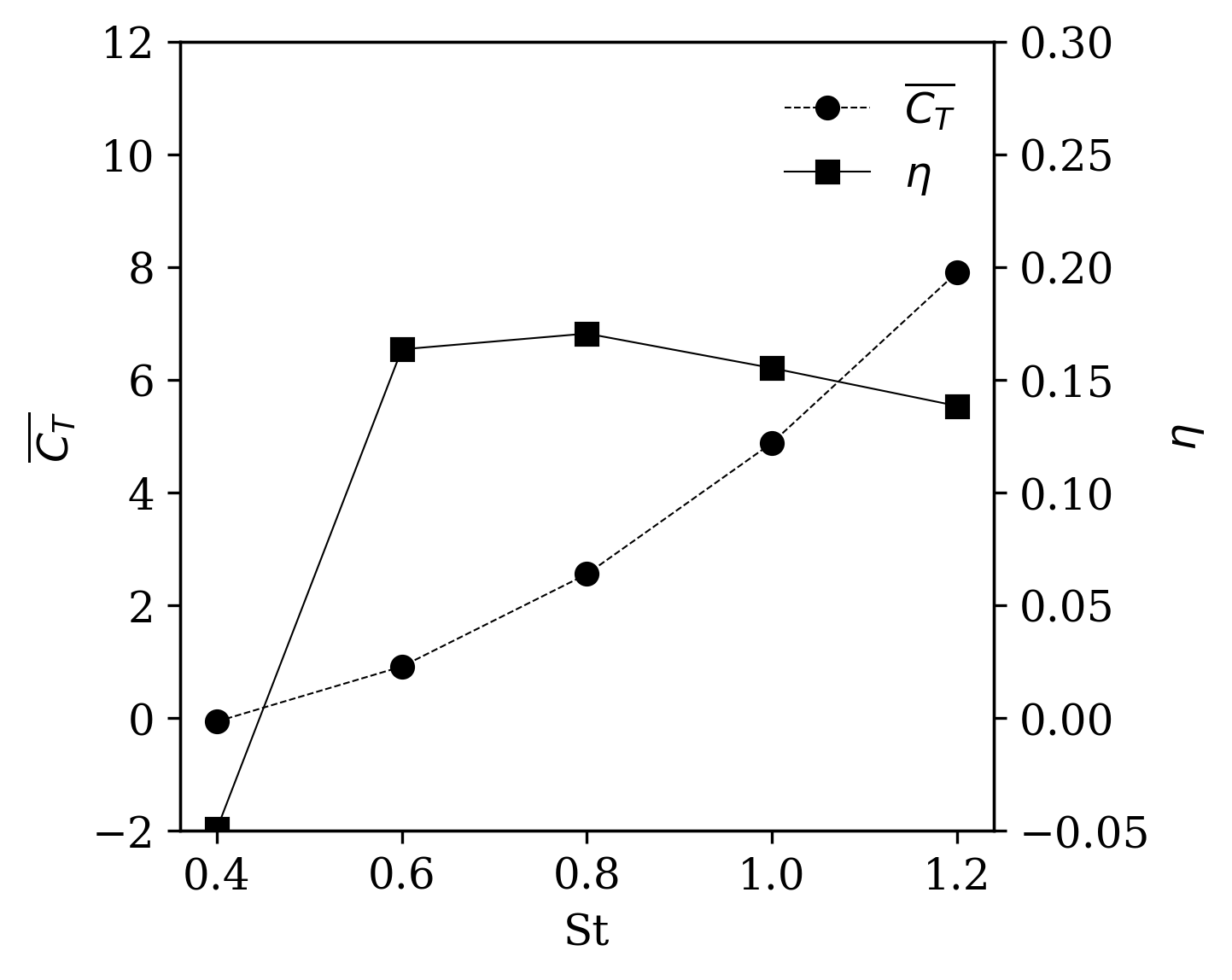}
  \caption{Cycle-averaged thrust coefficient ($\overline{C_T}$) and propulsive efficiency ($\eta$) for the baseline case as functions of the Strouhal number. Data were computed and averaged during the fifth flapping cycle. (See Fig.~11 of \citet{li_dong_2016} for comparison.)}
  \label{fig:strouhal_propulsive_efficiency}
\end{figure}

\cref{fig:strouhal_propulsive_efficiency} shows the mean thrust coefficient and propulsive efficiency for a range of Strouhal numbers.
The thrust coefficient monotonically increases with the Strouhal number.
We note a rapid increase in the propulsive efficiency, with a peak for $St = 0.8$, followed by a slow decrease.
The original study reported a maximum propulsive efficiency for $St = 0.6$, however their curve-fitting line shows that the optimal Strouhal lies between $0.6$ and $0.8$.
Here, we only report the values at the Strouhal numbers run as we do not know what interpolation procedure was done in the original study to generate the fitted line.

\subsection{Effect of the Reynolds number}

We also computed the three-dimensional flow at additional Reynolds numbers $100$ and $400$ for a circular flat plate ($AR = 1.27$) with Strouhal number $St = 0.6$ and a $90$-degree phase difference between the rolling and pitching motions.

\begin{figure}[!h]
  \centering
  \begin{subfigure}[c]{0.45\textwidth}
    \centering
    \includegraphics[width=\linewidth]{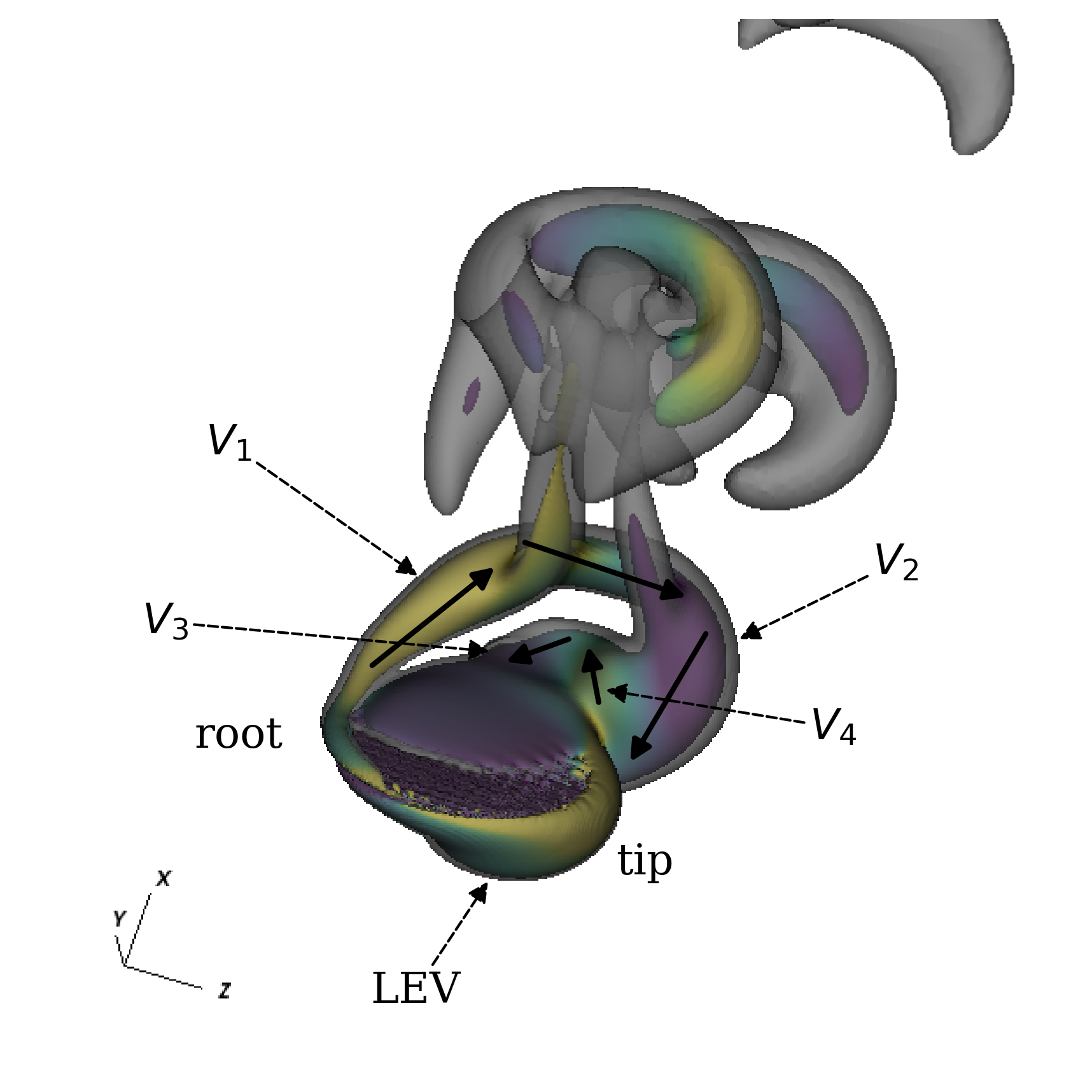}
    \caption{}
    \label{fig:reynolds_wake_topology:100_perspective}
  \end{subfigure}
  \hfill
  \begin{subfigure}[c]{0.45\textwidth}
    \centering
    \includegraphics[width=\linewidth]{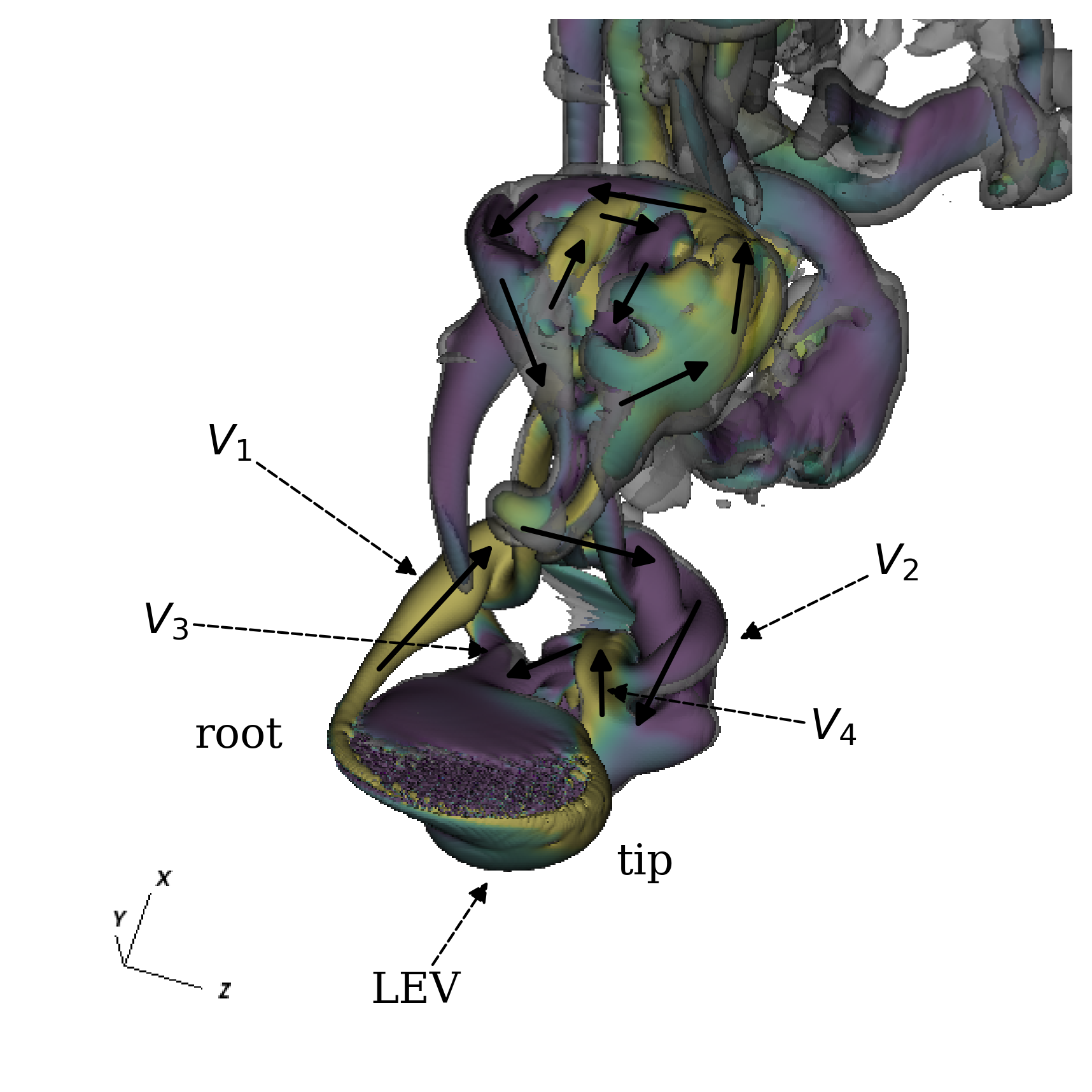}
    \caption{}
    \label{fig:reynolds_wake_topology:400_perspective}
  \end{subfigure}
  \caption{Vortex topology at $t / T = 4.25$ for a circular plate ($AR = 1.27$) with $St = 0.6$ and $\psi = 90^o$ at Reynolds numbers $Re = 100$ (a) and $Re = 400$ (b). We visualized vortical structures using the $Q$-criterion with $Q = 1$ (in gray) and $Q = 6$ (colored by the streamwise vorticity, $-5 \leq w_x \leq 5$). (See Fig.~12 of \citet{li_dong_2016} for comparison.)}
  \label{fig:reynolds_wake_topology}
\end{figure}

\cref{fig:reynolds_wake_topology} shows a perspective view of the near-wake topology at $t/T = 4.25$ at Reynolds numbers $100$ and $400$.
(\cref{fig:baseline_qcrit_perspective:0008500} shows the same view for $Re = 200$.)
For all Reynolds numbers investigated, we note the formation of a double-loop vortex around the trailing edge of the plate.
At lower Reynolds number ($Re = 200$), vortical structures dissipate more rapidly (due to increased viscous effects).
At the higher  Reynolds number ($Re = 400$), ``double-C''-shaped vortex rings are convected downstream.
As expected, vortex structures dissipate more rapidly for low Reynolds numbers.
At Reynolds number $Re = 400$, ``double-C''-shaped vortex rings propagate downstream.
Overall, we observe similar features as those published in \citet{li_dong_2016} and confirm that the flow dynamics of the low-aspect-ratio flapping wing is insensitive to the Reynolds number, at least, for the range investigated here.

\begin{figure}[!h]
  \centering
  \includegraphics[width=\linewidth]{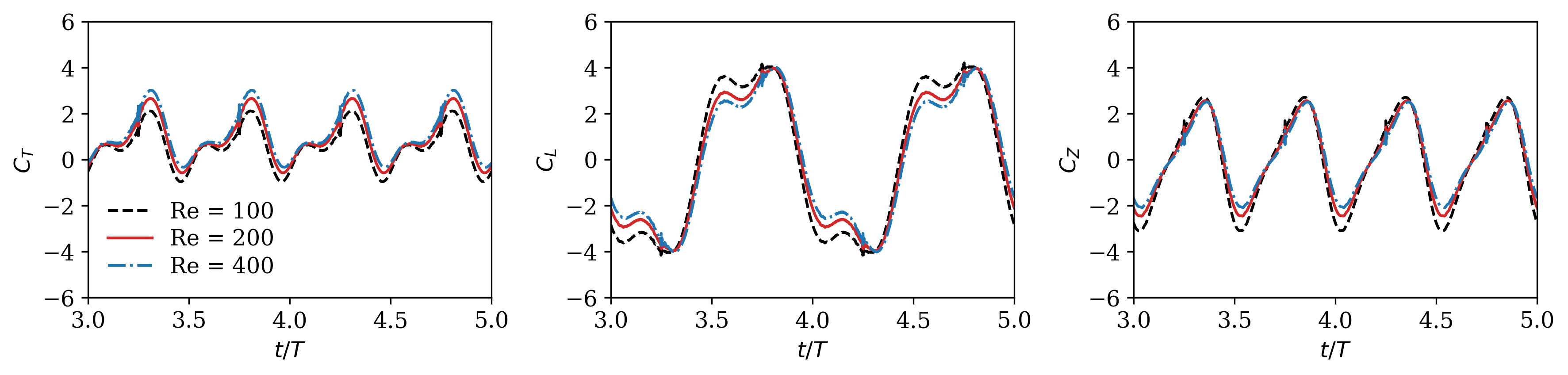}
  \caption{History of the thrust ($C_T$), lift ($C_L$), and spanwise ($C_Z$) coefficients over two flapping cycles for a circular flat plate ($AR = 1.27$) at Reynolds numbers $100$, $200$, and $400$. Strouhal number is $0.6$ with a phase difference of $90^o$ for all these cases. (See Fig.~13 of \citet{li_dong_2016} for comparison.)}
  \label{fig:reynolds_force_coefficients}
\end{figure}

\begin{table}[!h]
  \centering
  \begin{tabular}{cllll}
    \hline\hline
    \multirow{2}{*}{$Re$} &
      \multicolumn{2}{c}{Present} &
      \multicolumn{2}{c}{\citet{li_dong_2016}} \\
    & $\max \left( |C_T| \right)$ & $\overline{C_T}$ & $\max \left( |C_T| \right)$ & $\overline{C_T}$ \\
    \hline
    $100$ & $2.12$ & $0.58$ & $3.45$ & $1.17$ \\
    $200$ & $2.65$ ($+25\%$) & $0.91$ ($+57\%$) & $3.98$ ($+15\%$) & $1.46$ ($+25\%$) \\
    $400$ & $3.01$ ($+42\%$) & $1.14$ ($+97\%$) & $4.35$ ($+26\%$) & $1.65$ ($+41\%$) \\
    \hline\hline
  \end{tabular}
  \caption{Effect of the Reynolds number on the mean and peak thrust coefficient at $St = 0.6$ with $\psi = 90^o$ for a circular plate ($AR = 1.27$). Relative differences with respect to the case at $Re = 100$ are shown between parenthesis. For comparison, we also report values published in \citet{li_dong_2016}.}
  \label{tab:reynolds_thrust_stats}
\end{table}

\cref{fig:reynolds_force_coefficients} displays the history of the force coefficients over two flapping cycles for $Re =100$, $200$, and $400$.
\cref{tab:reynolds_thrust_stats} reports the mean and peak thrust coefficients comparing to the data published in \citet{li_dong_2016}.
Although we obtained different statistics in the force coefficients (mean and peak values), we report similar trends as in the original study.
First, the absolute peak value and mean value for the thrust coefficient increase with the Reynolds number.
Second, the lift force peaks twice every half cycle, for all Reynolds numbers.
Third, the negative peak in the spanwise force decreases with the Reynolds number.
Finally, the mean lift and spanwise coefficients remain approximately zero for all cases.

\subsection{Effect of the wing aspect ratio}

The baseline case investigated the flow dynamics produced by a circular wing ($AR = 1.27$).
Following the original study, we ran two additional simulations, for $AR = 1.91$ and $2.55$, to look at the effect of the wing aspect ratio on the wake topology and aerodynamic forces.
(Other parameters remained identical to the baseline case.)

\begin{figure}[!h]
  \centering
  \begin{subfigure}[c]{0.45\textwidth}
    \centering
    \includegraphics[width=\linewidth]{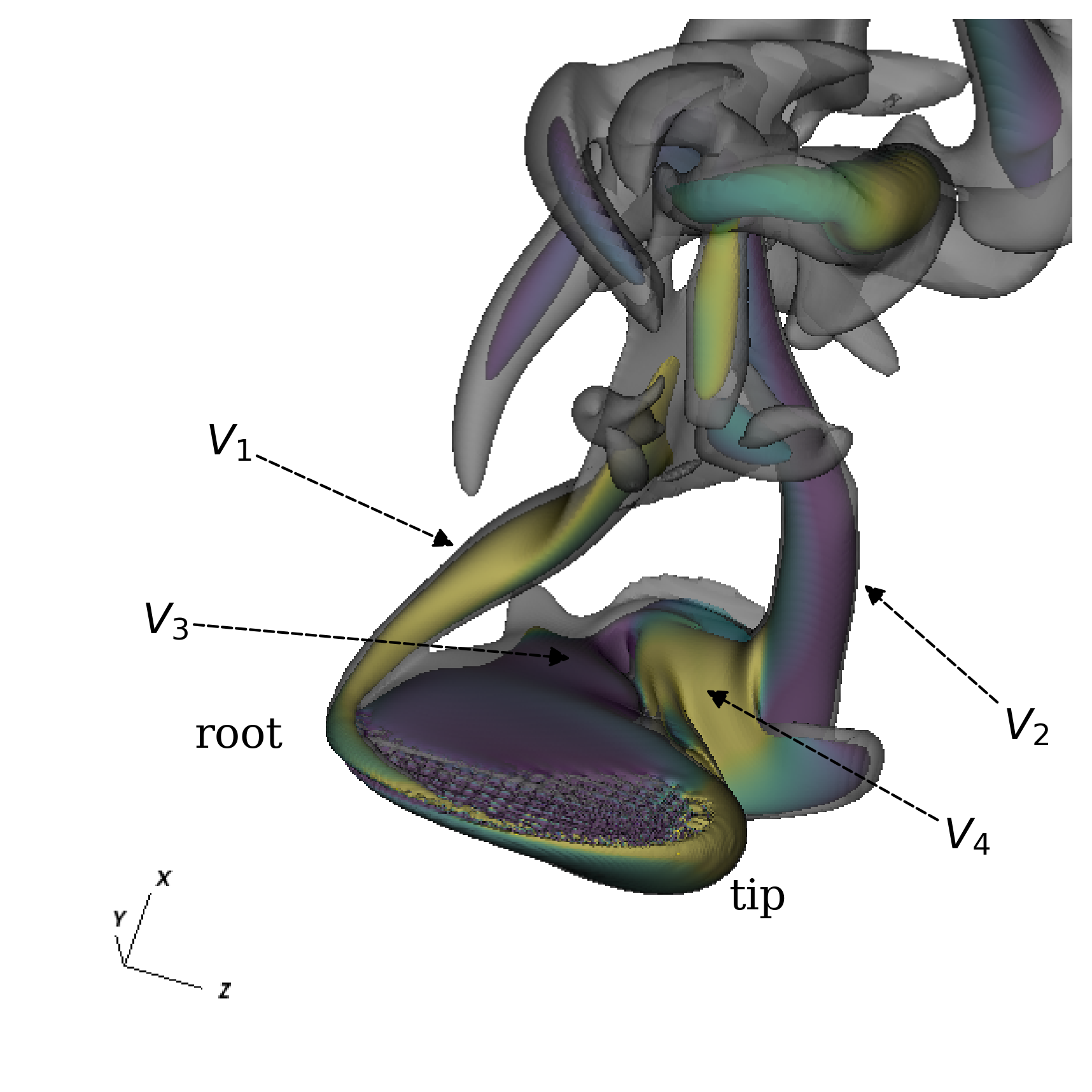}
    \caption{}
    \label{fig:ratio_wake_topology:1.91_perspective}
  \end{subfigure}
  \hfill
  \begin{subfigure}[c]{0.45\textwidth}
    \centering
    \includegraphics[width=\linewidth]{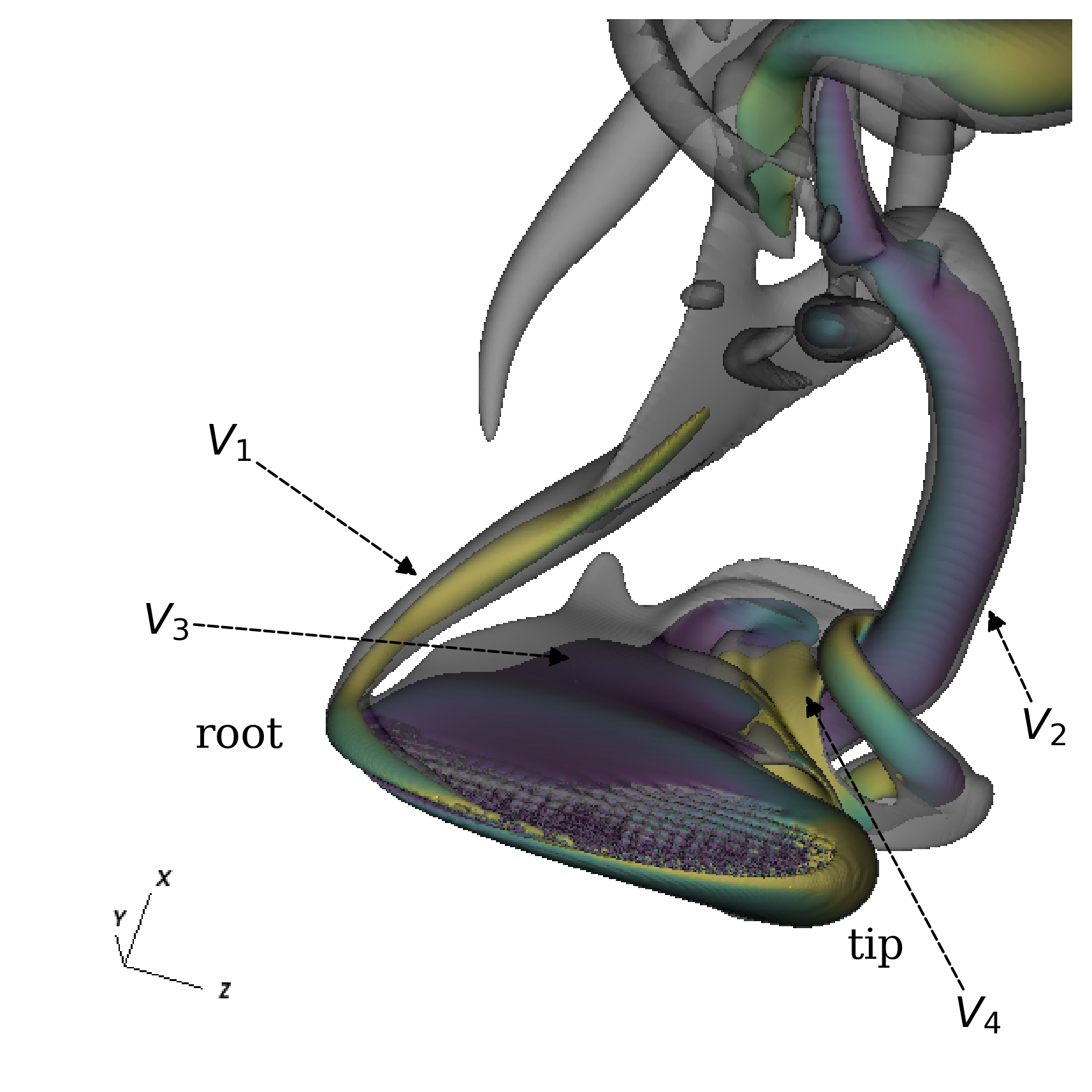}
    \caption{}
    \label{fig:ratio_wake_topology:2.55_perspective}
  \end{subfigure}
  \caption{Vortex topology at $t / T = 4.25$ for an elliptical plate with $AR = 1.91$ (a) and $AR = 2.55$ (b) for $Re = 200$, $St = 0.6$, and $\psi = 90^o$. We visualized vortical structures using the $Q$-criterion with $Q = 1$ (in gray) and $Q = 6$ (colored by the streamwise vorticity, $-5 \leq w_x \leq 5$). (See Fig.~14 of \citet{li_dong_2016} for comparison.)}
  \label{fig:ratio_wake_topology}
\end{figure}

\begin{figure}[!h]
  \centering
  \begin{subfigure}[c]{0.2\textwidth}
    \centering
    \includegraphics[width=\linewidth]{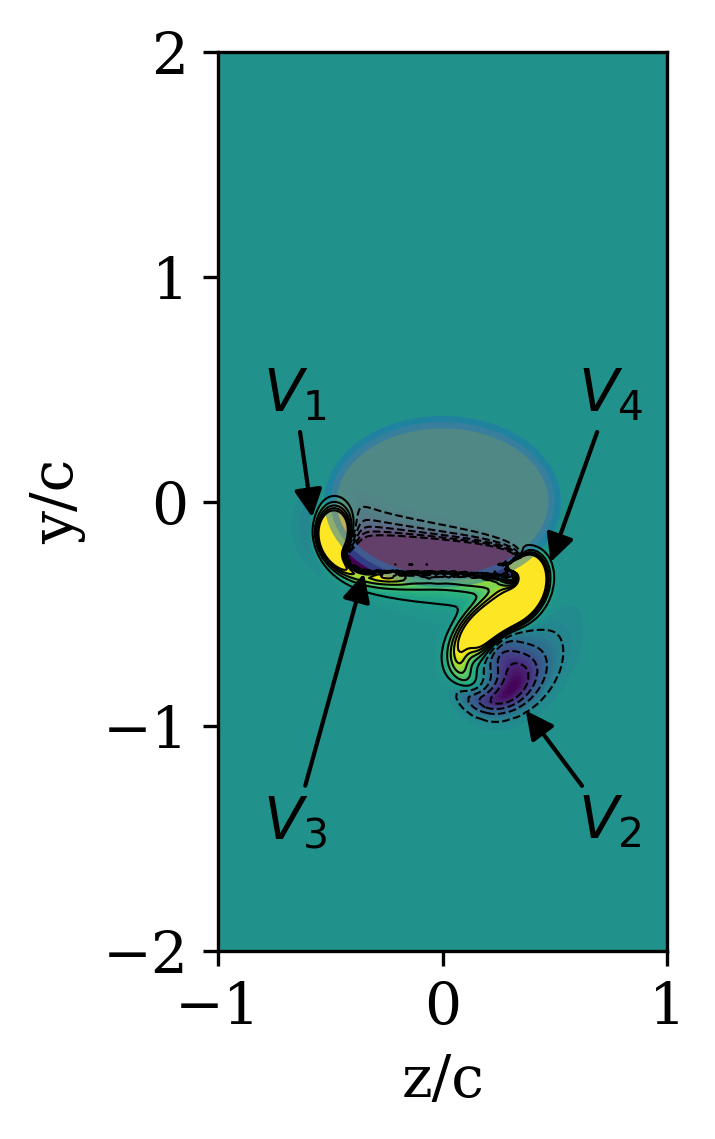}
    \caption{}
    \label{fig:ratio_vorticity_slices:1.27_wx}
  \end{subfigure}
  \hspace{3em}
  \begin{subfigure}[c]{0.2\textwidth}
    \centering
    \includegraphics[width=\linewidth]{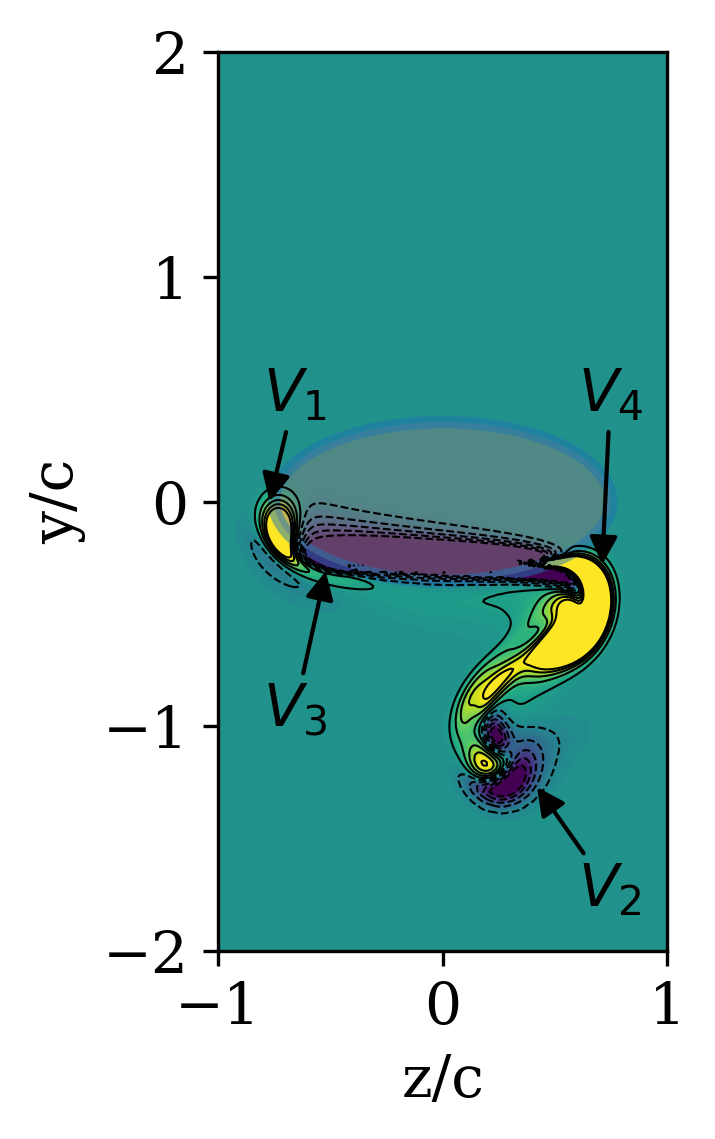}
    \caption{}
    \label{fig:ratio_vorticity_slices:1.91_wx}
  \end{subfigure}
  \hspace{3em}
  \begin{subfigure}[c]{0.23\textwidth}
    \centering
    \includegraphics[width=\linewidth]{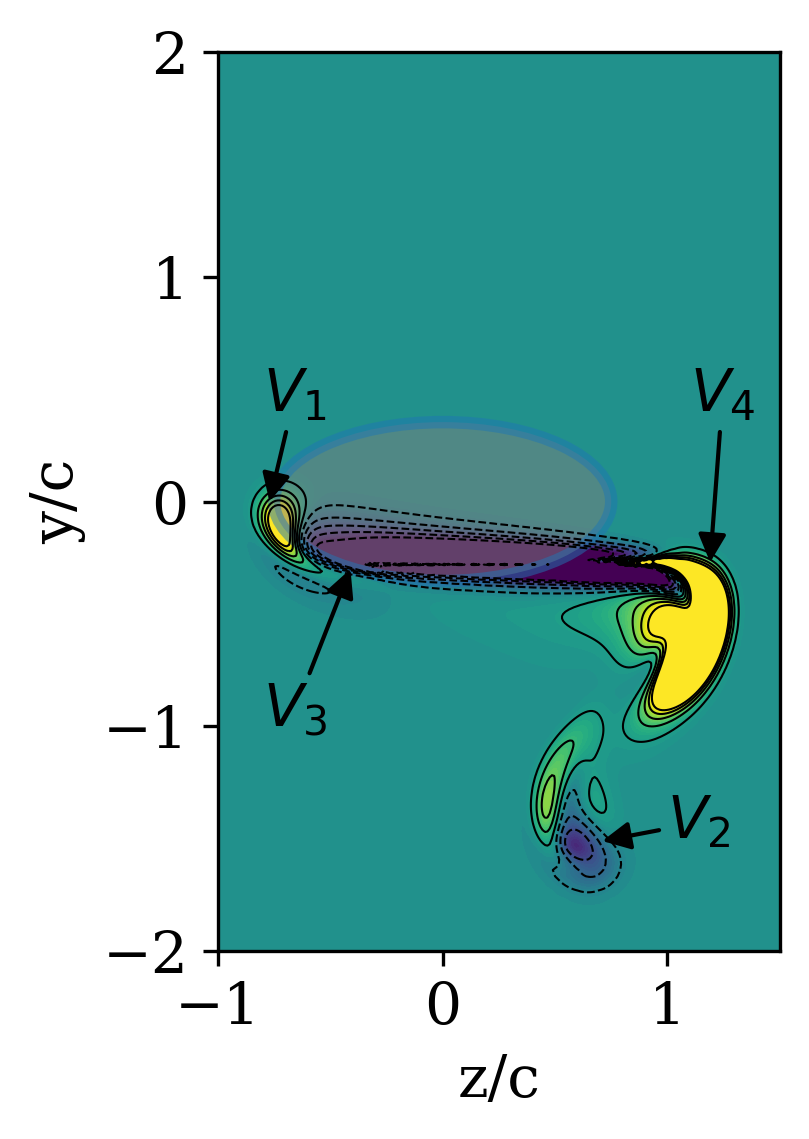}
    \caption{}
    \label{fig:ratio_vorticity_slices:2.55_wx}
  \end{subfigure}
  \vspace{0.5em}
  \begin{subfigure}[c]{0.32\textwidth}
    \centering
    \includegraphics[width=\linewidth]{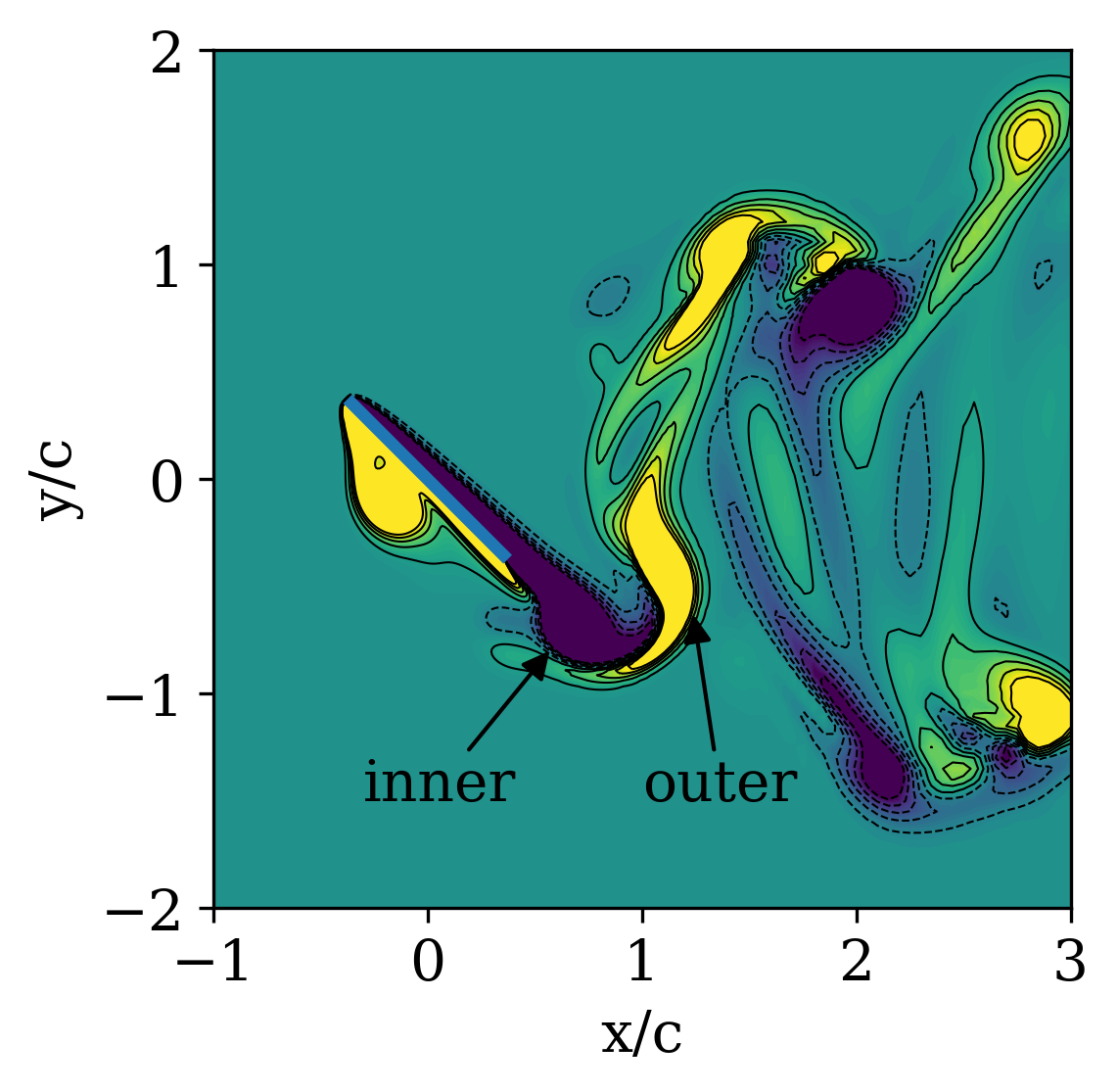}
    \caption{}
    \label{fig:ratio_vorticity_slices:1.27_wz}
  \end{subfigure}
  \hspace{0.2em}
  \begin{subfigure}[c]{0.32\textwidth}
    \centering
    \includegraphics[width=\linewidth]{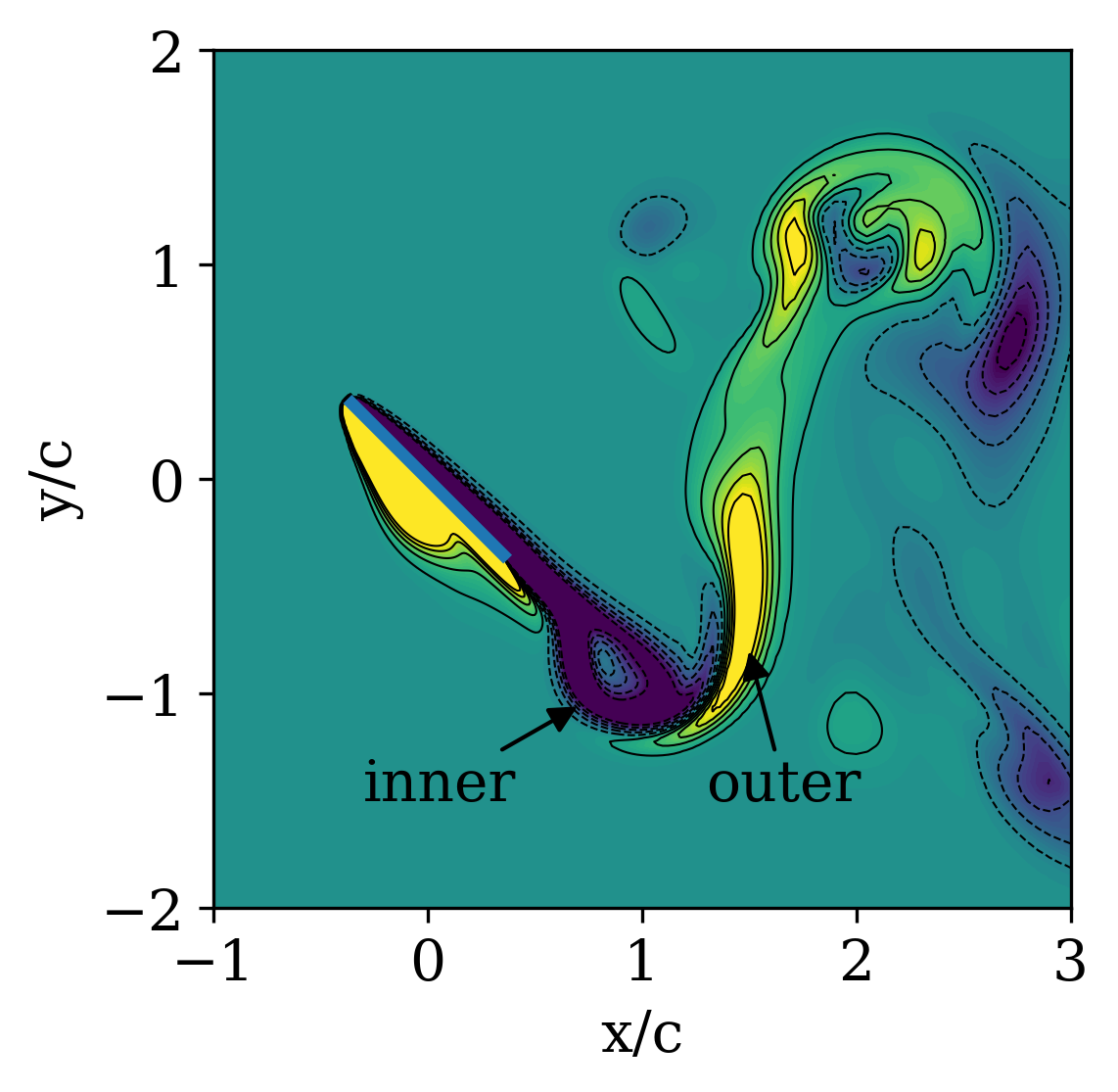}
    \caption{}
    \label{fig:ratio_vorticity_slices:1.91_wz}
  \end{subfigure}
  \hspace{0.2em}
  \begin{subfigure}[c]{0.32\textwidth}
    \centering
    \includegraphics[width=\linewidth]{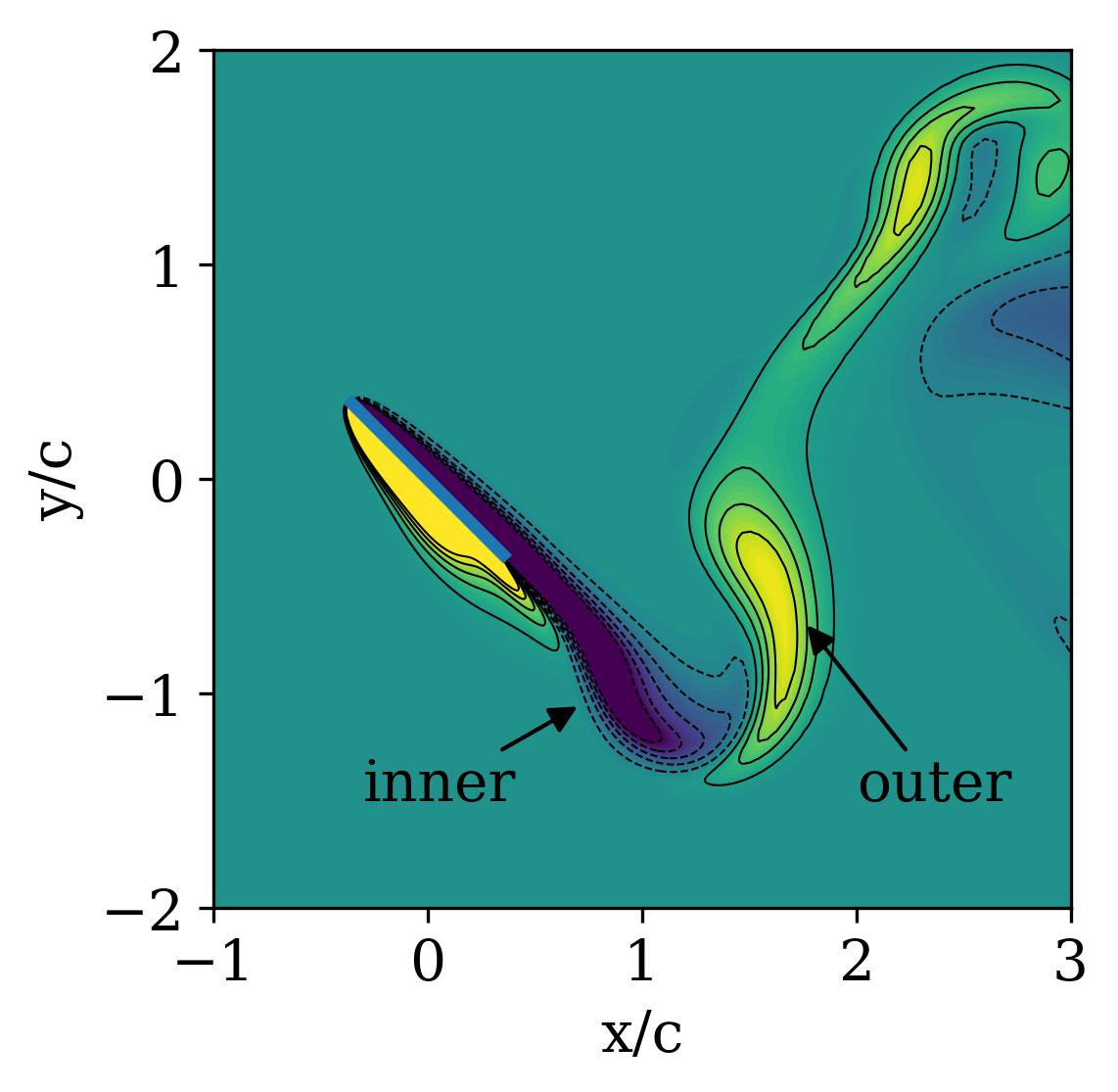}
    \caption{}
    \label{fig:ratio_vorticity_slices:2.55_wz}
  \end{subfigure}
  \caption{Comparison of the two-dimensional contours of the vorticity at $t/T = 4.25$ for elliptical plates with $AR = 1.27$ (a, d), $AR = 1.91$ (b, e), and $AR = 2.55$ (c, f). (a-c): streamwise vorticity ($-5 \leq w_x \leq 5$) in the $y/z$ plane at $x/c = 0.3$ (near wake). (d-f): spanwise vorticity ($-5 \leq w_z \leq 5$) in the $x/y$ plane at $z = S/2$ (mid-span). (See Fig.~15 of \citet{li_dong_2016} for comparison.)}
  \label{fig:ratio_vorticity_slices}
\end{figure}

\cref{fig:ratio_wake_topology} shows snapshots of the vortical structures at $t/T = 4.25$ generated by elliptical plates with aspect ratios $AR = 1.91$ and $2.55$.
\crefrange{fig:ratio_vorticity_slices:1.27_wx}{fig:ratio_vorticity_slices:2.55_wx} are two-dimensional slices of the streamwise vorticity in the near wake ($x/c = 0.3$) at the same instant in time.
\crefrange{fig:ratio_vorticity_slices:1.27_wz}{fig:ratio_vorticity_slices:2.55_wz} are slices of the spanwise vorticity at mid-span ($z = S/2$).
We stand by the observations made in the original study: as the aspect ratio increases, the size of the inner vortex loop (formed by vortices $V_3$ and $V_4$) increases in streamwise direction, while the magnitude of the spanwise vorticity decreases.

\begin{figure}[!h]
  \centering
  \includegraphics[width=\textwidth]{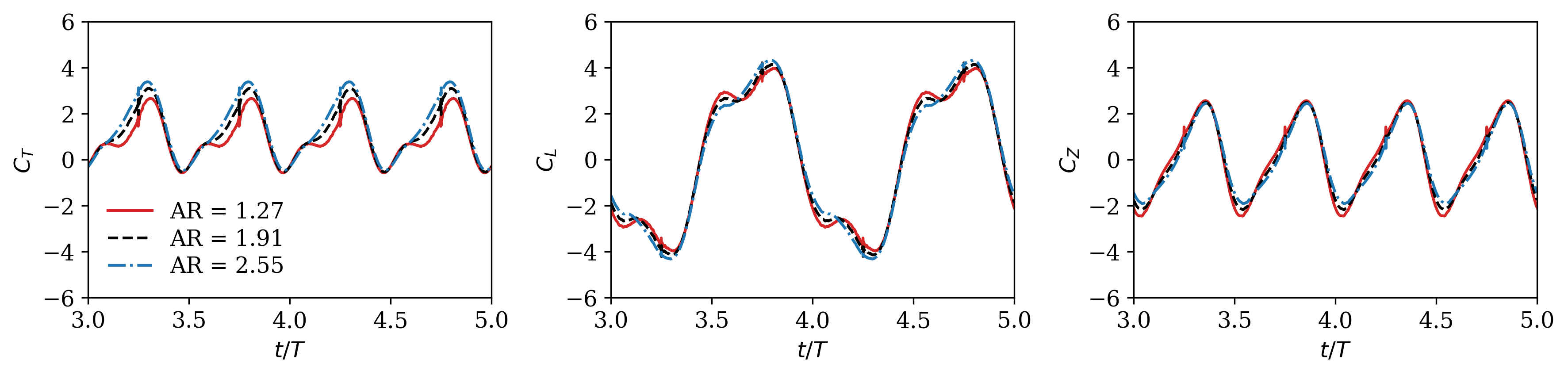}
  \caption{History of the thrust ($C_T$), lift ($C_L$), and spanwise ($C_Z$) coefficients over two flapping cycles for plates with different aspect ratios ($AR$) and with $St = 0.6$, $Re = 200$, and $\psi = 90^o$. (See Fig.~16 of \citet{li_dong_2016} for comparison.)}
  \label{fig:ratio_force_coefficients}
\end{figure}

\cref{fig:ratio_force_coefficients} displays the history of the force coefficients over two flapping cycles, obtained with aspect ratios $AR = 1.27$, $1.91$, and $2.55$.
The mean thrust coefficient increases with the wing aspect ratio ($1.17$ for $AR = 1.91$ and $1.37$ for $AR = 2.55$).
As mentioned in \citet{li_dong_2016}, this increase in the mean thrust coincides with the fact that larger aspect-ratio plates tend to have higher propulsive efficiency.
We also note a slight increase in the magnitude of the thrust peak as we increase the aspect ratio.
The authors of the original study reported a decrease in magnitude of the peak for the lift and spanwise coefficients.
However, our results show a slight increase in magnitude of the second peak (each half cycle) for the lift coefficient as the ratio increases.
Furthermore, the maximum value for the spanwise coefficient remains the same, while we note a decrease in the minimum value (as the ratio increases).

\subsection{Effect of the phase difference between pitching and rolling}

For all simulations reported so far, the phase-difference angle between the rolling and pitching motions was set to $\psi = 90^o$.
We now look at the effect of the phase difference on the wake topology and propulsive performance of a circular plate ($AR = 1.27$) at Reynolds number $Re = 200$ and Strouhal number $St = 0.6$.
We ran six additional simulations with phase-difference angles $\psi = 60^o$, $70^o$, $80^o$, $100^o$, $110^o$, and $120^o$.

\begin{figure}[!h]
  \centering
  \begin{subfigure}[]{0.45\textwidth}
    \centering
    \includegraphics[width=\linewidth]{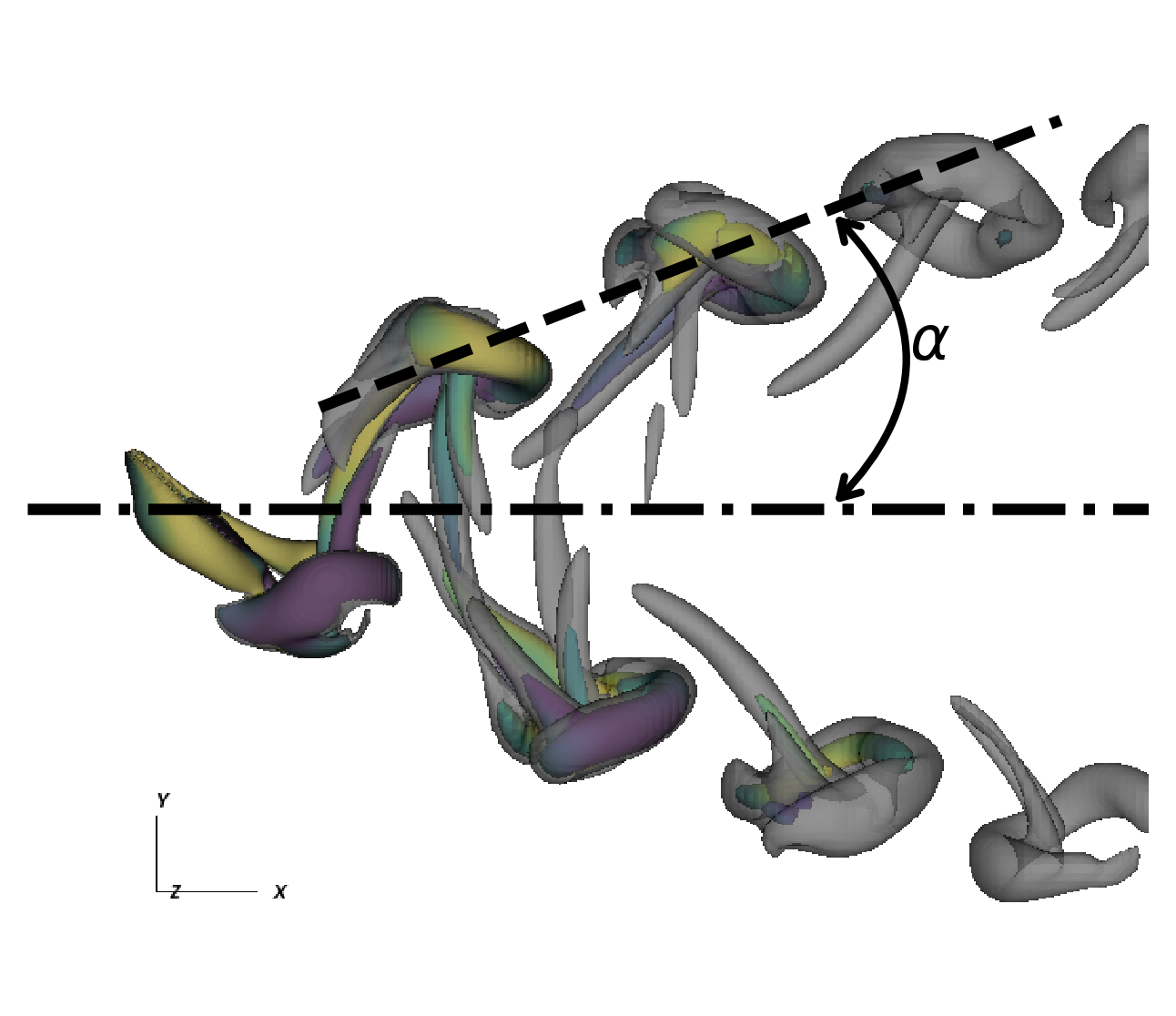}
    \caption{}
    \label{fig:phase_wake_topology:100_lateral}
  \end{subfigure}
  \hfill
  \begin{subfigure}[]{0.45\textwidth}
    \centering
    \includegraphics[width=\linewidth]{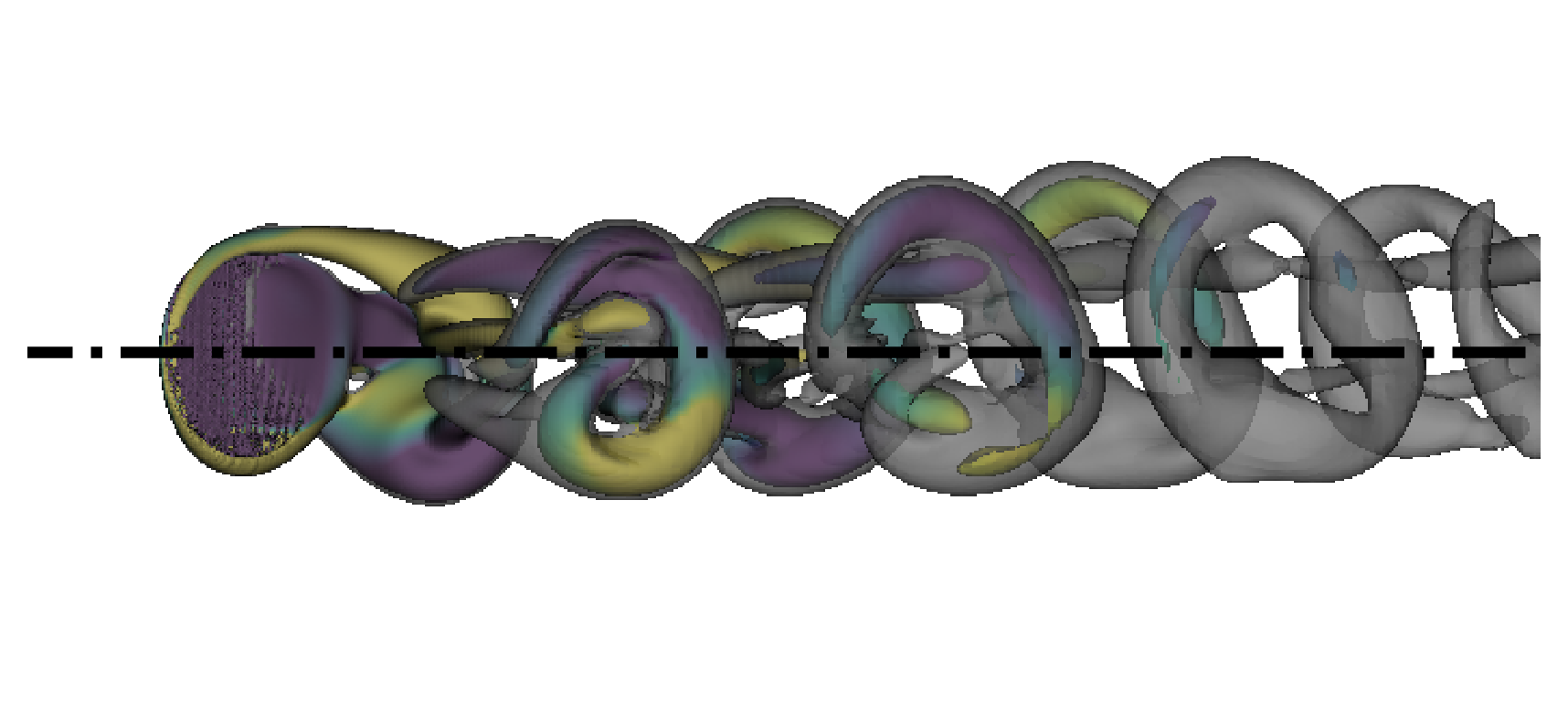}
    \caption{}
    \label{fig:phase_wake_topology:100_top}
  \end{subfigure}
  \vspace{1cm}
  \begin{subfigure}[]{0.45\textwidth}
    \centering
    \includegraphics[width=\linewidth]{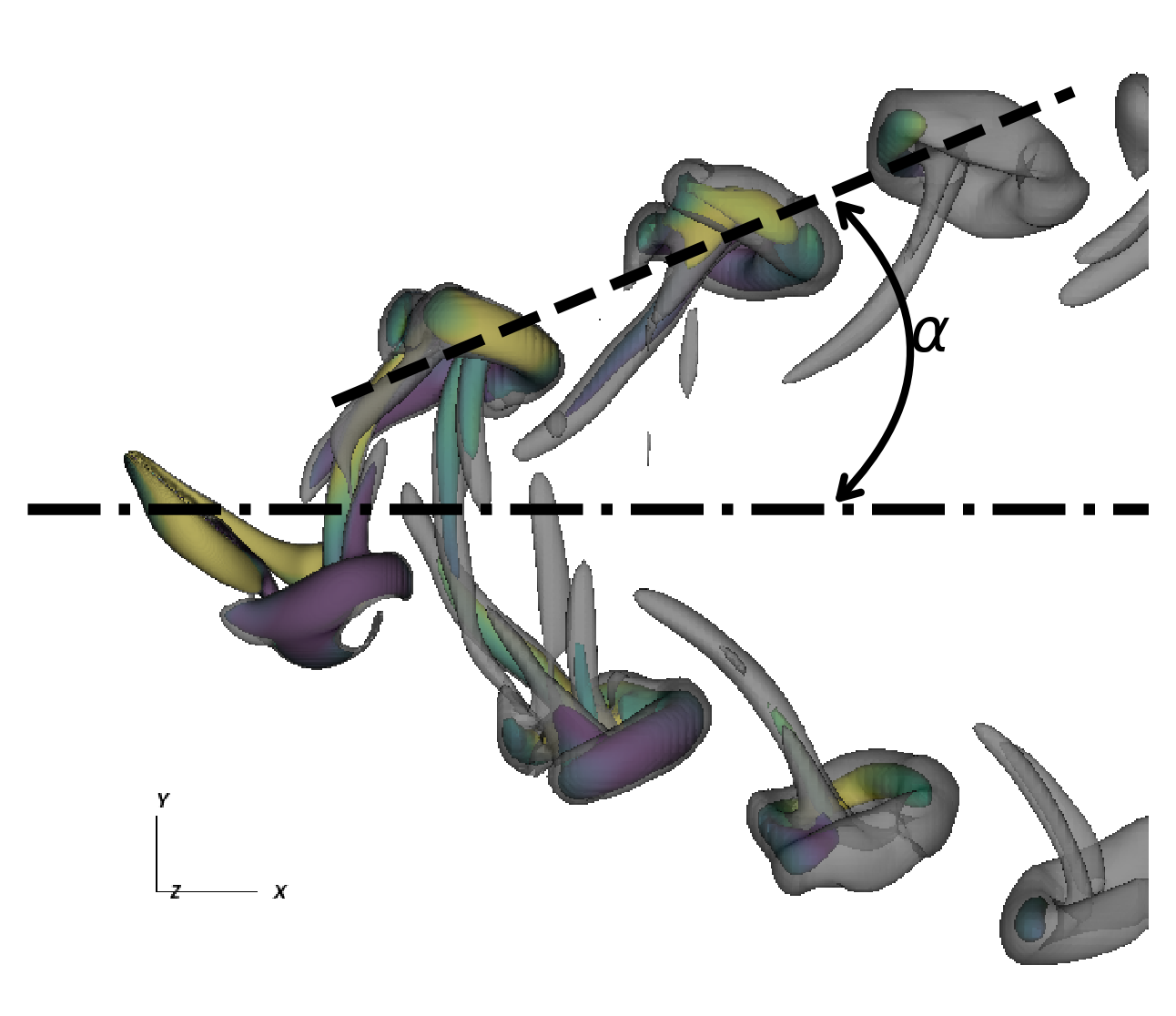}
    \caption{}
    \label{fig:phase_wake_topology:110_lateral}
  \end{subfigure}
  \hfill
  \begin{subfigure}[]{0.45\textwidth}
    \centering
    \includegraphics[width=\linewidth]{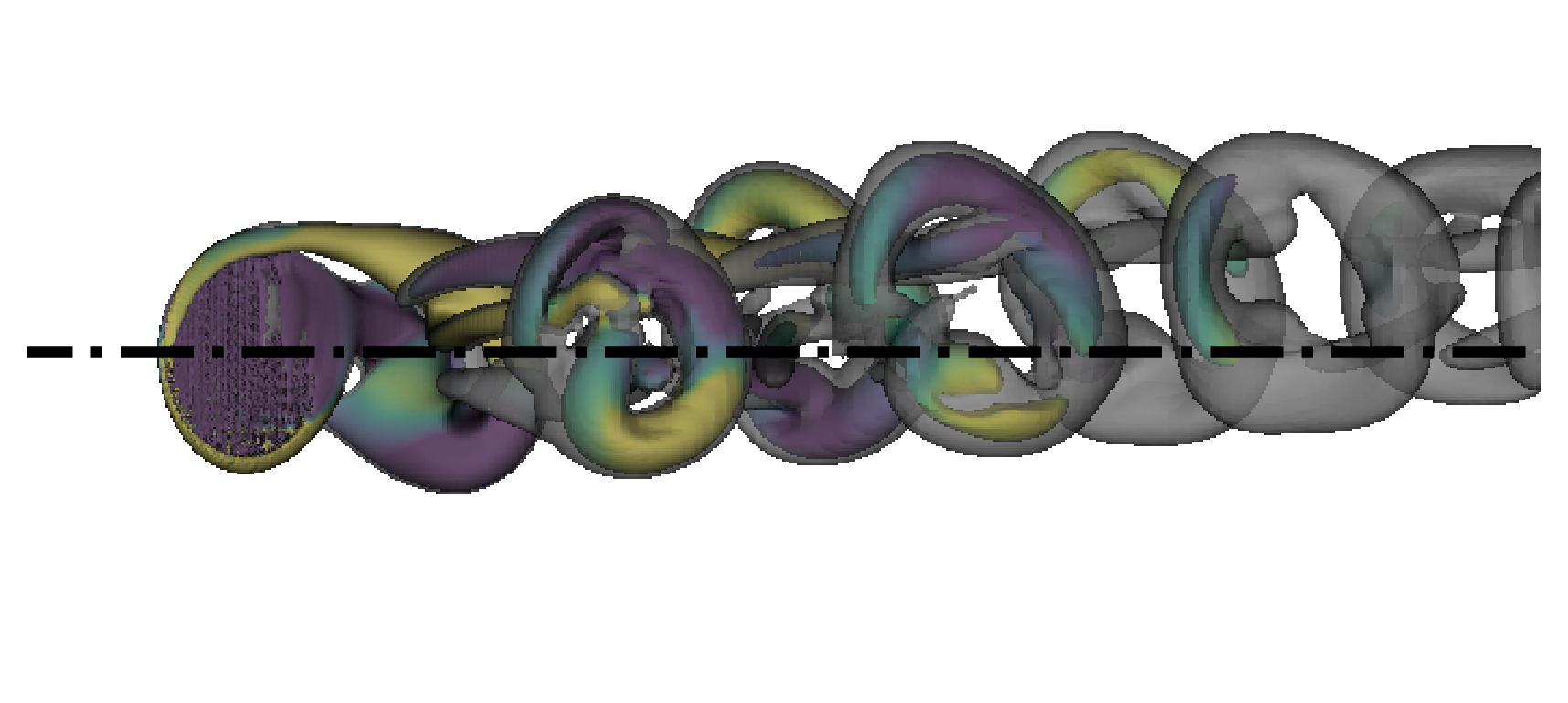}
    \caption{}
    \label{fig:phase_wake_topology:110_top}
  \end{subfigure}
  \vspace{1cm}
  \begin{subfigure}[]{0.45\textwidth}
    \centering
    \includegraphics[width=\linewidth]{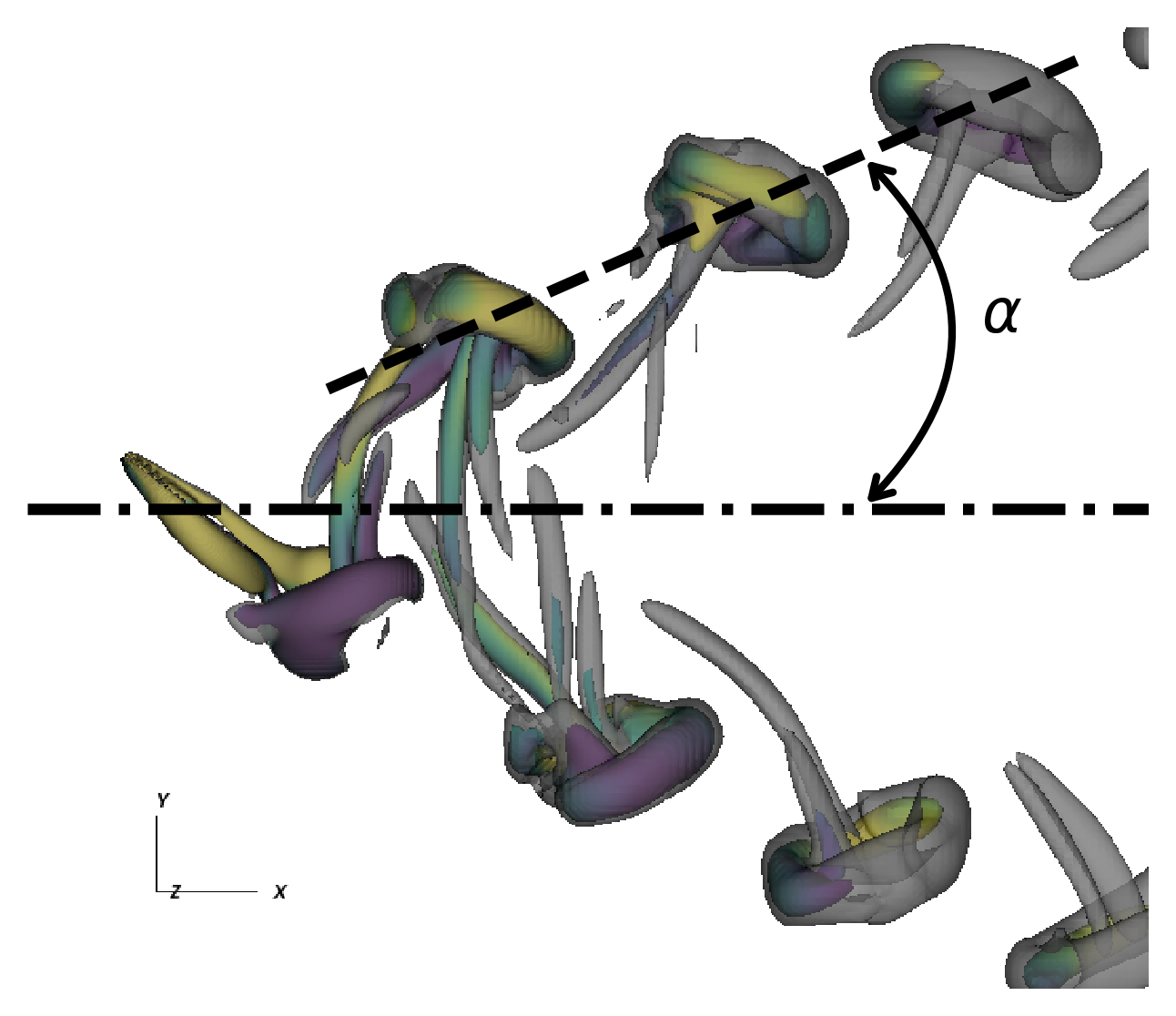}
    \caption{}
    \label{fig:phase_wake_topology:120_lateral}
  \end{subfigure}
  \hfill
  \begin{subfigure}[]{0.45\textwidth}
    \centering
    \includegraphics[width=\linewidth]{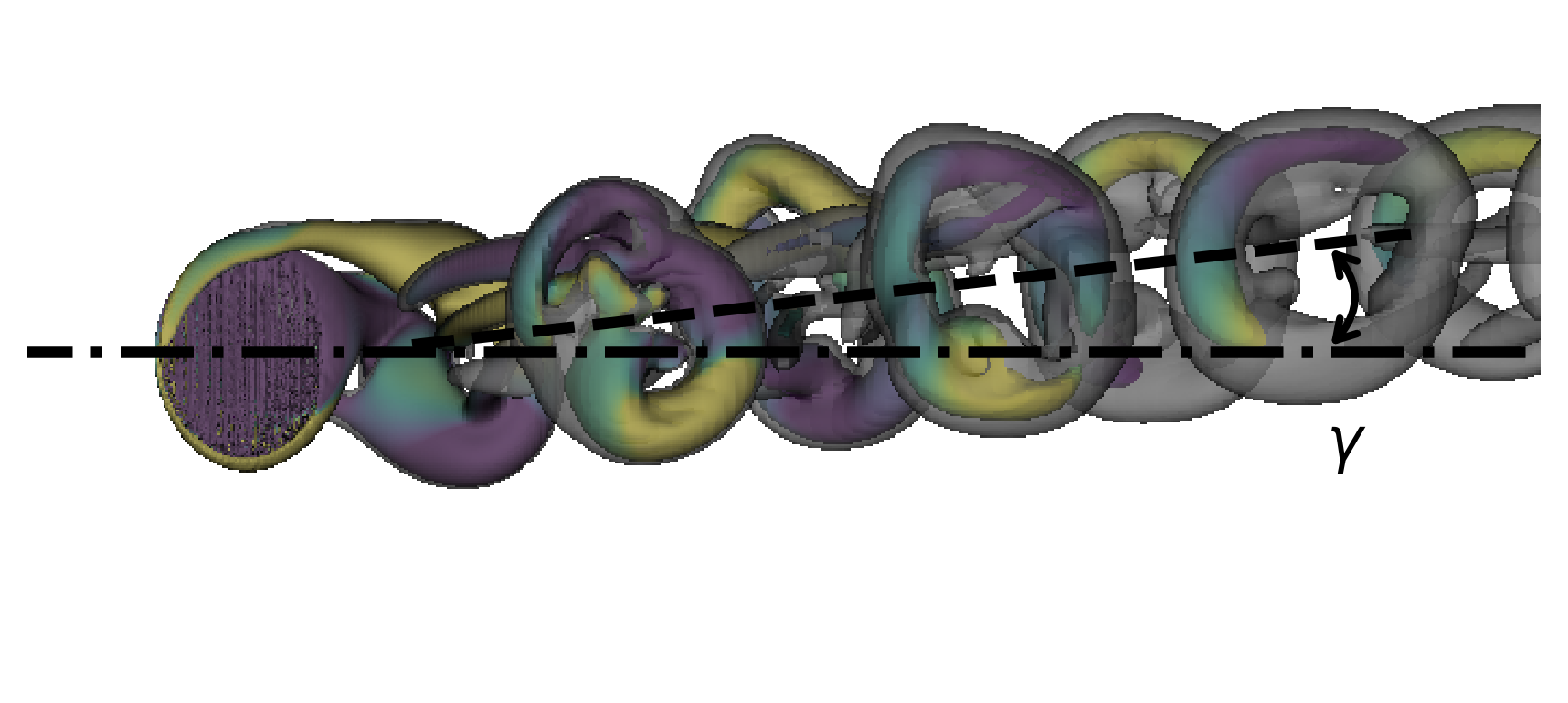}
    \caption{}
    \label{fig:phase_wake_topology:120_top}
  \end{subfigure}
  \caption{Wake topology captured at $t / T = 4.25$ (when the plate is at middle point during the upstroke) for a circular plate ($AR = 1.27$) at Reynolds number $200$, Strouhal number $0.6$, and phase differences $\psi = 100^o$ (top), $110^o$ (center), and $120^o$ (bottom). Wake structures are represented with the $Q$-criterion for $Q = 1$ (gray) and $Q = 6$ (colored with the streamwise vorticity, $-5 \leq w_x \leq 5$). (a,c,e) Lateral view of the wake; (b,d,f) top view. (See Fig.~17 of \citet{li_dong_2016} for comparison.)}
  \label{fig:phase_wake_topology}
\end{figure}

\begin{table}[!h]
  \centering
  \begin{tabular}{ccccc}
    \hline\hline
    \multirow{2}{*}{$\psi$ ($^o$)} &
      \multicolumn{2}{c}{Present} &
      \multicolumn{2}{c}{\citet{li_dong_2016}} \\
    & $\alpha$ ($^o$) & $\gamma$ ($^o$) & $\alpha$ ($^o$) & $\gamma$ ($^o$) \\
    \hline
    $90$ & $18$ & $-1$ & $21$ & $5$ \\
    $100$ & $21$ & $-4$ & $22$ & $-2$ \\
    $110$ & $23$ & $-6$ & $25$ & $-5$ \\
    $120$ & $24$ & $-6$ & $27$ & $-7$ \\
    \hline\hline
  \end{tabular}
  \caption{Effect of the phase difference on the wake oblique angle ($\alpha$) and wake deflection angle ($\gamma$) at $St = 0.6$ and $Re = 200$. For comparison, we also report values published in \citet{li_dong_2016}.}
  \label{tab:phase_angles}
\end{table}

\cref{fig:phase_wake_topology} shows snapshots (lateral and side views) of the wake topology at time $t/T = 4.25$ for phase-difference angles $\psi = 100^o$, $110^o$, and $120^o$.
(\cref{fig:baseline_wake_topology:lateral,fig:baseline_wake_topology:top} shows similar snapshots for $\psi = 90^o$.)
Looking at the top views, we note an increase in the deflection angle $\gamma$ as the phase-difference angle increases.
In other words, larger phase-difference angles lead to larger deflection of the wake from the tip towards the root.
We also note the wake oblique angle $\alpha$ increases with the phase-difference angles.
\cref{tab:phase_angles} reports the wake oblique angles and wake deflection angles computed at time $t/T = 4.25$ for solutions obtained with various phase-difference angles.
Although values for the oblique and deflection angles differ from the original study, the trends are similar.
(Note that we visually estimated these angles from the snapshots; we do not consider these measures to be very accurate but report them to show similar trend as in the original study.)

\begin{figure}[!h]
  \centering
  \includegraphics[width=\textwidth]{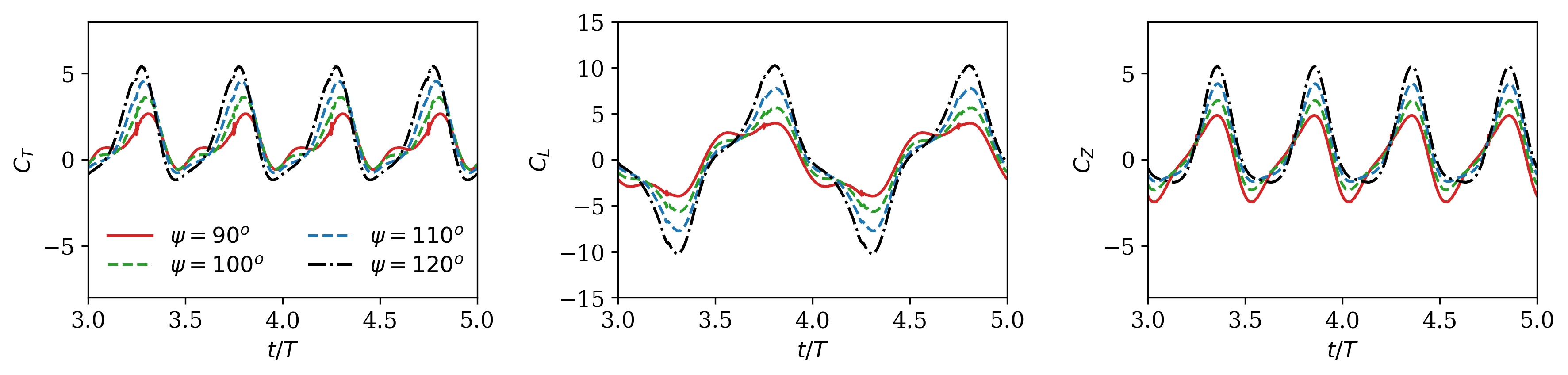}
  \caption{History of the thrust ($C_T$), lift ($C_L$), and spanwise ($C_Z$) coefficients over two flapping cycles for a circular plate ($AR = 1.27$) for various phase-difference angles ($\psi$) between the pitching and rolling motions. Strouhal number is $0.6$ and Reynolds number is $200$ for all these cases. (See Fig.~18 of \citet{li_dong_2016} for comparison.)}
  \label{fig:phase_force_coefficients}
\end{figure}

\begin{figure}[!h]
  \centering
  \includegraphics[width=0.5\textwidth]{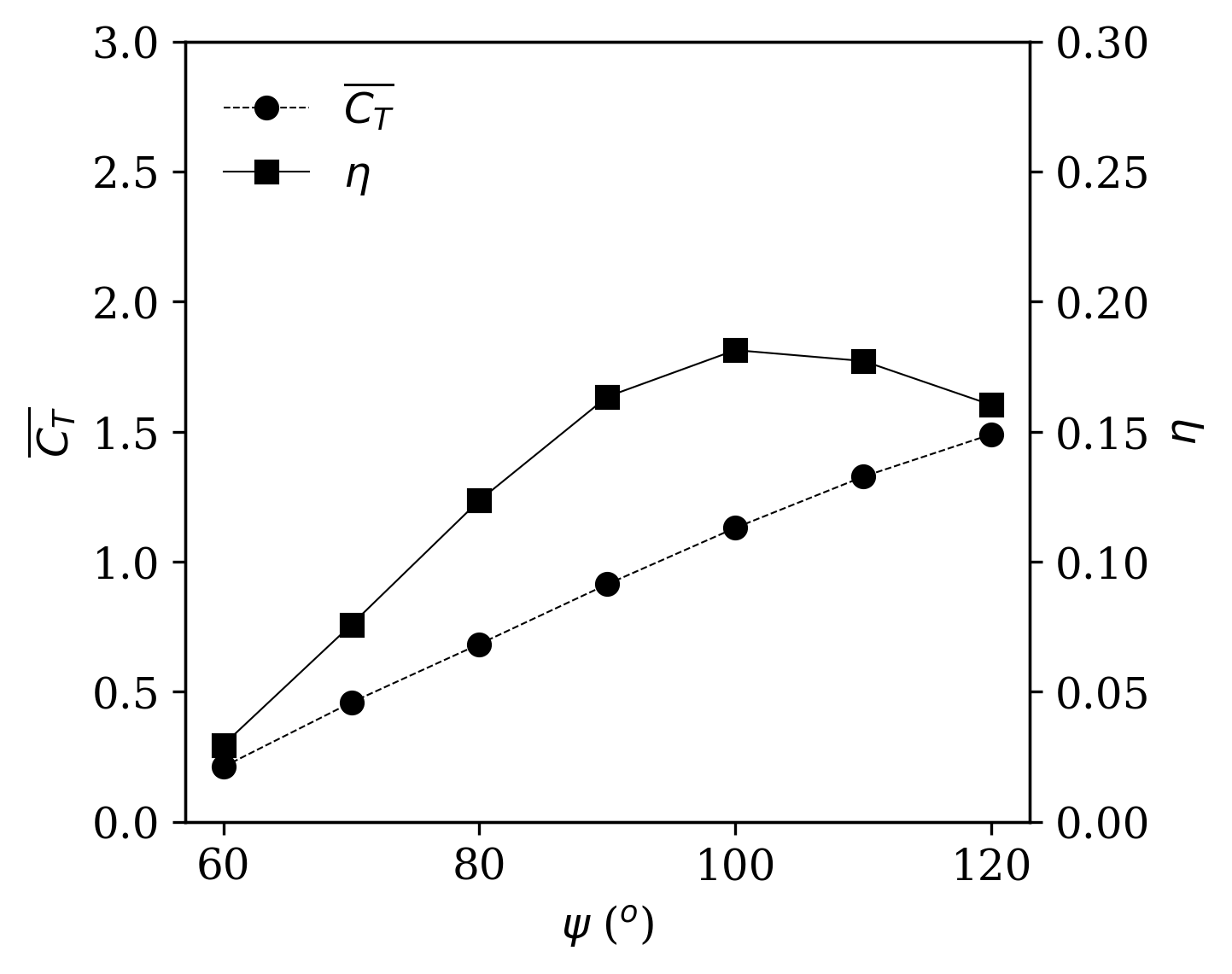}
  \caption{Cycle-averaged thrust coefficient ($\overline{C_T}$) and propulsive efficiency ($\eta$) for the baseline case as functions of the phase difference between the rolling and pitching motions. Data were computed and averaged during the fifth flapping cycle. (See Fig.~19 of \citet{li_dong_2016} for comparison.)}
  \label{fig:phase_efficiency}
\end{figure}

\cref{fig:phase_force_coefficients} shows the history of the force coefficients over two flapping cycles for a circular wing with phase-difference angles between $90^o$ and $120^o$.
We observe an increase in the peak magnitudes for all force components as the angle increases.
This increase coincides with the disappearance of the second peak in the thrust and lift forces that was reported for the baseline case ($\psi = 90^o$).
\cref{fig:phase_efficiency} reports the computed mean thrust coefficients and propulsive performance obtained for angles between $60^o$ and $120^o$.
The mean thrust coefficient increases as we increase the phase-difference angle.
As in the original study, we obtained a maximum propulsive efficiency for the case with $\psi = 100^o$.

\section{Conclusion}

In this study, we replicate the scientific findings published by \citet{li_dong_2016} using our own research software PetIBM,\supercite{chuang_et_al_2018}
which implements a different immersed boundary method than the one in the original study. We observe similar features in the force production, wake topology, and propulsive performance of a pitching and rolling wing.
Although our numerical values do not fully match those from the original study \citet{li_dong_2016}, we obtain the same trends and thus consider this replication attempt to be successful.

A CFD solver typically outputs the solution of primary variables.
For example, PetIBM outputs the pressure and velocity fields, as well as the body forces.
We often use multiple post-processing scripts to generate the final data and figures reported and analyzed in the manuscript; it involves computing secondary data, such as the vorticity field, the aerodynamic power and forces.
If the code is not made available, readers cannot inspect what has been done to produce these data; bugs introduced in these post-processing steps would go undetected.
If no code is available, we cannot explain discrepancies observed between our replication and the original study.
As Donoho and coworkers\supercite{donoho_et_al_2008} once said: ``The only way we'd ever get to the bottom of such a discrepancy is if we both worked reproducibly and studied detailed differences between code and data.''

We made our best efforts to ensure that our replication study is reproducible.
Our computational application makes use of fully open-source tools, and
we created a GitHub repository\footnote{PetIBM-rollingpitching: \url{github.com/barbagroup/petibm-rollingpitching}} for this study.
The repository contains the source code of the PetIBM application, as well as all input files of the simulations reported here, and pre- and post-processing Python scripts.
We adopted a reproducible workflow to run computational simulations; it makes use of Docker images and Singularity recipes to capture the computational environment.
With Singularity, we ran container-based jobs on our university-managed HPC cluster.
The GitHub repository also contains the job-submission scripts that were use to run the simulations on our cluster; they can be adapted to run on other platforms if readers are interested in reproducing our results.

Admittedly, not everyone has access to an HPC cluster with GPU nodes and with Singularity installed.
Lacking those resources, it becomes difficult to fully reproduce our workflow.
However, we made the effort to deposit on Zenodo\footnote{Repro-packs: \url{doi.org/10.5281/zenodo.4732946}} the primary data (directly output from our CFD solver) and post-processing scripts needed to reproduce the figures of the present manuscript.
Once the Zenodo repository is downloaded, readers should be able to spin up a Docker container and run a Bash script to compute the secondary data and generate the figures,
or generate different figures to explore the data in new ways.
The Docker images produced and used for this study are stored on DockerHub\footnote{DockerHub registry: \url{hub.docker.com/repository/docker/mesnardo/petibm-rollingpitching}}, under a basic free subscription.
In the event Docker adopts a policy to automatically purge inactive images (those who have not been recently downloaded) from the Hub, the Dockerfiles are version-controlled on the GitHub repository and can be used to re-build the images.

We spent time engineering a transparent and reproducible workflow to produce the artifacts of this replication study.
Surely, we cannot assert our steps will be fully reproducible in years from now; the software stack could very well become obsolete with new hardware generations.
While the likelihood of the study being reproducible may decrease with the years, the transparency of the steps we took to generate the data shall remain constant.

\hypersetup{linkcolor=black,urlcolor=darkgray}
\renewcommand\emph[1]{{\bfseries #1}}
\setlength\bibitemsep{0pt}
\printbibliography

\end{document}